\documentclass[12pt,a4paper,british]{iopart}
\usepackage{amssymb}
\usepackage{amsgen}
\usepackage{bbold}
\usepackage{color}
\usepackage[T1]{fontenc}
\usepackage[latin9]{inputenc}
\usepackage{babel}
\usepackage{graphicx}
\usepackage{young}
\usepackage[vcentermath]{youngtab}
\usepackage{epstopdf}
\usepackage[numbers,square,sort&compress]{natbib}
\usepackage{array}
\usepackage{verbatim}
\usepackage{amstext}
\usepackage{iopams}
\usepackage{young}
\usepackage{amsthm}
\expandafter\let\csname equation*\endcsname\relax
\expandafter\let\csname endequation*\endcsname\relax
\usepackage{amsmath}
\usepackage{amsfonts}
\usepackage{float}
\usepackage{upgreek}

\makeatletter

\renewcommand{\t}[1]{\textrm{#1}}

\newcommand{\bra}[1]{\langle #1|}
\newcommand{\ket}[1]{|#1\rangle}
\newcommand{\braket}[2]{\langle #1|#2\rangle}

\renewcommand{\t}[1]{\textrm{#1}}

\newcommand{\CH}{\mathcal C^{\t{H}}}
\newcommand{\cf}{\mathcal{X}} 
\newcommand{\CQ}{\mathcal{C}} 

\newcommand{\C}{\mathcal{C}} 
\newcommand{\G}{C}   
\newcommand{\cov}{\Sigma} 
\newcommand{\var}{\theta}   
\newcommand{\bvar}{{\boldsymbol{\var}}}  
\newcommand{\M}{M} 
\newcommand{\mH}{\mathcal{H}}

\newcommand{\gv}[1]{\ensuremath{\text{\boldmath$ #1 $}}}

\newcommand{\X}{\mathbf{X}}

\newcommand{\costs}{\mathcal{C}} 
\newcommand{\priorvar}{{\Delta^2\var}} 
\newcommand{\trace}[0]{\mathrm{Tr}} 
\newcommand{\tracep}[0]{{\mathrm{tr}}} 
\newcommand{\real}{\mathrm{Re}}
\newcommand{\imag}{\mathrm{Im}}

\newcommand{\openone}{\mathbb{1}}

\newcommand{\scal}[1]{\langle{#1}\rangle_{\rho_\bvar}}
\newcommand{\scalno}[1]{\langle{#1}\rangle}

\newcommand{\V}{Z[\bold{X}]}
\newcommand{\Vr}{Z[\bold{X}]}   
\newcommand{\bbr}{{\bold{r}}}
\newcommand{\GG}{\mathcal{C}}
\renewcommand{\ell}{m}

\renewcommand{\succeq}{\geq}

\newcommand{\bA}{\bold A}
\newcommand{\bB}{\bold B}

\newcommand{\lh}{\mathcal{L}} 

\begin{document}
\title{Multi-parameter estimation beyond Quantum Fisher Information}
\author{Rafa{\l} Demkowicz-Dobrza{\'n}ski$^1$, Wojciech G{\'{o}}recki$^1$, M\u{a}d\u{a}lin Gu\c{t}\u{a}$^2$}
\address{$^1$ Faculty of Physics, University of Warsaw, Pasteura 5, PL-02093 Warsaw, Poland}
\address{$^2$ University of Nottingham, School of Mathematical Sciences, University Park, NG7 2RD Nottingham, United Kingdom}

\begin{abstract}
This review aims at gathering the most relevant quantum multi-parameter estimation methods that go beyond the direct use of the Quantum Fisher Information concept. We discuss in detail the Holevo Cram\'er-Rao bound, the Quantum Local Asymptotic Normality approach as well as Bayesian methods. Even though the fundamental concepts in the field have been laid out more than forty years ago, a number of important results have appeared much more recently. Moreover, the field drew increased attention recently thanks to advances in practical quantum metrology proposals and implementations that often involve estimation of multiple parameters simultaneously. Since these topics are spread in the literature and often served in a very formal mathematical language, one of the main goals of this review is to provide a largely self-contained work that allows the reader to follow most of the derivations and get an intuitive understanding of the interrelations between different concepts using a set of simple yet representative examples involving qubit and Gaussian shift models.
\end{abstract}
\maketitle
\section{Introduction}
\subsection{Historical overview and motivation}


From the very beginning of quantum estimation theory \cite{Belavkin1972, Yuen1973, Belavkin1976, Holevo1973, Holevo1977, Helstrom1976, Holevo1982}
the simultaneous estimation of multiple parameters has been seen as a distinguished feature combining classical and quantum aspects of uncertainty. The pioneers of the newly emerging field realized that the non-commutativity of quantum theory lead to non-trivial trade-offs in multi-parameter estimation problems that are not present in classical as well as in single-parameter quantum models.

The introduction of the symmetric logarithmic derivative (SLD) quantum Cram{\'e}r-Rao (CR) bound \cite{Helstrom1967} and the related concept of the Quantum Fisher Information (QFI) may be regarded as the starting point of quantum estimation theory. Soon thereafter it became clear that the extension of the single-parameter SLD CR bound to the multi-parameter scenario cannot account for the potential incompatibility of measurements optimal for extracting information on different parameters.
 This observation led to a development of new multi-parameter bounds including a bound based on the right logarithmic derivative (RLD) \cite{Yuen1973, Belavkin1976} and most notably the
 Holevo Cram{\'e}r-Rao bound (HCR) \cite{Holevo1973}. In parallel,  multi-parameter quantum estimation problems have been
  analysed from a Bayesian perspective obtaining explicit solutions
  in case of some special cost functions and problems enjoying a sufficient symmetry \cite{Helstrom1976, Holevo1982}.

After this `golden age of quantum estimation theory'
 came the `golden age of quantum metrology' with
 the seminal proposal of utilizing non-classical states of light in order to increase the sensitivity of interferometric gravitational wave detectors \cite{Caves1981}. In quantum metrology one no longer assumes that the parameters are encoded in quantum states in a fixed way, but rather considers probe states which evolve under a parameter dependent dynamics and are later measured in order to extract information about the parameters of interest \cite{Giovaennetti2006, Paris2009, Toth2014, Demkowicz2015, Dowling2015, Pezze2018, Pirandola2018, Braun2018, Degen2017}.
 This is an appropriate framework to understand e.g. the potential of utilizing non-classical states of light in optical interferometry,
 but introduces an additional challenge of identifying the input probe state that yields the maximal information about the dynamical parameters.
  The initial studies in quantum metrology focused mainly on performance of particular estimation protocols utilzing standard error propagation formulas and some variants of Heisenberg uncertainty relation as a benchmark \cite{Yurke1986, Xiao1987, Holland1993}.
   Only a few years later, the field eventually incorporated the methods developed earlier by the founders of quantum estimation theory \cite{Sanders1995, Bollinger1996, Huelga1997, Sarovar2006, Shaji2007, Monras2007, Dorner2008}. This was to a large extent due to the paper by Braunstein and Caves \cite{Braunstein1994} which sparked the interest in the QFI as a natural operationally meaningful metric in the space of quantum states.

   Since the most relevant interferometric models considered at that time involved single parameter estimation problems,
 the QFI appeared to be the quantity of choice for the most studies.
 Thanks to its relatively simple structure, it was possible to
 develop efficient computational methods of optimization of optimal input states as well as derivation of universal fundamental bounds on the precision achievable in the most general quantum metrological protocols, not only in idealized noiseless models \cite{Lee2002, Giovaennetti2006} but also in presence of generic uncorrelated noise models \cite{Fujiwara2008, Escher2011, Demkowicz2012, Kolodynski2013, Demkowicz2014, Knysh2014, Demkowicz2017, zhou2018achieving} as well as some models involving noise correlations \cite{Jeske2014, Frowis2014, Layden2019, Chabuda2020}.

 While the quantum metrology field developed both experimentally and theoretically, it became clear that single-parameter models are often an oversimplification of real-life metrological setups \cite{Szczykulska2016}. Simultaneous estimation of phase \emph{and} loss in optical interferometric experiments \cite{Crowley2014, Gessner2018, Proctor2018}, phase and dephasing coefficient in atomic interferometry \cite{Knysh2013, Vidrighin2014}, waveform estimation \cite{Tsang2011}, quantum imaging \cite{Tsang2016, Lupo2016, Chrostowski2017, Rehacek2017, Zhou2019},
 multiple frequency estimation \cite{Gefen2019, Chen2019} or sensing of vector (e.g. magnetic) fields \cite{Baumgratz2016} are all problems that should be modelled within the multi-parameter estimation framework.
 Having a well developed quantum metrological toolbox based on the concept of the QFI at their disposal, researchers
 utilized it to address multi-parameter scenarios.  The main quantity of interest became the QFI matrix
 which helped to obtain a useful insight into a number of multi-parameter problems---see a review paper \cite{Liu2019} which focuses on the properties and use of the QFI matrix  in quantum metrology and beyond. This approach led to satisfactory results provided the issue of a measurement incompatibility was either absent or
 of marginal importance. In general, however, one may arrive at overly optimistic results by just focusing on the properties of the QFI matrix and in order to avoid it a more sophisticated approach may be required.

This prompted a renewed interest in the estimation methods developed over forty years ago, and also led to new theoretical results and tools \cite{Nagaoka1989, D'Ariano.3, Gill2000, masahito2005asymptotic, Bagan2006a, Petz&Jencova, Keyl&Werner, Suzuki2016, Albarelli2019, Albarelli2019a, Tsang2019, Tsang2019a} relevant for further developments in quantum metrology.
An area of significant current interest is that of asymptotic estimation for ensembles of independent, identically prepared systems. Similarly to the classical theory \cite{vanderVaart}, quantum central
limit plays an important role \cite{Hayashi2003} in understanding the statistical model in the limit of large ensembles. This led to the development of quantum local asymptotic normality (QLAN) theory, which provides a precise mathematical framework for describing the Gaussian approximation of multi-copy models \cite{GutaKahn, GutaJanssensKahn, KahnGuta, GutaJencova, KahnGuta2, Gill2011, ButuceaGutaNussbaum, yamagata2013quantum, Yang2019}. The upshot is an adaptive strategy for optimal estimation, with asymptotically normal errors, and a clear understanding of the significance of the HCR bound and its asymptotic achievability--- see also \cite{hayashi2008asymptotic,Yang2019}
for other approaches.

%
%


 This review aims at providing a comprehensive overview of the most important concepts and methods in quantum estimation theory that go beyond the standard  SLD CR bound and the related QFI matrix. Throughout the paper we assume a given quantum statistical model and focus solely on the measurement and estimator optimization problem. Hence, we stay within  the quantum estimation paradigm and do not  discuss the problem of identification of the optimal probe
 states which is a domain of quantum metrology.
 Since our understanding of multi-parameter metrological models is far from complete,
 we hope that collecting the state-of-the-art knowledge on multi-parameter quantum estimation in this review  will allow the reader to get a broader picture of the field as a whole, and appreciate the interrelations between ideas that are often discussed separately.
%
For example, even though the HCR bound has been around for quite a long time, a general understanding of its operational meaning became clearer thanks to QLAN theory as it was linked to
the saturability of the HCR bound for quantum Gaussian shift models. To our best knowledge there is no review that discusses these concepts together in a consistent and detailed way.


  This review has to large extent a self-contained and a bit pedagogical character, as the results we refer to are spread in the literature in publications where the mathematical formalism may sometimes be a challenge to a reader. We make an attempt to illustrate the
  concepts with examples which are chosen to be as simple as possible and yet provide a faithful representation of the
  interrelations between the concepts discussed. In particular, we highlight the examples where the discrepancies between the QFI based predictions and more informative approaches are the most pronounced. Note that recently there have appeared other review papers
  addressing closely related topics including already mentioned \cite{Liu2019} where the main object to interest is the QFI matrix,  which focuses on a geometric aspects of multi-parameter estimation \cite{Sidhu2019} as well as a perspective article focusing on the multi-parameter estimation in the context of quantum imaging \cite{Albarelli2019b}.

\subsection{Quantum estimation framework and notational conventions}
Before proceeding to the discussion of the actual concepts and results, let us first
describe in brief the quantum estimation framework both within the frequentist as well as Bayesian paradigms. This will
allow us to set up the stage as well as fix the notation that will be used throughout this paper.

Consider a family of quantum states $\rho_{\bvar}$ with encoded
values of $\mathfrak{p}$ real parameters which we will represent as a vector $\bvar = [ \theta_1,\dots,\theta_\mathfrak{p}]^T$. These states may be obtained as a result of the application of a $\bvar$ dependent quantum channel $\Lambda_\bvar$  to a fixed input state $\rho$, or simply be prepared by some quantum state preparation device.  A measurement, described by a set of positive operators $\{\M_m\}$  ($\M_m \succeq 0 $, $\sum_m M_m = \openone$) \cite{Nielsen2000}, is then performed on the system yielding a random measurement result $m$ with probability
\begin{equation}\label{eq.p(m)}
p_{\bvar}(m)
= \trace(\rho_\bvar \M_m).
 \end{equation}
 Based on the result $m$, one estimates the parameters using an estimator function $\tilde{\bvar}(m)$. Finally, one needs to specify a cost function $\C(\bvar,\tilde{\bvar}) \geq 0$,
that quantifies the `penalty' for the difference between the estimated value and the true one.
This leads to the final figure of merit representing the average estimation cost (or risk):
\begin{equation}
\mathcal{\C} = \sum_{m} p_\bvar(m) \C(\bvar, \tilde{\bvar}(m)).
\end{equation}
The goal of quantum estimation theory is to find the measurement $\{\M_m\}$ and the estimator $\tilde{\bvar}(m)$ that yield the minimal
average cost.

If $\bvar$ and $\tilde{\bvar}$ are sufficiently close to each other and the cost function is smooth, the latter can be approximated by the quadratic function
$\C(\bvar,\tilde{\bvar}) = (\bvar - \tilde{\bvar})^T \G (\bvar - \tilde{\bvar})$, where $\G$ is the Hessian of the cost function, which we will refer to as the cost matrix.
In this case the average cost can be written as:
\begin{equation}
\CQ = \tracep(\G\cov), \quad \cov = \sum_{m} p_{\bvar}(m) [\bvar - \tilde{\bvar}(m)][\bvar - \tilde{\bvar}(m)]^T,
\end{equation}
where $\cov$ is the covariance matrix of $\tilde{\bvar}$. Note that in order to avoid confusion, we will use $\tracep(\cdot)$ symbol to
denote the trace for matrices acting on the parameter space and the  $\trace(\cdot)$ to denote the trace with respect to the objects acting on the relevant Hilbert space of quantum states.

In the frequentist statistical paradigm the estimated parameter is considered to be unknown but fixed \cite{LehmanCasella1998}.
In order to have a non-trivial pointwise cost minimization problem, one imposes an unbiasedness
constraints on the allowed measurement and estimation strategies
in some region of parameter space $\Theta$:
\begin{equation}
\label{eq:lu1}
 \sum_m p_\bvar(m) \tilde{\bvar}(m) = \bvar, \qquad {\rm for~all~}\qquad \bvar \in \Theta
\end{equation}
or a weaker local unbiasedness (l.u.), which corresponds to the derivative of the above constraint at a fixed parameter value $\bvar=\bvar_0$:
\begin{equation}
\label{eq:lu}
\sum_m p_{\bvar}(m) \tilde{\bvar}(m) = \bvar, \quad \sum_m \boldsymbol{\nabla} p_\bvar(m)\tilde{\bvar}(m)^T = \mathcal{I},
\end{equation}
where $\boldsymbol{\nabla}$ denotes the gradient operator over parameters $\bvar$ while $\mathcal{I}$ is the $\mathfrak{p} \times \mathfrak{p}$ identity matrix---in what follows we use $\mathcal{I}$ to denote the identity in the parameter space, while $\openone$ denotes identity in the Hilbert space of quantum states.

The l.u. conditions assure that the estimator tracks the true value of the parameter faithfully up to the first order around point $\bvar_0$. This excludes pathological estimators, e.g. those which return a fixed value irrespective of the measurement outcome and thus appear to perform well when the true parameter coincides with this particular value. However, it is not clear how to  interpret the l.u. conditions operationally, and moreover the restriction may be regarded as imposing a serious limitation on estimation strategies.

An alternative solution is to consider a broader figure of merit, such as the maximum cost over all parameters $\bvar \in \Theta$. In this context, optimal estimators are called minimax \cite{LehmanCasella1998}. However, the problem of finding explicit minimax procedures is often intractable. Moreover, such estimators may be overly pessimistic with regards to the estimation cost around specific points of the parameter space, where they are outperformed by procedures which take such local information into account.
An even more refined notion of optimal estimator can be defined in the asymptotic setting where a large number of identical copies of the quantum state are available. A \emph{locally asymptotically minimax} cost, which will refer to as $\mathcal{C}_{\t{minmax}}$, captures the hardness of the estimation problem at any fixed point without making any unbiasedness assumptions.

When following the Bayesian approach \cite{RobertBayesianChoice2007} we will be considering the average Bayesian cost
defined as:
\begin{equation}
\label{eq:bayescost}
\overline{\C}  = \int \t{d}\bvar\, p(\bvar)\sum_{m} p_\bvar(m) \C(\bvar, \tilde{\bvar}(m)),
\end{equation}
where $p(\bvar)$ is the prior distribution which encodes our initial knowledge about the parameters to be estimated.
In this case no further constraint of unbiasedeness is imposed, and the task amounts to minimization $\overline{\C} $ over
$\{M_m\}$ and $\tilde{\bvar}(m)$.

The above optimization problems, are very challenging as they deal with optimization over the set of operators $\{M_m\}$ (with unconstrained number of elements) and estimator functions $\tilde{\bvar}(m)$. Furthermore, even if the single-parameter case is feasible in principle, the multi-parameter scenario may introduce further complications. Fortunately, in many cases one may avoid a brute-force optimization approach and either perform the optimization exactly thanks to the symmetry of the model or use  universal asymptotic properties of the problem
to derive informative asymptotic bounds. One of main points of the review is to show that in the asymptotic setting, the optimal estimation problem simplifies and the optimal costs of the different approaches to quantum parameter estimation agree with each other. In order to help the reader follow this review, we provide below an overview of the structure of the paper as well as highlight the most important results that are discussed in particular sections.

\subsection{Structure of the paper and main results}
    In Sec.~\ref{sec:hcr} we provide a comprehensive discussion of CR bounds with the main focus on the HCR bound. We discuss
  its different equivalent formulations, saturability, relation with the standard SLD CR bound
  as well practical ways to compute it. Below we list the main results discussed in this section:
  \begin{enumerate}
  \item{The HCR bound can be numerically computed via a semi-definite program (Sec.~\ref{sec:hcrsd}).}
  \item{In case of a full rank cost matrix $\G$, the HCR is equivalent to the SLD CR bound if and only if  $\trace(\rho_\bvar [L_i,L_j]) = 0$ for all $i,j$, where $L_i$ are the SLDs operators (Sec.~\ref{sec:SLDCR}).}
\item{If the cost matrix $\G$ is rank-one (all parameters except one are nuisance parameters) the SLD CR bound is always saturable (Sec.~\ref{sec:nuisance}).}
\item{The HCR bound is at most two times larger than the SLD CR bound (Sec.~\ref{sec:sldvshcr}).}
\item{In case of $\mathcal{D}$-invariant models the HCR bound conincides with the RLD bound (Sec.~\ref{sec:Dinvariance}).}
\item{The HCR bound is always saturable in case of pure state models,  $\rho_\bvar = \ket{\psi_\bvar}\bra{\psi_\bvar}$, even on the single copy level (Sec.~\ref{sec:saturability}).}
  \end{enumerate}

  Sec.~\ref{sec:examples} contains a detailed discussion of qubit estimation models
  illustrating the measurement incompatibility issue as well as the role of collective measurements in  saturation of asymptotic bounds. The second part of the section is devoted to the estimation theory of Gaussian shift models where the parameters are encoded  linearly in the mean of quantum Gaussian states with fixed covariance. The choice of the examples is intentional as it serves as  a `prelude' for the discussion of the QLAN theorem in Sec.~\ref{sec:qlan}, where these apparently unrelated qubit and Gaussian models are shown to be intimately connected.
  The quantitative results of this section are summarized in Tab.~\ref{tab:examples} where the HCR and SLD CR bounds are computed for all the models discussed. The key messages of this section are:
  \begin{enumerate}
  \item{Qubit models involving estimation of $(\theta,\varphi)$, $(r,\theta)$, $(r,\theta,\varphi)$ manifest respectively: fundamental measurement incompatibility, single copy measurement incompatibility which vanishes in the asymptotic limit and require collective measurements, fundamental measurement incompatibility where saturability of the HCR bound requires collective measurements (Sec.~\ref{sec:examplequbit}).}
  \item{The HCR and the SLD CR bounds for Gaussian shift models can be effectively computed (Sec~\ref{sec:examplegaussian}).}
  \item{The HCR bound is universally saturable for Gaussian shift models via application of linear measurement strategies (Sec~\ref{sec:examplegaussian}).}
  \end{enumerate}

  Sec.~\ref{sec:qlan} contains an overview of the QLAN theory, which shows that an estimation model involving large number of independent and identical copies of a finite dimensional quantum system may be approximated  by a Gaussian shift model, to which it converges in the asymptotic
  limit. The convergence holds for states in a shrinking neighbourhood of a fixed state
  $\rho_{\bvar_0}$ which can be parametrized in the `local' fashion as $\rho_{\bvar_0+ {\bf u}/\sqrt{n}}$, where $n$ is the sample size.
   The section includes a detailed discussion of qubit models as well as general $d$-dimensional models  highlighting the importance of the strong convergence approach in the QLAN which allows one to use the properties of Gaussian models to infer the corresponding
  properties for multiple-copy finite dimensional models in an operational fashion. The key results are:
  \begin{enumerate}
  \item{For pure-state multi-copy models, QLAN can be expressed in terms of the convergence of inner products of local product states towards the corresponding inner product of coherent state of a quantum Gaussian shift model (Sec.~\ref{sec:weakLAN}) }
  \item{The quantum central limit theorem (CLT) offers an intuitive understanding of the emergence of Gaussian shift model in QLAN for arbitrary states (Sec.~\ref{sec.clt.mixed})}
  \item
{The notion of strong convergence replaces the CLT argument with an operational way of comparing the models based on quantum channels, which extends the classical LAN theory developed by Le Cam \cite{LeCam} (Sec~\ref{sec:QLANstrong}). This provides a mathematically rigorous procedure for defining `optimal' (asymptotically locally minimax)  measurements (Sec.~\ref{sec:estimation.strategyLAN}) }
  \item{The key result of the whole section is that the HCR is asymptotically saturable on multiple copies of finite dimensional systems  thanks to the QLAN theorem and the tightness of the HCR for Gaussian shift models. In addition, the optimal estimators has asymptotically Gaussian distribution which allows to construct asymptotically exact confidence regions. }
  \end{enumerate}

  The considerations in the above mentioned sections fit into the frequentist estimation approach. Following this approach, both the HCR bound and the QLAN approaches were shown to be capable of resolving the incompatibility of measurement issue that affects the QFI based quantities.
  However, this approach is less effective in dealing with parameter estimation using finite resources (few copies of a quantum state) and does not take into account prior information about the parameters of interest.

  In order to remedy this, in Sec.~\ref{sec:bayes}, we turn to the Bayesian approach and present the  methods that allow us to obtain solutions
  that suffer from none of the above mentioned deficiencies.  Unfortunately these methods are capable of producing
  rigorous results only for a restricted class of metrological models, whereas in general one may obtain Bayesian CR type bounds
  which, unlike frequentist bounds, take into account the prior information and typically agree with the frequentist bounds in the asymptotic limit. The summary of the main results of this section is given below.
  \begin{enumerate}
  \item{Direct single- to multi-parameter generalization of the analysis of Bayesian models with a quadratic cost function does not yield
  a tight formula for the cost. For Gaussian priors it may be related with the QFI matrix and as such ignores the potential optimal measurement incompatibility issue (Sec.~\ref{sec:bayesquadratic}).}
  \item{For problems with symmetry, covariant measurements are optimal and may significantly simplify the search for a rigorous Bayesian  solution (Sec.~\ref{sec:bayescovariant}).}
  \item{Qubit multicopy models, when analysed using the Bayesian approach, yield  asymptotic formulas equivalent to the HCR bound averaged with the respective prior (Sec.~\ref{sec:bayesianqubit}).}
  \item{Bayesian CR-type bounds may be derived, that in particular show that in general the Bayesian cost may be asymptotically lower bounded by the average HCR bound (Sec.~\ref{sec:bayescr}). }
  \end{enumerate}

Finally, Sec.~\ref{sec:summary} summarizes the paper and provides an outlook on some open problems.

\section{Holevo Cram{\'e}r-Rao bound}
\label{sec:hcr}
\subsection{Classical CR bound}
We start with a brief reminder of the classical CR inequality for a generic statistical model $p_{\bvar}$ with probabilities $\{p_\bvar(m)\}$ depending smoothly on $\bvar\in \Theta \subset \mathbb{R}^{\mathfrak{p}}$. Given a sample from $p_\bvar$ one may lower bound the covariance of any
l.u. estimator $\tilde{\bvar}$ via the following matrix inequality \cite{Kay1993, LehmanCasella1998}
\begin{equation}
\cov\succeq F^{-1},\qquad F=\sum_{\ell} \frac{\bnabla p_{\bvar}(\ell) [\bnabla p_{\bvar}(\ell)]^T}{p_{\bvar}(\ell)},
\end{equation}
where $F$ is the (classical) Fisher Information (FI) matrix of $p_\bvar$ at
$\bvar$---we drop the explicit dependence of $F$ on $\bvar$ for notational compactness.
This implies the following bound on the effective estimation cost for
a given cost matrix $\G$:
\begin{equation}
\CQ = \tracep(\G\cov) \geq  \tracep (\G F^{-1}).
\end{equation}
The following remarks summarise the key features FI and the CR bound.
\begin{enumerate}
\item[(i)]{
The FI is additive for product probability distributions, i.e. if $p_\bvar(\ell_1,\ell_2)=p_{\bvar}(\ell_1)p_\bvar(\ell_2)$ then we have $F_{12} = F_{1}+F_2$. In particular for $n$ independent experiments the corresponding FI is $n$ times larger and
the bound scales inversely proportionally to $n$:  $\cov^{n}\succeq F^{-1}/n$.}
\item[(ii)]{
 If the true parameter $\bvar$ is close to some known value $\bvar_0$, we may look
 for locally unbiased estimators around this point; the following estimator saturates the CR bound and hence is optimal at $\bvar_0$
\begin{equation}\label{eq.one.step.estimator}
\tilde{\bvar}(m) = \bvar_0 + \frac{1}{p_{\bvar}(m)}F^{-1} \left.\bnabla p_\bvar(m)\right|_{\bvar= \bvar_0}.
\end{equation}
However, with the exception of models belonging to the class of exponential family of probability distributions \cite{Fend59}, the estimator will depend explicitly on
$\bvar_0$ and is not optimal away from this point. This drawback can be remedied in a scenario where many independent samples are available, where the following two stage adaptive procedure can be applied \cite{Barndorff2000}:
a `reasonable' preliminary estimator $\bvar_0$ is computed on a subsample, while the remaining samples are used to compute the final estimator by using the above formula. Alternatively, the estimator \eqref{eq.one.step.estimator} can be seen as one step of the Fisher scoring algorithm for computing the maximum likelihood estimator \cite{Demidenko}. }
\item[(iii)]{
If the measurement data consists of $n$ independent and identically distributed (i.i.d.) samples $(m_1,\dots,m_n)$ from $p_\bvar$,  then under mild regularity conditions, the maximum likelihood estimator $\tilde\bvar^{n}_{\t{ML}}(\ell_1,...,\ell_n)\equiv {\rm arg\,max}_\bvar p_\bvar(\ell_1)...p_\bvar(\ell_n)$ is asymptotically unbiased  and achieves the CR bound:
\begin{equation}
\lim_{n\rightarrow\infty}n\cov^{n}_{\t{ML}}=F^{-1},
\end{equation}
where $\cov^{n}_{\t{ML}}$ is the covariance matrix of $\tilde\bvar^{n}_{\t{ML}}$ \cite{LehmanCasella1998, Kay1993}. Most importantly, unlike the estimator discussed in (ii), it is asymptotically normal, performs optimally for all parameter values and depends solely on the observed data and the probabilistic model involved.
As a result it is one of the most widely used estimator in practical applications.}
\end{enumerate}

To summarize: the CR is asymptotically achievable and the optimal cost scales as
$ \tracep(\G F^{-1})/n$, where $n$ is the sample size.
The multi-parameter aspect of the problem does not introduce any additional difficulties compared with the single parameter case apart from the fact that the CR bound involves matrices rather than scalars.

\subsection{Quantum SLD CR bound}
Let us move now to the the quantum case where $p_{\bvar}(\ell)= \trace(\rho_\bvar \M_\ell)$ and the optimization is performed not only over estimators $\tilde\bvar(\ell)$, but also over measurements $\{M_\ell\}$. In this case the covariance matrix of an arbitrary l.u. estimator may be lower bounded by the inverse of the QFI matrix $F_Q$ \cite{Helstrom1976, Braunstein1994, Liu2019}
\begin{equation}
\label{clq}
\cov\succeq F_Q^{-1},\quad F_{Q}=\tfrac{1}{2}\trace(\rho_\bvar \{\mathbf{L},\mathbf{L}^T\}),
\end{equation}
where $\mathbf{L}=(L_1,\dots,L_{\mathfrak{p}})^T$ are SLDs satisfying
\begin{equation}\label{eq:SLD}
\bnabla \rho_\bvar=\frac{1}{2} \{ \mathbf{L}, \rho_\bvar\},
\end{equation}
 and $\{\cdot,\cdot\}$ denotes the anticommutator.
We will refer to this bound as the SLD CR bound, due to the fact that it involves the choice of the SLD as an operator generalization of
the logarithmic derivative.
When the cost matrix $\G$ is given, this implies the following bound on the effective cost:
\begin{equation}
\CQ = \tracep(\G \Sigma ) \geq \tracep(\G F_Q^{-1}) =: \mathcal{C}^{\t{SLD}}.
\end{equation}

Intuitively, QFI quantifies the amount of information about the parameter $\bvar$ potentially available in
a state $\rho_\bvar$. Similarly to the FI, the QFI is additive for models consisting of product states. In particular, for $n$ copies of a quantum system $\rho_\bvar^{\otimes n}$ the corresponding QFI matrix is $n F_Q$.

On a formal level, the issue of saturability of the SLD CR bound amounts to the  question of the existence of a measurement $\{\M_m\}$ for which the corresponding
probabilistic model $p_\bvar(m) = \trace(\rho_\bvar M_m)$ yields the FI matrix $F$ equal to  $F_Q$.
In the single parameter case $\var\in \mathbb{R}$, it may be verified that  the classical  FI corresponding to measuring the SLD operator is equal to the QFI, and hence this measurement is optimal. Although the SLD generally depends on the unknown parameter $\var$, this problem can be addressed by using the two-stage adaptive procedure described in point (ii) above, when a large number of independent copies of the state $\rho_\bvar$ are available.
The achievability of the SLD CR bound for correlated states needs to be treated separately. In particular, in quantum metrology, where the `samples' are typically correlated, an indiscriminate use of the QFI as a figure of merit may lead to some unjustified claims regarding the actually achievable asymptotic bounds \cite{Hall2012, Jarzyna2015, Gorecki2019}.

The multi-parameter case is  in general more involved. If all the SLDs corresponding to different parameters  commute one may saturate the bound by performing  a joint  measurement of the SLDs.
However, if the SLDs do not commute, it may happen that measurements that are optimal for different parameters are fundamentally incompatible. In this case, the measurement minimizing the total cost may strongly depend on a particular cost matrix $\G$.
Therefore, while classically we may say that $\Sigma=F^{-1}$ is the `optimal achievable covariance matrix' (independently on the choice of $C$), in the quantum case different cost matrices may correspond to different optimal covariance matrices, for which in general it might not be possible to say which is larger or smaller as the matrix ordering is only partial.
From that one may see that any fundamental saturable quantum bound cannot have a form of a matrix inequality analogous to Eq.~\eqref{clq}---it needs to be based on the minimization of the scalar cost $\tracep(C\Sigma)$,
as the problem of minimization of  $\Sigma$ itself is ill defined from the very beginning.

An important tool for studying the achievable cost in multi-parameter estimation problems is the HCR bound which is an extension of the SLD CR bound and will be the focus of the following section. In Sec.~\ref{sec:qlan}
we will show how the asymptotically achievability of the HCR bound follows from the general theory of QLAN.



\subsection{Formulation of the HCR bound}
\label{sec:hcrderive}
Among different equivalent formulations of the HCR, we will start with the one that is the most tractable computationally. It lower bounds the cost of a locally unbiased estimator as \cite{Nagaoka1989, hayashi2008asymptotic}
\begin{equation}
\label{HCR}
\CQ = \tracep(\G \cov)\geq\CH =\min_{\bold{X},V}\left(\tracep(\G V)\,\big|\,V\succeq \V, \trace\left(\bnabla \rho_{\bvar} \mathbf{X}^T\right)=\mathcal{I}\right),
\end{equation}
where $\X = [X_1,\dots,X_\mathfrak{p}]^T$ is a vector representing a collection of $\mathfrak{p}$ Hermitian matrices acting on the system's Hilbert space, $V$ is a $\mathfrak{p} \times \mathfrak{p}$ real matrix while $\V = \trace(\rho_\bvar \mathbf{X} \mathbf{X}^T)$ is a $\mathfrak{p} \times \mathfrak{p}$ complex matrix. At a first sight, this bound appears rather technical and not obvious to calculate. Still, as shown later on in the paper not only it can be efficiently calculated, but also plays a fundamental role in the whole quantum estimation theory as it is actually  \emph{the} asymptotically tight bound for general multi-copy estimation models.

\emph{Proof of the HCR bound.}  We present a proof largely based on \cite{Nagaoka1989},
 which provides the necessary intuition required to grasp the physical content of the bound.


For any measurement $\{M_\ell\}$, estimator $\tilde{\bvar}(\ell)$,  and some fixed $\bvar$, we define a vector of Hermitian matrices $\X = [X_1,\dots, X_\mathfrak{p}]^T$:
\begin{equation}
\label{eq:Xi}
\X:=\sum_\ell(\tilde\bvar(\ell)-\bvar)M_\ell.
\end{equation}
If $\tilde{\bvar}(\ell)$ is a l.u.  estimator then by Eqs.~(\ref{eq:lu1},\ref{eq:lu}), the operators $\X$ need to satisfy the conditions
\begin{equation}
\trace(\rho_\bvar \mathbf{X})=0, \quad  \trace(\bnabla \rho_\bvar \mathbf{X}^T)=\mathcal{I}
\end{equation}
at $\bvar=\bvar_0$. If the measurement $\{M_\ell\}$ is projective (i.e. $M_\ell M_{\ell'}=\delta_{\ell\ell'}M_\ell$) then the following equality holds
\begin{equation}
\cov=\sum_{\ell}(\tilde\bvar(\ell)-\bvar)(\tilde\bvar(\ell)-\bvar)^T\trace(M_\ell\rho_{\bvar})=\trace (\rho_\bvar \X \X^T).
\end{equation}
Although for non-projective measurements the equality generally fails, we will now show
that it can be replaced by an inequality.

Let us define an extended  Hilbert space $\mathbb{C}^{\mathfrak{p}}\otimes\mH$, where the Hilbert space  $\mH$ of the system is tensored
with a $\mathfrak{p}$ dimensional space of parameters. Consider a linear operator on
$\mathbb{C}^{\mathfrak{p}}\otimes\mH$

\begin{equation}
{\bf R} := \sum_\ell
[(\tilde\bvar(\ell)-\bvar)\openone-\X]
M_\ell
[(\tilde\bvar(\ell)-\bvar)\openone-\X]^T,
\end{equation}
which by construction is a positive operator.
This implies that the following partial trace (in accordance with our previous convention $\trace$ in the formulas that follow denotes the trace over $\mathcal{H}$ only) is also positive
\begin{align}
&\trace (( \mathcal{I}\otimes\rho_{\boldsymbol{\var}}) {\bf R}) =
\sum_{\ell}(\tilde\bvar(\ell)-\bvar)\trace(M_\ell\rho_{\bvar}) (\tilde\bvar(\ell)-\bvar)^T
\nonumber
\\
&- \trace\bigg[\rho_\bvar\bigg(
\sum_\ell(\tilde\bvar(\ell)-\bvar)M_\ell \X^T+\X\sum_\ell(\tilde\bvar(\ell)-\bvar)^T M_\ell-\X\sum_\ell M_\ell \X^T\bigg)\bigg]
\nonumber
 \\
&=\cov-\trace(\rho_\bvar \X \X^T) \succeq 0,
\end{align}
where in last step we have used \eqref{eq:Xi} and the identity $\sum_\ell M_\ell=\openone$.
Hence we arrive at the following matrix inequality which holds for any measurement $M_\ell$
\begin{equation}
\label{matrixineq}
\cov\succeq \V,   \quad \V = \trace(\rho_\bvar \mathbf{X} \mathbf{X}^T),
\end{equation}
where $\V$ is a  $\mathfrak{p}\times \mathfrak{p}$ Hermitian matrix.
Now, we can trace the above inequality with a given cost matrix $\G$ to obtain scalar inequality
 \begin{equation}
 \label{eq:ineqcost}
 \C = \tracep(\G \cov ) \geq \tracep(\G \V)
 \end{equation}
 and since the above  depends on the measurement and estimators only via $\mathbf{X}$,
  we will obtain a universally valid bound if we minimize the r.h.s. over $\mathbf{X}$ keeping in mind the l.u. conditions
   $\trace(\rho_\bvar \mathbf{X})=0$, $\trace\left(\bnabla\rho_{\bvar}\mathbf{X}^T \right)=\mathcal{I}$. Note that without the
 l.u. conditions we would get a trivial bound
 $\tracep(\G \cov) \geq 0$.

This procedure, however, is not in general the
optimal way to obtain a scalar inequality from a matrix inequality \eqref{matrixineq}. Since $\V$ is in general a complex matrix, application of the  trace with the real symmetric cost matrix $\G$ causes the information hidden in the imaginary part of $\V$
to be lost. To remedy this issue, we may introduce a $\mathfrak{p} \times \mathfrak{p}$ real matrix $V$ satisfying $V\succeq \V$ and we
end up with the stronger HCR bound:
\begin{equation}
\label{HCR}
  \CQ = \tracep(\G \cov)\geq\CH :=\min_{\X,V}\left(\tracep(\G V)\big|V\succeq \V, \, \trace\left(\bnabla \rho_{\bvar}\mathbf{X}^T\right)=\mathcal{I}\right),
\end{equation}
where we have kept only the second of the previously mentioned l.u. conditions, as the first condition may be dropped without affecting the result. To see this, let $\trace(\rho_\bvar \mathbf{X})=\mathbf{c}$. Then we may redefine $\tilde{\mathbf{X}} = \mathbf{X} - \mathbf{c} \openone$, for which the first l.u. condition is satisfied and at the same time the second l.u. condition is not affected. Finally, such a transformation will  also lower the r.h.s. of \eqref{eq:ineqcost} (by the standard argument involving the inequality between the variance and the second moment) and hence the result of minimization with or without the first l.u. condition is the same.
\qed

\subsection{Numerical evaluation}
\label{sec:hcrsd}
There are a number of equivalent formulation of the HCR bound \cite{hayashi2008asymptotic} but before presenting them let us discuss an efficient numerical algorithm for calcualting the HCR bound which is based on the above formula. Interestingly, despite the well established position of the HCR bound in the quantum estimation literature, an explicit formulation of the algorithm which allows to efficiently calculate the HCR bound numerically in terms of a linear semi-definite program was proposed only recently \cite{Albarelli2019}.

In order to write the HCR bound as a linear semi-definite program one needs to express the condition $V\succeq \V$ in a way that it is linear in both $V$ and $X_i$.  Let $\{\Lambda_a\}$ be a  basis of $\lh(\mH)$ (Hermitian operators acting on $\mH$), orthonormal according to the Hilbert-Schmidt inner product, i.e. $\trace(\Lambda_a\Lambda_b)=\delta_{ab}$. We may now represent matrices $X_i$ and $\rho_\bvar$ as vectors of coefficients $\boldsymbol{x}_i,\boldsymbol{s}_\bvar\in\mathbb R^{(\dim\mH)^2}$ with respect to the basis $\{\Lambda_a\}$. Since $\trace(X_iX_j\rho_{\bvar})$ may be seen as a non-negative defined bilinear form on $\lh(\mH)$, it may also be written as:
\begin{equation}
\trace(X_iX_j\rho_{\bvar})=\boldsymbol{x}_i^TS_\bvar \boldsymbol{x}_j=\boldsymbol{x}_i^TR_\bvar^\dagger R_\bvar \boldsymbol{x}_j,
\end{equation}
where $S_\bvar$ is a positive semi-definite matrix and
$R_\bvar$ is an arbitrary matrix satisfying $S_\bvar=R_\bvar^\dagger R_\bvar$ (e.g. the Cholesky decomposition).
Note, that according to the above formula $S_{\bvar}$ and $\boldsymbol{s}_\bvar$ are related
\begin{equation}
\left(S_{\bvar}\right)_{a b}  = \sum_c \trace(\Lambda_a \Lambda_b \Lambda_c) (\boldsymbol{s}_\bvar)_c.
\end{equation}
 Introducing $\bold{x}=[\boldsymbol{x}_1,...,\boldsymbol{x}_{\mathfrak{p}}]$ (no transposition here is intentional) we may rewrite the above equality in a compact way
\begin{equation}
\V = \bold{x}^T R_\bvar^\dagger R_\bvar \bold{x},
\end{equation}
where $R_\bvar \bold{x} = [R_\bvar\boldsymbol{x}_1,...,R_\bvar\boldsymbol{x}_{\mathfrak{p}}]$ is in fact
a $(\dim \mH)^2 \times \mathfrak{p}$ matrix.
Then, we use a general fact that for any matrices $A,B$ the following are equivalent
\begin{equation}
A-B^\dagger B\succeq 0 \quad \Longleftrightarrow\quad
\begin{bmatrix}
A& B^\dagger\\
B& \openone
\end{bmatrix}\succeq 0
\end{equation}
so that we may rewrite \eqref{HCR} as a linear semi-definite problem:
\begin{equation}
\min_{V,\bold{x}}\tracep(\G V),\quad
{\rm subject~to:}
\begin{bmatrix}
V&  \bold{x}^T R_\bvar^\dagger \\
 R_\bvar \bold{x} & \openone
\end{bmatrix}\succeq 0,\quad \boldsymbol{x}_i^T\frac{\partial \boldsymbol{s}_\bvar}{\partial\var_j}=\delta_{ij},
\end{equation}
where $\openone$ is the identity on $\mathbb{R}^{(\dim \mH)^2}$.
The above semi-definite program may be easily implemented numerically.

\subsection{Equivalent formulations of the HCR bound}
\label{sec:hcrequivalent}
Below we show that for a given $\mathbf{X}$ minimization over $V$ in \eqref{HCR} may be performed directly. This leads us to a more explicit form of the HCR bound \cite{Nagaoka1989, hayashi2008asymptotic}. However, even though this form appears more informative from an analytical point of view, at the same time it is less suitable for  numerical implementation.

First, for any cost matrix $\G$, the inequality $\sqrt{\G}V\sqrt{\G}\succeq \sqrt{\G}\V\sqrt{\G}$ is still valid after transposition operation is applied $\sqrt{\G}V\sqrt{\G}\succeq\sqrt{\G}\V^T\sqrt{\G}$. For Hermitian matrices the transposition operation leaves the real part of the matrix unchanged and changes the sign of the imaginary part. Therefore, given any column vector $\gv v_i$, these two inequalities lead to
\begin{equation}
\label{step1}
\gv v_i^T\sqrt{\G}\left(V-{\rm Re}\V\right)\sqrt{\G}\gv v_i \geq\gv v_i^T\left(\pm i\sqrt{\G}\imag \V \sqrt{\G}\right)\gv v_i.
\end{equation}
By summing over vectors $\gv v_i$, which form the eigenbasis of $ i\sqrt{\G}\imag \V \sqrt{\G}$, we get a trace variant of the above inequality:
\begin{equation}
\label{step2}
\tracep(\G V)\geq \tracep(\G \real\V)+\tracep(|\sqrt{\G}\imag \V\sqrt{\G}|),
\end{equation}
where the absolute value of an operator $|B|:=\sqrt{B^\dagger B}$ appears as a result of $\pm$ on the r.h.s. of the inequality.
The last inequality may always be saturated by taking
$V=\real \V +\sqrt{C^{-1}}|\sqrt{C}\imag \V\sqrt{C}|\sqrt{C^{-1}}$.
As a result the HCR bound may be written equivalently as~\cite{Nagaoka1989, hayashi2008asymptotic}:
\begin{equation}
\label{HCRtracenorm}
\CH:=\min_{\X}\left(\tracep(\G \real \V)+\|\sqrt{\G}\cdot {\rm Im}\V\cdot\sqrt{\G}\|_1
\ \big| \
\trace\left(\bnabla \rho_{\bvar}\mathbf{X}^T\right)=\mathcal{I}\right),
\end{equation}
where $\|B\|_1:=\tracep(|B|)$ is the trace norm.
The last term is often written in literature as  $\tracep[{\rm abs}(\G\cdot {\rm Im}\V)]$ \cite{Suzuki2016,Holevo1982,hayashi2008asymptotic},  where $\tracep(\t{abs}(\cdot))$ is the sum of absolute values of eigenvalues, and note that for non-Hermitian matrices is not the same as $\tracep| \cdot|$.

Finally, we present yet another formulation of the HCR bound, originally proposed by Matsumoto only for the pure states \cite{matsumoto2002new}, and here generalized to arbitrary density matrices.
This formulation has proven particulary suitable when discussing saturability of pure state models, as shown in  Sec.~\ref{sec:saturability} and, moreover,  it has been successfully employed in designing the optimal quantum error correction protocols in multi-parameter quantum metrology \cite{Gorecki2019}.

The essential feature that makes the HCR bound stronger than the SLD CR bound, but at the same time makes this bound harder to compute is the fact that $\V$ may be complex. This is related with incompatibility of measurements which are optimal from the point of view of estimation of different parameters.

This issue may be approached by formally considering matrices  $Y_i\in\lh(\mathcal H\oplus \mathbb{C}^{\mathfrak{p}})$
acting on a properly extended space  instead of $X_i\in\lh(\mathcal H)$, but with an additionally restriction $\imag Z[\bold{Y}]=0$, which reflects the requirement that the measurements on this extended subspace will no longer suffer from the incompatibility issue.

Let us decompose $Y_i\in\lh(\mathcal H\oplus \mathbb{C}^\mathfrak{p})$ into $Y_i=X_i+\tilde{X}_i$, where $P_{\mathcal H}X_i P_{\mathcal H}=X_i$  and $P_{\mathcal H}\tilde{X}_i P_{\mathcal H}=0$ ($P_{\mathcal H}$ is the projection onto $\mathcal H$).
Now, one can see that using this decomposition we have $Z[\bold{Y}]=Z[\bold{X}+\bold{\tilde{X}}]=
Z[\bold{X}]+Z[\bold{\tilde{X}}]$ and since both $Z[\bold{X}]$ and $Z[\bold{\tilde{X}}]$ are positive semi-definite, then
$V\succeq Z[\bold{X}+\bold{\tilde{X}}]$ implies $V\succeq Z[\bold{X}]$.
Therefore, for any  fixed $\bold{X}$ we have:
\begin{equation}
\min_{V,\mathbf{\tilde{X}}:{\rm Im}Z[\bold{X}+\bold{\tilde{X}}]=0}\left(\tracep(\G V):V\succeq Z[\bold{X}+\bold{\tilde{X}}]\right)\geq\min_{V}\left(\tracep(\G V):V\succeq Z[\bold{X}]\right).
\end{equation}
The above inequality will be saturated if we find $\bold{\tilde{X}}$ satisfying
\begin{equation}
Z[\bold{\tilde{X}}]=\sqrt{\G^{-1}}|\sqrt{\G}{\rm Im}Z[\bold{X}]\sqrt{\G}|\sqrt{\G^{-1}}-i{\rm Im}Z[\bold{X}].
\end{equation}
Indeed, such  $\bold{\tilde{X}}$ always exist as the r.h.s. is a positive semi-definite matrix and in general for any positive semi-definite
$\mathfrak{p}\times \mathfrak{p}$ matrix $A$ there exists $\bold{\tilde{X}}$, such that $Z[\bold{\tilde{X}}]=A$. To see this let $A=\sum_{k=1}^\mathfrak{p} a_k\ket{a_k}\bra{a_k}$ and let $\ket{\lambda}$ be an arbitrary non-zero eigenvector of $\rho_\bvar$. Consider $\tilde{X}_i$ of the form $\tilde{X}_i=\sum_{k=1}^\mathfrak{p} \frac{1}{\sqrt{\lambda}}(\bar{\alpha}_{ik}\ket{k}\bra{\lambda}+\alpha_{ik}\ket{\lambda}\bra{k})$,
where $\ket{k}$ is a basis in $\mathbb{C}^\mathfrak{p}$---note that this operator satisfies the requirement
$P_{\mathcal H}\tilde{X}_i P_{\mathcal H}=0$. Then
$Z[\bold{\tilde{X}}]=\sum_{k=1}^\mathfrak{p}
\begin{bmatrix}
\alpha_{1k}&
\cdots&
\alpha_{\mathfrak{p}k}\\
\end{bmatrix}^T
\cdot
\begin{bmatrix}
\bar{\alpha}_{1k}&\hdots&\bar{\alpha}_{\mathfrak{p}k}\\
\end{bmatrix}.$
 Setting $[\alpha_{1k},..,\alpha_{\mathfrak{p}k}]^T=\sqrt{a_k}\ket{a_k}$ we have $Z[\bold{\tilde{X}}]=A$.
 This all implies that the HCR may be alternatively formulated as:
\begin{equation}
\label{matsu}
\CH:=\min_{Y_i\in\lh(\mH\oplus\mathbb C^\mathfrak{p})}\left(\tracep(\G Z[\bold{Y}])\ \big| \  \trace\left(\bnabla \rho_{\bvar}\mathbf{Y}^T\right)=\mathcal{I}, \imag Z[\bold{Y}]=0\right).
\end{equation}

\subsection{Relation with the standard SLD CR bound}
\label{sec:SLDCR}
While deriving the HCR bound in Sec.~\ref{sec:hcrderive} we have
mentioned that the bound \eqref{eq:ineqcost}, obtained naively by applying $\tracep(\G\cdot)$ to the matrix inequality \eqref{matrixineq}, is in general not the optimal way to obtain a scalar bound from a matrix inequality.
If, nevertheless, we  pursue this line of derivation, it turns out that the  bound
corresponds exactly to the standard SLD CR bound $\mathcal{C}^{\t{SLD}}$:
\begin{equation}
\label{HCRSLD}
\mathcal{C}^{\t{SLD}}=\min_{\X}\left(\tracep(\G \V)\ \big|\ \trace
\left(\bnabla \rho_{\bvar}\mathbf{X}^T\right)=\mathcal{I}\right).
\end{equation}
In order to prove this fact, and also establish the relation between the SLD CR bound and the HCR bound, we need to introduce some more mathematical tools.

Any Hermitian matrix $X$ acting on $\mH$ may be written down using following block structure:
\begin{equation}
X=
\begin{bmatrix}
X^\t{R}& X^{\t{RK}}\\
X^{\t{KR}}& X^\t{K}
\end{bmatrix}
\end{equation}
where $X^{\t{R}}\in\lh({\rm Range}( \rho_\bvar))$ and $X^{\t{K}}\in\lh({\rm Ker} (\rho_\bvar))$,
 where $\t{Range}$ and $\t{Ker}$ denote the range and the kernel of an operator. Since $X_i^{\t{K}}$ does not affect $\trace(X_iX_j\rho_\bvar)$, we may restrict ourselves to the subspace of matrices $X$ for which $X^{\t{K}}=0$---more formally we deal with elements of the space $\lh(\mH)/\lh({\rm Ker}(\rho_\bvar))$ (which is \emph{not} equivalent to $\lh({\rm Range}(\rho_\bvar))$, as off-diagonal blocks $X^{\t{RK}},X^{\t{KR}}$ are still important here). We define a scalar product on this subspace:
\begin{equation}
\label{eq:scal}
\scal{X,Y}:=\trace\left(\rho_\bvar\tfrac{1}{2}\{X,Y\}\right),
\end{equation}
for which the l.u. condition take a very concise form:
\begin{equation*}
\scal{\mathbf{L},\mathbf{X}^T}=\mathcal{I}.
\end{equation*}
In particular, it means that if we write $X_i=X_i^{\parallel}+X_i^{\perp} \in\lh(\mH)$, where $X_i^{\parallel}\in{\rm span}_{\mathbb R}\{L_1,...,L_\mathfrak{p}\}$ and $X_i^{\perp}\perp{\rm span}_{\mathbb R}\{L_1,...,L_\mathfrak{p}\}$, then the l.u. condition implies that the parallel part is $\mathbf{X}^{\parallel}=F_Q^{-1} \mathbf{L}$ and there is no restriction for $\mathbf{X}^\perp$.
Next, one may see that:
\begin{equation}
\real(\V)=\scal{\mathbf{X},\mathbf{X}^T} =\real Z[\X^\parallel]+\real[\X^\perp],
\end{equation}
where
\begin{equation}
\real Z[\X^\parallel]=\real\trace[\rho_\bvar  F_Q^{-1} \mathbf{L} \mathbf{L}^T (F_Q^{-1})]=
F_Q^{-1}\real\trace(\rho_\bvar \mathbf{L} \mathbf{L}^T) F_Q^{-1}=F_Q^{-1}.
\end{equation}
From that it is clear that in order to minimize \eqref{HCRSLD} one should choose $\X^\perp=0$ and then the SLD CR bound is recovered.
The HCR bound may now be rewritten in the form:
\begin{multline}
\label{eq:hcrperp}
\CH:=\mathcal{C}^{\t{SLD}}+
\min_{\X^\perp}\left(\tracep(\G \real Z[\X^\perp])+\|\sqrt{\G}\cdot {\rm Im}Z[\X^\perp+\X^\parallel]\cdot\sqrt{\G}\|_1
\ \big| \
\mathbf{X}^\parallel=F_Q^{-1} \mathbf{L}\right).
\end{multline}
We see that the HCR bound is identical to the SLD CR bound if and only if $\sqrt{\G}\imag Z[\X^\parallel]\sqrt{\G}=0$, which
for full rank $\G$ is equivalent to:
\begin{equation}
\trace(\rho_\bvar [L_i,L_j])=0, \qquad {\rm for~all~} i,j.
\end{equation}
While this last condition has appeared  in a number of papers \cite{Genoni2013a, Vidrighin2014, Crowley2014, Suzuki2016}, the fact that this is indeed  a necessary \emph{and} sufficient  condition for the equality between the SLD CR and the HCR bounds was not obvious and it was stated explicitly in \cite{Ragy2016}.

\subsection{Scalar function estimation in the presence of nuisance parameters}
\label{sec:nuisance}
In quantum state tomography the usual figure of merit is derived from a proper distance function on quantum states,
whose quadratic approximation has a strictly positive cost matrix $\G$. Here, we look in more detail at the opposite situation where $\G$ is a rank-1 matrix, so that $\G = \mathbf{c}\mathbf{c}^T$ for some real valued vector $\mathbf{c}$. This occurs when the aim is to estimate a particular scalar function of the parameter, even though one deals with a multidimensional parameter manifold; locally, the parameter can be separated in the component along $\mathbf{c}$ which needs to be estimated, and other components which are regarded as nuisance parameters;
see \cite{Suzuki2019, Suzuki2019a, Tsang2019, Yang2019} for a more general discussion of estimation in presence of nuisance parameters. The setup is also related to that semi-parametric estimation, where the estimation problem is often non-parametric (i.e. infinite dimensional parameter as in homodyne tomography of a cv state) but we are interested in a finite dimensional function of the parameter (e.g. the expectation value of certain observables). This setup is also relevant for the distributed sensing scenarios \cite{Ge2018, Sekatski2019},
interferometry \cite{Jarzyna2012}, field gradient sensing \cite{Altenburg2017, Apellaniz2018} and many others.

%


Even though this may appear as a single parameter estimation problem, the uncertainty about the nuisance parameters  leaves multi-parameter hallmark on the solution. Nevertheless, the argument below shows that this effect is fully captured by the SLD CR bound as in this case the HCR and the SLD CR bound coincide. To see this let us inspect the HCR bound in the form \eqref{HCRtracenorm} and notice that
\begin{equation}
i\sqrt{C}  \t{Im}\V \sqrt{C}  \propto \mathbf{c}  \mathbf{c}^T (i \t{Im}\V) \mathbf{c} \mathbf{c}^T.
\end{equation}
Since $\V$ is a hermitian matrix, $i \t{Im}\V$ is a purely imaginary Hermitian matrix and  the expectation
$\mathbf{c}^T (i \t{Im}\V) \mathbf{c}$ is equal to zero for any real vector ${\bf c}$.
By comparing with formula \eqref{HCRSLD} we conclude that
\begin{equation}
\CH = \C^{\t{SLD}},   \quad \t{for } \G = \mathbf{c}\mathbf{c}^T.
\end{equation}

\subsection{Maximal discrepancy between the SLD and the HCR bounds}
\label{sec:sldvshcr}
Interestingly, while the HCR bound is in general tighter than the SLD CR bound it will at most provide a factor of $2$ improvement over the SLD CR bound---a simple fact that has not been pointed out explicitly until very recently \cite{Albarelli2019a, Carollo2019} (see also
\cite{Tsang2019a} were a weaker bound was derived). This can be shown as follows.  For any $\X$ the matrix $\V$ is positive semi-definite.
 Now, adopting the reasoning that led to equation \eqref{step2}, namely: start with $\sqrt{\G}\V  \sqrt{\G} \succeq 0$; take the transpose $\sqrt{\G} \V^T \sqrt{\G}\succeq 0$; add and subtract the two inequalities; separate the real and imaginary parts and take the trace on both sides;  we arrive at:
 \begin{equation}
\V\succeq 0\Rightarrow\tracep(\G \real \V)\geq\|\sqrt{\G}\cdot {\rm Im}\V\cdot\sqrt{\G}\|_1.
\end{equation}
Next, applying it to the second formulation of the HCR bound \eqref{HCRtracenorm} and using \eqref{HCRSLD}:
\begin{equation}
\CH=\min_{\X}\tracep(\G \real \V)+\|\sqrt{\G}\cdot {\rm Im}\V\cdot\sqrt{\G}\|_1
\leq 2\min_{\X}\tracep(\G \real \V)=2 \mathcal{C}^{\t{SLD}}
\end{equation}
we prove the statement.

Since, as will be discussed further on, the HCR bound is asymptotically saturable on many copies, the factor of $2$ represents the maximal asymptotic impact that measurement incompatibility can have on the optimal estimation of multiple parameters. This factor can also be understood from the perspective of the QLAN theory discussed in Sec.~\ref{sec:qlan}. Indeed, QLAN shows that the quantum estimation problem with many identical copies is asymptotically equivalent to estimating the
mean in a Gaussian shift model. The factor $2$ stems from the fact that in a Gaussian shift model, one can group the coordinates of the cv system into two families (positions and momenta of individual modes) such that the means of each family can be estimated optimally by simultaneously measuring all coordinates in the family. We will come back to this point in Sec.~\ref{sec:qlan}.


%
%
%

\subsection{$\mathcal{D}$-invariance and the RLD CR bound}\label{sec:Dinvariance}
Using the notations and the concept of scalar product introduced in Sec.~\ref{sec:SLDCR}, the real part of $\V$ and the l.u. conditions read $\real\V=\scal{\mathbf{X},\mathbf{X}^T}$ and $\scal{\mathbf{L},\mathbf{X}^T}=\mathcal{I}$. In order to write the imaginary part $\imag\V$ in a analogous way let us introduce a commutation superoperator $\mathcal{D}$ \cite{Holevo1982, Gill2011, hayashi2008asymptotic,yamagata2013quantum} satisfying\footnote{Its existence and uniqueness may be shown using the eigenbasis of $\rho_\bvar$: $\braket{i|\mathcal D(X)}{j}=\frac{i(\rho_{ii}-\rho_{jj})}{\rho_{ii}+\rho_{jj}}\braket{i|X}{j}$. Here we use the definition introduced in \cite{yamagata2013quantum}, which differs from the one from \cite{Holevo1982} by a factor $2$.}:
\begin{equation}
\label{eq:dinv}
\{\mathcal D (X),\rho_\bvar\}=i[X,\rho_\bvar],\quad \mathcal D (X)\in    \lh(\mH)/\lh({\rm ker}\rho_\bvar).
\end{equation}
Then we have:
\begin{equation}
\label{eq:Ddecomp}
\trace(\rho_\bvar X_i X_j)=\scal{X_i,X_j}+i\scal{\mathcal D(X_i),X_j}.
\end{equation}
Now we will prove that when looking for the optimal $X_i$ we may always restrict ourselves to $X_i$ which belong to the subspace $\mathcal{T}\subseteq\lh(\mH)/\lh({\rm ker}\rho_\bvar)$, which is the smallest $\mathcal D$-invariant subspace containing $\t{span}_{\mathbb{R}}\{L_1,...,L_\mathfrak{p}\}$; in other words this is a subspace obtained by sequential actions of $\mathcal{D}$ starting with operators from $\t{span}_{\mathbb{R}}\{L_1,...,L_\mathfrak{p}\}$. Let us denote by $P(X)$  and $P^\perp(X)$ the orthogonal projections of an operator $X$ onto respectively $\mathcal{T}$ and its orthogonal complement $\mathcal{T}^\perp$. According to \eqref{eq:Ddecomp} we can write:\begin{equation}
\trace(\rho_\bvar P(X_i)P^\perp(X_j))=\scal{P(X_i),P^\perp(X_j)}+i\scal{\mathcal D(P(X_i)),P^\perp(X_j)}.
\end{equation}
The first term on the r.h.s. is zero by definition of
$P(X)$  and $P^\perp(X)$. The second term on the r.h.s is zero as well since $\mathcal{T}$ is $\mathcal{D}$-invariant and hence
$\mathcal{D}(P(X_i)) \in \mathcal{T}$.
As a result $\trace(\rho_\bvar P(X_i)P^\perp(X_j))=0$.
Thanks to this we have
\begin{equation}
Z(\bold{X})=Z(P(\bold{X}))+Z(P^{\perp}(\bold{X}))\geq Z(P(\bold{X})).
\end{equation}
Now, since  $\mathcal{T}$ subspace contains
$\t{span}_{\mathbb{R}}\{L_1,...,L_\mathfrak{p}\}$ operators,
 then if the l.u. condition $\scal{\mathbf{L},\mathbf{X}^T}=\mathcal{I}$ is satisfied for $\mathbf{X}$ then it is also satisfied for
 $\scal{\mathbf{L},P(\mathbf{X})}=\mathcal{I}$.  Therefore, projecting onto $\mathcal{T}$ is always advantageous in performing the minimization. This proves that we may restrict to tuples ${\bf X}$ having all components in $\mathcal{T}$.
 Note, that since $\trace{(\bnabla \rho_\bvar)}=0$ then $\scal{\mathbf{L},\openone} = 0$,
 and this equality remains unchanged under the action of $\mathcal{D}$ operator on $L_i$. As a result, $ \scal{X,\openone}=0$ for all ${X\in \mathcal{T}}$.

In particular, if $\mathcal{T}={\rm span}_{\mathbb R}\{L_1,...,L_\mathfrak{p}\}$ we will say that the model is $\mathcal{D}$-invariant.
 In this case it follows from \eqref{eq:hcrperp} that the result of minimization over $\mathbf{X}$ is given analytically as $\mathbf{X}=F_Q^{-1} \mathbf{L}$ and we have:
\begin{equation}
\CH=\tracep(\G F_Q^{-1})+\frac{1}{2}\|\sqrt{\G}\cdot F_Q^{-1}\trace(\rho_\bvar [\mathbf{L},\mathbf{L}^T])F_Q^{-1}\cdot\sqrt{\G}\|_1,
\end{equation}
where we have used the fact that $\imag(\mathbf{L}\mathbf{L}^T) = \frac{1}{2i}[\mathbf{L},\mathbf{L}^T]$.
It is also worth noting that the above equation may be written in an equivalent form, if one introduces the RLD $\bnabla \rho_\bvar=\rho_\bvar\mathbf{L_R}$  and the corresponding RLD bound \cite{Belavkin1976}:
\begin{equation}
\cov\succeq {F}_R^{-1},\quad{\rm where}\quad F_R=\trace(\rho_\bvar \mathbf{L_R}\mathbf{L_R}^T).
\end{equation}
In contrast to the standard QFI, $ F_R$ is not necessary real, and using the reasoning similar to the one presented in
Sec~\ref{sec:hcrequivalent} the RLD scalar bound takes the form \cite{Holevo1982}:
\begin{equation}
\CQ = \tracep(\G\cov)\geq \tracep(\G\real{{F}_R^{-1}})+\|\sqrt{\G}\cdot \imag{{F}_R^{-1}}\cdot \sqrt{\G}\|_1,
\end{equation}
Next, it may be shown \cite{Holevo1982} that $\real{{F}_R^{-1}}=F_Q^{-1}, \imag {{F}_R^{-1}}=\frac{1}{2}F_Q^{-1}\trace(\rho_\bvar\imag(\mathbf{L}\mathbf{L}^T))F_Q^{-1}$ and therefore for
$\mathcal{D}$-invariant models the HCR bound is equivalent to the RLD bound.

Since the $\mathcal{D}$-invariance property may at a first sight appear like an non-intuitive mathematical concept, let us
provide here some more operational description of it in case of unitary parameter estimation.
Imagine a quantum model where the parameters are being encoded
in a unitary way via a set of generators $\mathbf{G} = [G_1,\dots,G_\mathfrak{p}]^T$:
\begin{equation}
\rho_{\bvar} = e^{-i \mathbf{G}^T \bvar} \rho_0 e^{i \mathbf{G}^T \bvar}.
\end{equation}
If we consider estimation around $\bvar=0$ point, the potential non-commutativity of $G_i$ does not affect the form of the first derivatives
which read:
\begin{equation}
\left.\boldsymbol{\nabla} \rho_\bvar\right|_{\bvar=0} = i [ \rho_0, \mathbf{G}]
\end{equation}
and as a result the SLDs satisfy the following equation:
\begin{equation}
i [ \rho_0, \mathbf{G}] = \frac{1}{2}\{\rho_0, \mathbf{L}\}.
\end{equation}
Inspecting the definition of the $\mathcal{D}$ operator \eqref{eq:dinv} we see that up to the 1/2 factor $\mathcal{D}(G_i)$ is $L_i$.
The $\mathcal{D}$ invariance property, may now be understood as follows.
If we take the resulting SLDs and plug them into the definition of the model as new generators $\tilde{G}_i = L_i$,
the resulting new SLDs $\tilde{L}_i$ should be spanned by the original ones so $\tilde{L}_i \in \t{span}_{\mathbb{R}}\{L_1,...,L_\mathfrak{p}\}$.
Therefore, the $\mathcal{D}$-invariance property amounts to a statement that if we treat the orignal SLDs as additional generators of the unitary transformation the resulting span of the SLDs should not change.

\subsection{The HCR bound on multiple copies}

\label{sec:collectiveproof}
In this subsection we show, that similarly to
the SLD CR bound the HCR bound on multiple copies equals $1/n$
of the single copy formula \cite{yamagata2013quantum, hayashi2008asymptotic}:
 \begin{equation}
 \label{eq:additivity}
 n \CH(\rho_\bvar^{\otimes n}) = \CH(\rho_\bvar).
 \end{equation}
This fact is crucial, as it implies that when the HCR bound is calculated for a single copy it already provides information on the scenario where collective measurements are performed on many copies.

 Consider an $n$-fold tensor space $\mH^{\otimes n}$ and a quantum state that represents $n$ copies  $\rho_\bvar^{\otimes n}$ of a system.
  For any matrix $A$ we define:
\begin{equation}
A^{(n)}:=\sum_{k=1}^n\openone^{\otimes k-1}\otimes A \otimes \openone^{\otimes n-k}.
\end{equation}
In particular, in the $n$-copy model the SLDs
are given as $L_i^{(n)}$, where $L_i$ are the single copy SLDs. Note also, that:
$\mathcal{D}(A^{(n)})=(\mathcal D(A))^{(n)}$ and hence
$\mathcal{T}^{(n)}=\{X^{(n)}: X\in \mathcal{T}\}$.
Next, note that
\begin{multline}
\trace(\rho_\bvar^{\otimes n}A^{(n)}B^{(n)})
=\trace\left(\rho_\bvar^{\otimes n}\left(\sum_{k=1}^n\openone^{\otimes k-1}\otimes A \otimes \openone^{\otimes n-k}\right)\left(\sum_{l=1}^n\openone^{\otimes l-1}\otimes B \otimes \openone^{\otimes n-l}\right)\right)\\
=n\trace(\rho_\bvar AB)+n(n-1)\trace(\rho_\bvar A)\trace(\rho_\bvar B).
\end{multline}
Moreover, since $\scal{X_i,\openone} = 0$ for all ${X_i \in T}$, from the above formula we have $\trace(\rho_\bvar^{\otimes n}X_i^{(n)}X_j^{(n)})=n\trace(\rho_\bvar X_iX_j)$ as all the cross-terms vanish. Therefore, if $\mathbf{X}$ minimizes the Holevo bound for a single copy of the system, then
$\frac{1}{n}\mathbf{X}^{(n)}$ minimizes it for the $n$ copies. Indeed, note that the l.u. condition for $\mathbf{X}$, $\scal{\mathbf{L},\mathbf{X}^T}=\mathcal{I}$, implies that the $n$ copy variant of the l.u. condition will be satisfied for $\frac{1}{n}\mathbf{X}^{(n)}$:
$\scal{\frac{1}{n}\mathbf{X}^{(n)}, \mathbf{L}^{(n)T}} = \frac{1}{n} n \scal{\mathbf{X},\mathbf{L}^T} = \mathcal{I}$. This proves \eqref{eq:additivity}.

\subsection{Saturability}
\label{sec:saturability}
 Having proven the HCR bound and showing its $1/n$ scaling when applied to multi-copy models, we now turn to discuss its saturability.

In this section we will show, that for pure state models there always exists a measurement saturating the HCR bound already on the single copy level ~\cite{matsumoto2002new}. In case of mixed states, the HCR bound is saturable in general only asymptotically, and this in general requires  collective measurements performed on many copies. A discussion of this fact will be postponed until Sec.~\ref{sec:qlan} where it will be addressed using the QLAN perspective.

Let us focus on the HCR bound in the variant derived in \eqref{matsu}.
Let $\mathbf{Y}$ be the operators resulting from the minimization in \eqref{matsu} for $\rho_{\boldsymbol{\var}}=\ket{\psi_{\boldsymbol{\var}}}\bra{\psi_{\boldsymbol{\var}}}$.
Let us define  $\ket{y_i}:=Y_i\ket{\psi_{\boldsymbol{\var}}}$.
 As $\braket{\psi_{\boldsymbol{\var}}}{y_i}=0$ and $\braket{y_i}{y_j}\in \mathbb{R}$ for all $i,j$, one may choose a basis $\{\ket{b_i}\}$ of the $\t{span}\{\ket{\psi_{\boldsymbol{\var}}},\ket{y_1},\ldots,\ket{y_\mathfrak{p}}\}$ satisfying:
$\braket{\psi_{\boldsymbol{\var}}}{b_i}\in \mathbb{R}\backslash\{0\}$ and $\braket{y_i}{b_j}\in \mathbb{R}$ for all $i,j$.
Then one can define a projective measurement on $\mH\oplus\mathbb C^{\mathfrak{p}}$:
\begin{gather}
M_\ell=\ket{b_\ell}\bra{b_\ell}\; (\ell=1,\ldots,p+1),\quad
M_0=\openone-\textstyle{\sum}_{\ell=1}^{\mathfrak{p}+1}\ket{b_\ell}\bra{b_\ell},
\end{gather}
with the corresponding estimator:
\begin{equation}
\tilde\var_i(\ell)=\frac{\braket{b_\ell}{y_i}}{\braket{b_\ell}{\psi_{\boldsymbol{\var}}}}+\var_i,\;\ell\geq 1,\quad \tilde\var_i(0)=0,
\end{equation}
which is l.u. at the fixed point $\bvar$ and satisfies
\begin{equation}
\ket{y_i}=\sum_{\ell=0}^{\mathfrak{p}+1}(\tilde\var_i(\ell)-\var_i)M_\ell\ket{\psi_{\boldsymbol{\var}}}\Rightarrow \Sigma_{ij}=\braket{y_i}{y_j}.
\end{equation}
Any projective measurement on $\mH\oplus\mathbb C^{\mathfrak{p}}$ clearly defines a general measurement on $\mH$. Therefore, for pure states the HCR bound is saturable in a single-shot measurement and no collective measurement can further boost the estimation precision in such case.

For mixed states this is no longer the case in general and as mentioned before saturability will be guaranteed only asymptotically when measurements are performed on many-copies.
Since, as shown in Sec.~\ref{sec:collectiveproof}, the HCR bound for an $n$-copy model is equal to the $1/n$ of the single copy HCR bound, we can summarize the results on saturability via the following chain of inequalities:
 \begin{equation}
\CQ^{n}=\tracep(\G\cov^{n})\geq \tfrac{1}{n}\mathcal \CH(\rho_\bvar) \geq \tfrac{1}{n} \mathcal{C}^{\t{SLD}}(\rho_\bvar),
\end{equation}
where $\cov^{n}$ is a covariance matrix corresponding to l.u. estimation strategy performed on $n$-copy state.
The first inequality is always saturable for pure states and any $n$, while for mixed states it is guaranteed to be saturated asymptotically as $n\rightarrow \infty$.

As discussed in Sec.~\ref{sec:SLDCR}, for full rank $C$ the second inequality  becomes equality if and only if $\trace(\rho_\bvar [L_i ,L_j])=0$ for all $i,j$, in which case the SLD CR bound is equivalent to the HCR bound.
In the light of the saturability conditions of the HCR bound this also implies that the measurement incompatibility is not affecting the achievable precision in the asymptotic limit involving many copies, whereas for pure state models this statement is valid also for any finite $n$.

\subsection{Estimating functions of parameters}
\label{sec:function}
Assume that we have analyzed the estimation problem and the corresponding CR bounds using $\bvar$ parametrization of quantum states.
It might happen, that in some physical situation it might be more natural to think in terms of estimation of certain functions of $\bvar$, i.e.
\begin{equation}
\bvar^\prime = \mathbf{f}(\bvar),
\end{equation}
where we assume that $\mathbf{f}$ is an invertible vector function of parameters.
It is now straightforward to write the relevant quantities in the new parametrization provided they are known in the old parametrization.
 All we need to is to replace all the gradient operators $\boldsymbol{\nabla}$ with
$\boldsymbol{\nabla^\prime} = (J^T)^{-1} \cdot \boldsymbol{\nabla}$, where $J$ ($\t{det} J \neq 0$) is the derivative matrix of the $f$ function taken at the estimation point $\bvar_0$: $J_{ij} = \left.\tfrac{\partial \var^\prime_i}{\partial \var_j}\right|_{\bvar = \bvar_0}$. As a result
the corresponding SLD operators and the inverted QFI matrices will transform:
\begin{equation}
\mathbf{L}^\prime =  (J^T)^{-1} \mathbf{L}, \quad F_Q^{\prime-1} = J  F_Q^{-1} J^T,
\end{equation}
wheras the  objects that appear in the computation of  the HCR bound transform as:
\begin{equation}
\mathbf{X}^\prime =  J \mathbf{X}, \quad \V^\prime = J \V J^T.
\end{equation}

Taking a `dual' point of view we may also say that, since all the scalar bounds  are obtained by some variants of tracing the matrices $F_Q^{-1}$, $\V$ together with the cost matrix,
therefore, when calculating a scalar bound within the new parametrization using a cost matrix $C^\prime$,
 this bound may be always calculated using the objects obtained in the old parametrization, provided we replace the $C^\prime$ matrix with
\begin{equation}
C =  J^T C^\prime J.
\end{equation}

\section{Examples}
\label{sec:examples}
In order to illustrate the concepts intorduced in Sec.~\ref{sec:hcr}, we discuss two classes of examples. In Sec.~\ref{sec:examplequbit} we discuss qubit estimation examples, while in Sec.~\ref{sec:examplegaussian}
we discuss the Gaussian shift model examples. These examples encompass all non-trivial features that may appear in estimation problems including non-compatibilty of optimal measurements, as well as the potential advantage offered by collective measurement. The discussion of these two classes will also be helpful in understanding the general concept of QLAN presented in Sec.~\ref{sec:qlan}, where generic many-copy estimation  models become asymptotically equivalent to the Gaussian shift models.

\subsection{Qubit models.}
\label{sec:examplequbit}
In this section we use the standard Bloch ball parametrization of qubit states \cite{Nielsen2000}:
\begin{equation}
\rho_{\textbf{r}}
=\tfrac{1}{2}\left(\openone+\boldsymbol{\sigma} \cdot \bold{r}\right),
\end{equation}
where $\boldsymbol{\sigma}$ is a vector of Pauli matrices and
$\bold r$  is the Bloch vector with polar coordinates $(r, \theta, \varphi)$.

\subsubsection{Two parameter pure state model.}
\label{sec:examplequbit1}
First, let us consider a problem of estimation of an unknown pure qubit state, where the state is
parametrized with angles $(\theta,\varphi)$ and we set $r=1$:
\begin{equation}
\rho_{(\theta,\varphi)} = \ket{\psi_{(\theta,\varphi)}}\bra{\psi_{(\theta,\varphi)}},\   \ket{\psi_{(\theta,\varphi)}}=\cos(\theta/2)\ket{0}+\sin(\theta/2)e^{i\varphi}\ket{1}.
\end{equation}
We choose the cost matrix $\G$ in a way
that it corresponds to the natural metric on the sphere (coinciding with the Fubini-Study metric \cite{Bengtsson2006})
\begin{equation}
\G= \t{diag}[1,\sin^2(\theta)].
\end{equation}
For simplicity, thanks to the rotational symmetry we may focus on estimation around the point $(\theta,\varphi) = (\pi/2,0)$.
We have:
\begin{equation}
\rho_{(\theta,\varphi)}=\tfrac{1}{2}(\openone+\sigma_x),\quad
\partial_\theta\rho_{(\theta,\varphi)}=-\tfrac{1}{2}\sigma_z, \quad
\partial_\varphi\rho_{(\theta,\varphi)}=\tfrac{1}{2}\sigma_y.
 \end{equation}
In order to calculate the HCR bound we first apply the l.u. conditions on the $\mathbf{X}$ operators:  $\trace\left(\bnabla\rho_{(\theta,\varphi)}\mathbf{X}^T \right)=\mathcal{I}$, $\trace(\rho_{(\theta,\varphi)} \mathbf{X})=0$---according to the discussion in Sec.~\ref{sec:hcrderive} the second condition is not necessary as it does not affect the final result of the minimization but we impose it nevertheless to reduce the number of free parameters and simplify the reasoning. As a result we get
\begin{equation}
X_\theta=-\sigma_z+\alpha_\theta(\openone - \sigma_x),\quad X_\varphi=\sigma_y+\alpha_\varphi(\openone - \sigma_x),\quad \alpha_{\theta/\varphi}\in\mathbb R.
\end{equation}
Note, that $Z[X]$ does not depend on $\alpha_{\varphi/\theta}$ (as $1-\sigma_x\in \mathcal L(ker \rho_{(\theta,\varphi)}$). Therefore without loss we may set $\alpha_{\varphi/\theta}=0$:
\begin{equation}
X_\theta=-\sigma_z,\quad X_\varphi=\sigma_y,\quad \Vr=
\begin{bmatrix}
1&i\\
-i&1
\end{bmatrix},
\end{equation}
for which the corresponding HCR bound is:
\begin{equation}
\label{eq:hcrqubit1}
\CH_{(\theta,\varphi)}=4.
\end{equation}
Using the formula \eqref{HCRSLD} we obtain
\begin{equation}
\label{eq:sldqubit1}
\mathcal{C}^{\t{SLD}}_{(\theta,\varphi)}= \min_{\bf X}\tracep(\G\Vr) = 2,
\end{equation}
without the need to compute the actual SLDs. Still for completeness, we provide below the explicit form of the QFI matrix
and the SLDs (note that since the state is pure the SLDs are not unique):
\begin{equation}
  F_Q =\t{diag}[1,\sin^2(\theta)], \
L_\theta = \boldsymbol{\sigma}\cdot \partial_\theta \mathbf{r}, \
   L_\varphi = \boldsymbol{\sigma}\cdot \partial_\varphi \mathbf{r}
\end{equation}
and it is clear from the above that indeed $\mathcal{C}^{\t{SLD}}_{(\theta,\varphi)} = \tracep{(C F_Q^{-1})} = 2$.

We see that the the HCR bound is twice as large as the SLD CR,
which corresponds to the maximal possible discrepancy, as discussed in Sec.~\ref{sec:sldvshcr}. It means that the measurements optimal for both of these parameters are `maximally' incompatible---the hallmark of this is the noncommutativity of the SLDs. A measurement for which the corresponding classical FI matrix $F$ yields $\tracep(C F^{-1})$  saturating the HCR bound may be constructed by combining the optimal measurements for the two parameters with equal weights:
\begin{equation}
\{M_\ell\}=
\Big\{\tfrac{1}{2}\ket{+}\bra{+}_y, \,\tfrac{1}{2}\ket{-}\bra{-}_y, \, \tfrac{1}{2}\ket{+}\bra{+}_z, \,  \tfrac{1}{2}\ket{-}\bra{-}_z\Big\}.
\end{equation}

\subsubsection{Two parameter mixed state model. }
Let us now consider a mixed state qubit model with fixed $\varphi=0$, where the parameters $(r,\theta)$ correspond to the length (representing the purity of the state) and the latitude $\theta$ of the Bloch vector:
\begin{equation}
\rho_{(r,\theta)}=\tfrac{1}{2}(\openone+r\sin(\theta)\sigma_x+r\cos(\theta)\sigma_z).
\end{equation}
Unlike in the pure state model there is no natural choice for the cost matrix for this problem, as the $r$ parameter is not associated with any
group action in the space of quantum states.
Therefore we only assume that the cost matrix is diagonal in $(r,\theta)$ and consider
\begin{equation}
\G=\t{diag}[c(r),r^2],
\end{equation}
where $c(r)>0$ determines the character of the cost function with respect to the $r$ parameter and the $r^2$ cost in case of $\theta$ we choose for convenience in order to stay in agreement with the spherical coordinate conventions. In particular, the Euclidean metric corresponds to the choice $c(r)=1$, while a more natural Bures metric  \cite{Bures, Bengtsson2006} corresponds (up to a constant) to $c(r) = 1/(1-r^2)$.
Without loss of generality, we consider estimation around the point $(r,\theta) = (r,0)$, in which case we have:
\begin{equation}
\rho_{(r,\theta)}=\tfrac{1}{2}(\openone+r\sigma_z),\quad \partial_r\rho_{(r,\theta)}=\tfrac{1}{2}\sigma_z,\quad \partial_\theta\rho_{(r,\theta)}=\tfrac{1}{2}r\sigma_x,
\end{equation}
and the l.u. conditions imply that
\begin{equation}
X_r=\sigma_z- r \openone+\alpha_r\sigma_y,\quad X_\theta=\tfrac{1}{r}\sigma_x+\alpha_\theta\sigma_y,\quad \alpha_{r/\theta} \in\mathbb R.
\end{equation}
Direct minimization of the cost leads to $\alpha_{r/\theta}=0$:
\begin{equation}
X_r=\sigma_z-r\openone,\quad X_\theta=\tfrac{1}{r}\sigma_x,\quad\Vr=
\t{diag}[1-r^2, \tfrac{1}{r^2}]
\end{equation}
and the final HCR bound reads:
\begin{equation}
\label{eq:hcrqubit2}
\CH_{(r,\theta)}=c(r)(1-r^2) +1.
\end{equation}
Interestingly, the SLD CR bound $\min_{\bf X}\tracep(\G\Vr)$ yields the same result:
\begin{equation}
\mathcal{C}^{\t{SLD}}_{(r,\theta)}=\CH_{(r,\theta)},
\end{equation}
which can also be independently confirmed using the explicit form of the SLDs and the QFI matrix:
\begin{equation}
 F_Q = \t{diag}[\tfrac{1}{1-r^2}, r^2]
,\
L_r = \tfrac{1}{1-r^2}\left( \boldsymbol{\sigma} \cdot \mathbf{r}   - r \openone \right), \
 L_\theta = \boldsymbol{\sigma} \cdot \partial_\theta \mathbf{r}.
\end{equation}
From the above form of SLDs we find that $\trace(\rho_\bbr[L_r,L_\theta])=0$, so  according to the discussion from Sec.~\ref{sec:SLDCR}
the two bounds must indeed be equal. Note however, that the SLDs do not commute as operators $[L_r,L_\theta]\neq 0$.
In fact, as discussed in detail in \cite{Bagan2004, Vidrighin2014} in this case there is no local single qubit measurement that saturates the CR bound and
 hence collective measurements prove advantageous.

To shed more light in this problem, one may refer to the the Hayashi-Gill-Massar bound (HGM) \cite{Gill2000, masahito2005asymptotic,yamagata2011efficiency} which is valid for
qubit estimation models and is always saturable using local measurement.
It states that:
\begin{equation}
\GG_{(r,\theta)} \geq \GG^{\t{HGM}}_{(r,\theta)}:=\left(\tracep\left[ \sqrt{\sqrt{F_Q^{-1}}\G \sqrt{F_Q^{-1}}}\right]\right)^2=
[1+\sqrt{c(r)(1-r^2)}]^2.
\end{equation}
It is worth noticing, that this bound may also be saturated by using  weighted measurements optimal for both parameters:
\begin{equation}
\{M_\ell\}=
\Big\{ p_z\ket{+}\bra{+}_z, \, p_z\ket{-}\bra{-}_z, \, p_x \ket{+}\bra{+}_x, \, p_x \ket{-}\bra{-}_x\Big\},
\quad p_z+p_y=1,
\end{equation}
with  weights chosen so to optimize the corresponding classical CR bound  $\tracep(\G F^{-1})=\frac{c(r)(1-r^2)}{p_z}+\frac{1}{p_x}$.

\begin{figure*}[t!]
 \begin{center}
 \includegraphics[width=0.8 \textwidth]{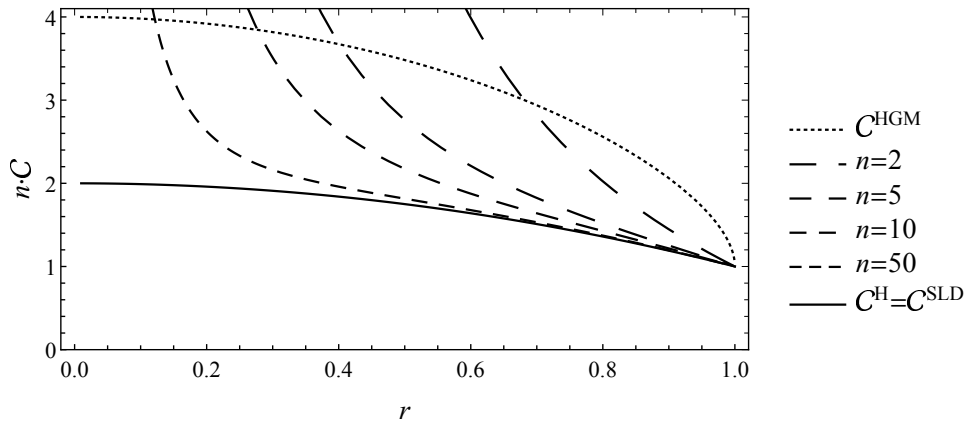}
  \end{center}
\caption{Comparison of the HCR bound (which in this case coincides with the SLD CR bound, solid line) and the HGM bound (saturable using local measurements, dotted line) for the total cost (normalized by the number of qubits) corresponding to the Euclidean distance, $c(r)=1$, in the estimation of the Bloch vector parameters $(r,\theta)$ as a function of $r$.
  Dashed lines represent the performance of the exemplary collective measurement which approaches the HCR bound with an increasing number of copies of the system $n$.}
\label{fig:collective}
\end{figure*}

 These bounds are compared in Fig.~\ref{fig:collective} (for $c(r)=1$) from which it is clear that the HGM bound is significantly larger than the HCR bound everywhere except the border of the Bloch sphere. This implies that collective measurement allow to achieve a better precision in comparison with the local measurements---note that for the Bures distance cost $c(r)=1/(1-r^2)$, $C^{\t{H}}_{(r,\theta)} = 2$,
 $C^{\t{HGM}}_{(r,\theta)} = 4$ are parameter independent and hence the advantage of collective approach is the same irrespectively of the value of $r$.

From a practical point of view it is important to understand what is the structure of a collective measurement that yields the maximal information on the length of the  Bloch vector $r$ without loosing information on the angle $\theta$.
 It can be checked by direct computation that $\rho_{(r,\theta)}$ can be written as
 $\rho_{(r,\theta)}=e^{-i \sigma_y \theta/2}\frac{e^{2 \sigma_z \beta/2}}{2 \cosh \beta}e^{i\sigma_y\theta/2}$,
 where $\tanh(\beta)=r$.
 Hence, the tensor product of $n$ copies will have an analogous form:
\begin{equation}
\rho_{(r,\theta)}^{\otimes n}=e^{-iJ_y\theta}\frac{e^{2 J_z\beta}}{[2\cosh\beta]^n}e^{iJ_y\theta},
\end{equation}
where  $J_i=\frac{1}{2}\sum_{k}\sigma^{(k)}_i$ are the total angular momentum operators.
Now, instead of measuring $r$ directly (which would correspond to measuring $J_z$), one may perform
 a projection onto subspaces with a well defined value of the total angular momentum---then no information about
 $\theta$ is lost, since the rotation $e^{-iJ_y\theta}$ commutes with the total angular momentum operator.
 Moreover, it turns out \cite{Keyl2001, hayashi2008asymptotic,Bagan2006a}  that in the limit of $n \rightarrow \infty$  such a measurement gives the same precision of estimating $r$ as the optimal direct measurement, provided $r>0$. Finally,
 the optimal measurement to extract the information on $\theta$ is performed---the $J_x$ measurement.
 The performance of this collective measurement strategy is depicted in Fig.~\ref{fig:collective}, where
 a visible improvement  with the increase number of copies involved is visible, and the precision achieved will approach the asymptotic bound for $n \rightarrow \infty$. We will see a generalization of this measurement strategy in the discussion of the QLAN in Sec.~\ref{sec:qlan}.

\subsubsection{Three parameter mixed state model.}
\label{sec:examplequbit3}
\begin{figure*}[t!]
\begin{center}
\includegraphics[width=0.8 \textwidth]{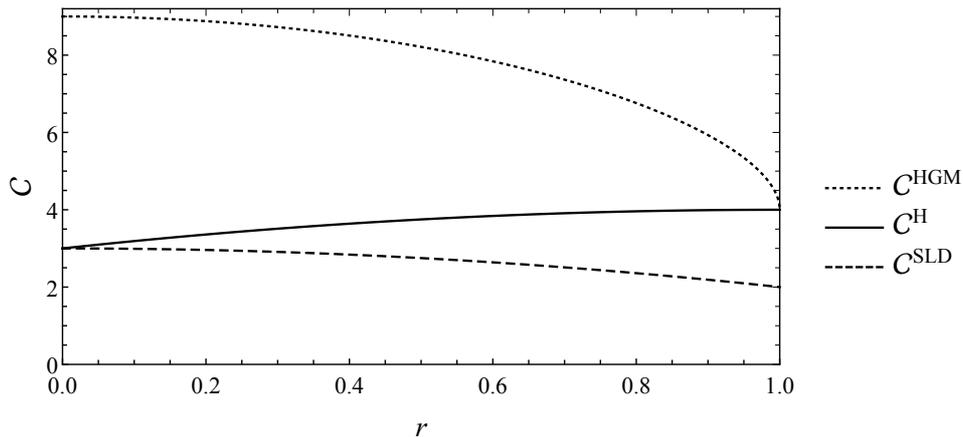}
\end{center}
\caption{Comparison of HGM, HCR and the SLD CR bounds for the total estimation Euclidean cost  in case of the estimation of a completely unknown mixed state of a qubit, as a function of Bloch's vector's length.}
\label{fig:comp}
\end{figure*}

Finally, let us consider the most challenging qubit estimation problem, namely estimation of a completely unknown qubit state.
Following the line of reasoning from the previous examples we will consider the cost matrix to be
\begin{equation}
\label{eq:3dcost}
\G  = \t{diag}[c(r), r^2, r^2 \sin^2(\theta)].
\end{equation}
In order to obtain the HCR bound, it will be more convenient to switch from spherical $(r, \theta, \varphi)$ to Cartesian coordinates
where we write the Bloch vector as $\bold{r}=[r_x,r_y,r_z]^T$,
\begin{equation}
\rho_{\bold{r}}=\tfrac{1}{2}(\openone+\boldsymbol{\sigma} \cdot \bold{r}).
\end{equation}
In this parametrization the partial derivatives over the parameters are  $\partial_i\rho_{\bold{r}}=\frac{1}{2}\sigma_i$,
and the l.u. conditions lead to $X_i=\sigma_i - r_i\openone$ with no free parameters to optimize over.
We can therefore write:
\begin{equation}
\V_{ij}=\trace\left[\tfrac{1}{2}(\openone+\boldsymbol{\sigma}\cdot \mathbf{r})(\sigma_i-r_i\openone)(\sigma_j-r_j\openone) \right]=
\delta_{ij} - r_i r_j + i \sum_k \varepsilon_{ijk} r_k,
\end{equation}
where $\varepsilon_{kij}$ is the Levi-Civita symbol.
In order to calculate the cost using the cost matrix \eqref{eq:3dcost} defined for spherical coordinates, we can use
the general approach presented in Sec.~\ref{sec:function}, and transform the above $\V$ written in Cartesian coordinates
to spherical coordinates:
\begin{equation}
\V^\prime = J \V J^T =
\begin{bmatrix}
 1-r^2 & 0 & 0 \\
 0 & \frac{1}{r^2} & \frac{i}{r \sin(\theta)} \\
 0 & -\frac{i}{r \sin(\theta)} & \frac{1}{r^2 \sin^2(\theta)}
 \end{bmatrix},
\end{equation}
where $J$ is the derivative of the standard transformation from Cartesian to spherical coordinates, which we do not write here explicitly.
We may now use \eqref{HCRtracenorm} to compute the HCR bound for the cost matrix \eqref{eq:3dcost}:
\begin{equation}
\label{eq:hcrqubit3}
\mathcal{C}^{\t{H}}_{\mathbf{r}}= 2 + c(r)(1-r^2) + 2r,
\end{equation}
where the last term comes from the imaginary part of the $\V$ matrix.
The QFI matrix is in fact the inverse of the real part of the $\V$ matrix and reads
\begin{align}
F_Q &= \t{diag}[\tfrac{1}{1-r^2},r^2,r^2 \sin^2(\theta)], \
L_r = \tfrac{1}{1-r^2}\left( \boldsymbol{\sigma} \cdot \mathbf{r}   - r \openone \right), \,
L_\theta =  \boldsymbol{\sigma} \cdot \partial_{\theta} \mathbf{r}, \
L_{\varphi} =  \boldsymbol{\sigma} \cdot \partial_{\varphi} \mathbf{r},
\end{align}
where we also have provided an explicit form of the SLDs for completness.
Therefore, the SLD CR and the HGM bounds read:
\begin{equation}
\mathcal{C}^{\t{SLD}}_{\mathbf{r}}=2+ c(r)(1-r^2), \quad \GG^{\t{HGM}}_{\mathbf{r}}=(2+\sqrt{c(r)(1-r^2)})^2.
\end{equation}
In Figure \ref{fig:comp} we present the comparison of all the three bounds and its dependence on length of Bloch vector $r$ for the Euclidean distance $c(r)=1$ case.

In order to get a better intuition in preparation for the QLAN discussion in Sec.~\ref{sec:qlan}, let us return to the Cartesian parametrization  and consider estimation around the point $\bold{r}=[0,0,r]^T$. Then, locally the two  parameters $r_x,r_y$ may be interpreted as rotations of the Bloch vector and the third one, $r_z$, as its length. At this point the QFI matrix and the corresponding SLDs read:
\begin{equation}
F_Q = \t{diag}[1,1,\tfrac{1}{1-r^2}], \ L_x=\sigma_x ,\  L_y=\sigma_y ,\ L_z = \t{diag}[\tfrac{1}{1+r},-\tfrac{1}{1-r}].
\end{equation}
Let us notice the following properties:
\begin{equation}
\label{incom}
\trace(\rho_\bold{r}[L_x,L_y])=r,\quad\trace(\rho_\bold{r}[L_y,L_z])=0,\quad\trace(\rho_\bold{r}[L_z,L_x])=0.
\end{equation}
Taking into account the discussion in Sec.~\ref{sec:SLDCR}, we see, that only $r_x,r_y$ are fundamentally incompatible---the third one may be effectively measured independently of the others (at least in the asymptotic limit utlizing collective measuremenrs).
For $|r|=1$ we recover the pure state case discussed in the first example of this section where the HCR bound and the HGM bound coincide, as local measurements saturate the HCR bound in case of pure states. In general the optimal local measurements (saturating the HGM bound) have a similar structure as in the previous example:
\begin{equation}
\{M_\ell\}=\bigcup_{k\in\{x,y,z\}}\Big\{ p_k\ket{+}\bra{+}_k, p_k\ket{-}\bra{-}_k\Big\},\quad \sum_{k\in\{x,y,z\}}p_k=1,
\end{equation}
with $p_x,p_y,p_z$ chosen to minimize $\tracep(\G F^{-1})=\frac{1}{p_x}+\frac{1}{p_y}+\frac{c(r)(1-r^2)}{p_z}$, as at this point the cost matrix in Cartesian coordinates reads $\G=\t{diag}[1,1,c(r)]$.

Finally, the fundamental measurement incompatibility vanishes only at $|r|=0$ (SLD CR bound coincides with the HCR  bound), but since the HGM is still larger at this point it implies that the collective measurements are necessary to obtain the optimal performance.


\subsection{Estimation for general quantum Gaussian shift models}
\label{sec:examplegaussian}
In this section we consider a general problem of estimating the parameters of a quantum Gaussian shift model, which is a special class
of general Gaussian estimation models \cite{Gao2014, Nichols2018}. Aside from the mathematical interest and practical importance, the problem is directly relevant for the QLAN theory described in Sec.~\ref{sec:qlan}. In a nutshell, QLAN shows that that any model consisting of an ensemble of finite dimensional identically prepared systems is asymptotically equivalent in a statistical sense to a Gaussian shift model which encodes the local `tangent space' structure of the original one. In particular, each qubit model discussed in the preceding section will have a corresponding Gaussian model. A key property of Gaussian shift models is that the HCR bound is always saturable in a single-shot scenario. Combined with the QLAN theorem this will provide the proof of the asymptotic
saturability of the HCR in the multi-copy setting.

Consider a continuous variable system consisting of $\mathfrak{q}$ modes with canonical coordinates
$(Q_{i}, P_{i})$, satisfying the commutation relations \cite{WeedbrokReviewGaussian2012}
\begin{equation}
 [Q_{i}, P_{j}] = i\delta_{i,j} \mathbb{1} , \qquad i,j=1,\dots, \mathfrak{q}.
\end{equation}
The joint system can be represented on the tensor product space $\mathcal{F}^{\otimes \mathfrak{q}}$ such that the pair $(Q_{i}, P_{i})$ acts on $i$-th copy of the one-mode Fock space $\mathcal{F}$. Since it will be relevant for the QLAN formulation, we also allow for $\mathfrak{c}$ `classical real valued variables'
$(Z_{1},\dots , Z_{\mathfrak{c}}) $  which commute with each other and with all $(Q_{i}, P_{i})$. These can be represented as position observables on $\mathfrak{c}$ additional copies of $\mathcal{F}$, whose affiliated algebra is $L^\infty(\mathbb{R}^\mathfrak{c})$.
We put all canonical observables together as  a column vector
\begin{equation}
{\bf R}:= [R_{1},\dots, R_{\mathfrak{r}}]^T\equiv [ Q_{1}, P_{1},\dots , Q_{\mathfrak{q}},  P_{\mathfrak{q}}, Z_{1},\dots, Z_{\mathfrak{c}}]^T,\quad \mathfrak{r}=2\mathfrak{q}+\mathfrak{c},
\end{equation}
and write their commutation relations as
\begin{equation}
[ R_{i}, R_{j}] = iS_{i,j} \mathbb{1},  \quad  [ \mathbf{R}, \mathbf{R}^T] = i S  \mathbb{1},
\end{equation}
where $S$ is the $\mathfrak{r} \times \mathfrak{r}$ block diagonal symplectic matrix of the  form
\begin{equation}
S= {\rm diag}[\Omega, \dots ,\Omega, 0,\dots 0 ], \quad
\Omega=
\begin{bmatrix}
  0&1\\
-1 &0
 \end{bmatrix}.
\end{equation}

A state  of this hybrid quantum-classical system is described in terms of its density
matrix $\varrho$ (in this review we use $\varrho$ to represent continuous variable system states, in particular Gaussian states, in order to differentiate it from finite dimensional states $\rho$) which is a positive and normalised element of $\mathcal{T}^{1}(\mathcal{F}^{\otimes \mathfrak{q}})\otimes L^{1}(\mathbb{R}^{\mathfrak{c}})$, where $\mathcal{T}^1$ denotes the space of trace-class linear operators  and $L^{1}$ the space of absolutely integrable functions.
For any state $\varrho$ let us define its characteristic function
\begin{equation}
\cf_{\rho}(\boldsymbol{\xi}):=
\trace\left( \varrho\, e^{i \boldsymbol{\xi}^T {\bf R}} \right) \equiv \scalno{e^{i \boldsymbol{\xi}^T {\bf R}},\openone}_\varrho, \quad \boldsymbol{\xi} \in \mathbb{R}^\mathfrak{r}.
\end{equation}
where the symbol `${\rm Tr}$'  is understood as taking trace over the quantum part and integrating over the classical part.
We will say that a state $\varrho$ is Gaussian if and only if its characteristic function $\cf$ is Gaussian:
\begin{equation}
\label{eq:characteristic}
\cf_{\varrho}(\boldsymbol{\xi}) = e^{i \boldsymbol{\xi}^T {\bf r} } e^{-   \boldsymbol{\xi}^T V \boldsymbol{\xi}/2 },
\end{equation}
where
\begin{equation}
\mathbf{r} := \scalno{\mathbf{R},\openone}_\varrho,
\quad V :=  \scalno{({\bf R}- {\bf r}\mathbb{1} ),({\bf R}- {\bf r} \mathbb{1})^T}_\varrho
\end{equation}
are the mean and the covariance matrix of the state respectively---note that we have used the previously introduced notation
involving the scalar product as defined in equation \eqref{eq:scal}.
The positivity of the density matrix imposes a restriction on the allowed covariance matrices, as  expressed by the matrix Heisenberg uncertainty relation \cite{Simon1994, WeedbrokReviewGaussian2012}:
\begin{equation}
\label{eq:heisenberggaussian}
V \succeq \frac{i}{2} S.
\end{equation}
It is worth stressing that the opposite implication holds as well---to any covariance matrix satisfying \eqref{eq:heisenberggaussian} there corresponds a unique zero-mean Gaussian state $\varrho$.


A \emph{Gaussian shift model} with parameters $\bvar\in \mathbb{R}^{\mathfrak p}$ is a family of Gaussian states
 $\varrho_{\bvar}$ with some \emph{fixed} covariance matrix $V$ and mean depending linearly on $\bvar$
 \begin{equation}
 \mathbf{r} = A \bvar,
 \end{equation}
 with $A:\mathbb{R}^{\mathfrak{p}} \to\mathbb{R}^{\mathfrak{r}}$ a given injective linear map.
For purely quantum models with no classical degrees of freedom ($\mathfrak{c}=0$), the states $\varrho_{\bvar}$ can be obtained by applying unitary shift operators to the mean zero Gaussian state with covariance matrix $V$
\begin{equation}
\varrho_{\bvar} = e^{-i \mathbf{R}^T G \bvar} \varrho_0 e^{i \mathbf{R}^T G \bvar},\qquad G = S A.
\end{equation}

Thanks to the fact that the parameters enter linearly into the mean of the Gaussian state and the covariance matrix is fixed, the SLDs of a Gaussian shift model are linear combinations of the canonical coordinates \cite{Holevo1982, Monras2013}.
To see this, consider the characteristic function of $\varrho_{\bvar}$
\begin{equation}
\cf_{\bvar}(\boldsymbol{\xi}) =  e^{i \boldsymbol{\xi}^T A {\bf \bvar} } e^{-   \boldsymbol{\xi}^T V \boldsymbol{\xi}/2 },
\end{equation}
and take derivatives over $\theta_i$ to get
\begin{equation}
\label{eq:sldgausstheta}
\partial_i \cf_{\bvar}(\boldsymbol{\xi}) = i (A^T \boldsymbol{\xi})_i \cf_{\bvar}(\boldsymbol{\xi}).
\end{equation}
On the other hand, using the definition \eqref{eq:SLD} of the SLDs, the derivative may be expressed as
\begin{equation}
\label{eq:sldgauss1}
\partial_i \cf_{\bvar}(\boldsymbol{\xi}) = \trace\left(\partial_i \varrho_{\bvar}  e^{i \boldsymbol{\xi}^T {\bf R}}\right) = \trace\left(\tfrac{1}{2}\{L_i,\varrho_{\bvar } \} e^{i \boldsymbol{\xi}^T {\bf R}} \right).
\end{equation}
Making use of the following algebraic property
$\tfrac{1}{2}\{R_i, e^{i \boldsymbol{\xi}^T \mathbf{R}}\} = \tfrac{1}{i}\tfrac{\partial}{\partial \xi_i} e^{i \boldsymbol{\xi}^T \mathbf{R}} $,
which can be proven using  the standard BCH formula, we get
\begin{equation}
\label{eq:anticomm}
\trace\left(\tfrac{1}{2}\{R_i, \varrho_{\bvar}\} e^{i \boldsymbol{\xi}^T \mathbf{R}} \right)  = \trace\left(\tfrac{1}{2}\{R_i,  e^{i \boldsymbol{\xi}^T \mathbf{R}}\} \varrho_{\bvar} \right)  = \frac{\partial}{i \partial \xi_i} \cf_{\bvar} = (A \bvar + i V \boldsymbol{\xi})_i \cf_{\bvar}.
\end{equation}
Now, if we take
\begin{equation}
{\bf L} = [L_1, \dots , L_{\mathfrak{p}}]^T, \quad
\mathbf{L} = A^T V^{-1} \mathbf{R} + A^TV^{-1}A \bvar\openone,
\end{equation}
and substitute into the r.h.s. in \eqref{eq:sldgauss1} we obtain \eqref{eq:sldgausstheta}
 and hence we see that this a correct formula for the  SLDs operators. For simplicity and without loss of generality, from now on we will  consider estimation around $\bvar=0$ in which case the formula for the SLDs simplifies to
\begin{equation}
\label{eq:sldgauss}
\mathbf{L} = A^T V^{-1} \mathbf{R}
\end{equation}
and the the QFI matrix has the same expression as its classical counterpart
\begin{equation}
\label{eq:gaussfisher}
F_Q = \scalno{{\bf L}, {\bf L}^T}_{\rho_{\bvar}} = A^T V^{-1} A.
\end{equation}
We are now in position to derive the HCR bound for the Gaussian shift model. Recall, from Sec.~\ref{sec:collectiveproof} that when performing the minimization in the formula \eqref{HCR} for the HCR bound, we may always restrict the class of operators $\mathbf{X}$ to belong to the smallest $\mathcal{D}$-invariant subspace $T$ that contains $\t{span}_{\mathbb{R}}\{L_1,\dots,L_{\mathfrak{p}}\}$. Using the fact that in the Gaussian shift model the SLDs are linear functions of canonical variables, we show below that $T\subseteq \t{span}_{\mathbb{R}}\{R_1,\dots,R_{\mathfrak{r}}\}$ (for $\bvar \neq 0$ we need to include $\openone$ in the span as well).

To see this, note that the characteristic function of $[R_i, \rho_{\bvar}]$ is equal to
\begin{equation}
\label{eq:DinR}
\trace\left(i[R_i, \varrho_{\bvar}]e^{i \boldsymbol{\xi}^T \mathbf{R}} \right)  =i \trace\left(\varrho_{\bvar} [e^{i \boldsymbol{\xi}^T \mathbf{R}},R_i] \right)=i \trace\left( \varrho_{\bvar} [i\boldsymbol{\xi}^T \mathbf{R},R_i]e^{i \boldsymbol{\xi}^T \mathbf{R}} \right)
=i( S\boldsymbol{\xi} )_i \cf_{\bvar}(\boldsymbol{\xi})
\end{equation}
which corresponds to the original $\cf_{\bvar}$ multiplied by some linear transformation of $\boldsymbol{\xi}$ with imaginary coefficients. Next, equation \eqref{eq:anticomm} with $\bvar =0$ reads
\begin{equation}
\trace\left(\{R_j, \varrho_{\bvar}\} e^{i \boldsymbol{\xi}^T \mathbf{R}} \right)=2i(V \boldsymbol{\xi})_j \cf_{\bvar} (\boldsymbol{\xi}).
\end{equation}
 For the time being, we can restrict ourselves to quantum modes only since the $\mathcal{D}$ operator is trivial
 for classical variables. In this case $V$ is a strictly positive real matrix, and hence ${\rm Range}(V) = \mathbb{R}^\mathfrak{r}$. Since any operator is in one-to-one correspondence with its characteristic function, this means that
 ${\rm span}_\mathbb{R}\left\{\ \{R_j, \varrho_{\bvar}\},\, j=1, \dots , \mathfrak{r}\right\}$ contains
 $i[R_i, \varrho_{\bvar}]$ for any $i$.
 Taking into account the definition \eqref{eq:dinv} of the  operator $\mathcal{D}$, this implies that $\mathcal{D}(R_i)$ may be written as linear combination of components of ${\bf R}$, $\mathcal{D}(R_i) = \frac{1}{2}(S V^{-1} \mathbf{R})_i$ , and hence $\t{span}_{\mathbb{R}}\{  R_1,\dots,R_{\mathfrak{r}}\}$ is
 $\mathcal{D}$ invariant which was to prove.





Therefore, when calculating  the HCR bound for the Gaussian shift model we may restrict the minimization to
$\mathbf{X}$ operators of the form $\mathbf{X} = B \mathbf{R}$,
where $B$ is a linear map $B:\mathbb{R}^{\mathfrak{r}} \to\mathbb{R}^{\mathfrak{p}}$. Moreover, taking into account the explicit form of the SLD operators given by equation \eqref{eq:sldgauss}, the l.u. condition may be equivalently written as:
\begin{equation}
 \mathcal{I}= \trace(\bnabla \varrho_{\bvar} \mathbf{X}^T) =
 \trace\left(\varrho_{\bvar}\tfrac{1}{2}\{A^T V^{-1}\mathbf{R}  ,(B \mathbf{R})^T \}   \right) = A^T V^{-1} V B^T = (B A)^T.
\end{equation}
Additionally,
\begin{equation}
\t{Im}\V  =
\frac{1}{2i} \trace(\varrho_\bvar [\mathbf{X}, \mathbf{X}^T])= \frac{1}{2 i  } B \trace(\varrho_\bvar [\mathbf{R},\mathbf{R}^T])  B^T =
 \frac{1}{2} B S  B^T.
\end{equation}
and therefore the HCR bound
\begin{equation}
\CH=\min_{\X=B \mathbf{R}}\left(\tracep(\G \real \V)+\|\sqrt{\G}\cdot {\rm Im}\V\cdot\sqrt{\G}\|_1
\ \big|\  \, \trace\left(\bnabla \varrho_{\bvar}\mathbf{X}^T\right)=\mathcal{I}\right)
\end{equation}
may be written directly as the minimization over the linear map $B$
\begin{equation}
\label{Bholevo}
\CH=\min_{B}\left(\tracep(\G B V B^T)+\|\sqrt{\G}\cdot B S B^T\cdot\sqrt{\G}\|_1
\ \big|\   BA = \mathcal{I}  \right).
\end{equation}

In general there is no closed analytical formula for the solution of this minimization problem.
However, in a special case when the number of parameters of interest is maximal, $\mathfrak{p}=\mathfrak{r}$,
 the  operator $A$ has a unique inverse and we obtain an explicit bound by simply substituting $B=A^{-1}$:
\begin{equation}
\label{eq:hcrgauss}
\CH =
\tracep\left(  \G A^{-1}  V (A^{-1})^T \right)
+ \frac{1}{2}\left\|  \sqrt{\G}A^{-1} S (A^{-1})^T \sqrt{\G}  \right\|_1.
\end{equation}
While the first term is identical to the  cost of the corresponding classical Gaussian estimation problem, the second term in \eqref{eq:hcrgauss} represents the additional contribution due to non-commutativity.
This model is also $\mathcal{D}$-invariant, since $\t{span}_{\mathbb{R}}\{L_1,\dots,L_{\mathfrak{p}}\}$
corresponds to the span of all canonical variables which is $\mathcal{D}$-invariant.
Therefore, as discussed in Sec.~\ref{sec:Dinvariance}, the above bound coincides with the RLD CR bound.

On the other hand when $\mathfrak{p}=1$, i.e. when we estimate only a single scalar variable, the
HCR reduces to the SLD CR bound by the same arguments as given in Sec.~\ref{sec:nuisance}, where it was shown that this is a general feature of multi-parameter estimation problems with rank-1 cost matrix.

Finally, we show that for the Gaussian shift models the HCR bound is always saturable (on the single copy level!).
For simplicity, let us again assume the absence of classical degrees of freedom as the saturability issue is trivial for them---there is no measurement issue involved at all.
For this, it will be enough to consider the so called linear measurement \cite{Holevo1982}.
 A linear measurement can be implemented by coupling the system with an independent ancillary system and measuring a commuting family of coordinates of the joint system.
Let  $\tilde{\bf R}:= [\tilde{R}_{1}, \dots ,\tilde{R}_{\mathfrak{r}}]^T$ be the coordinates of the ancillary system with the same number of modes and a symplectic matrix $\tilde{S}$. We assume that the joined system+ancillary state is $\rho_\bvar \otimes \tilde{\rho}$ where $\tilde\rho$ is a fixed zero mean Gaussian state with covariance matrix $\tilde{V}$. 
Let $B$ the result of optimization \eqref{Bholevo}. The measurement is defined by a $\mathfrak{p}-$tuple ${\bf Y} = [Y_1, \dots ,Y_\mathfrak{p}]^T$ of {\it mutually commuting} variables of the form
$$
{\bf Y} ={\bf X} + \tilde{\bf X} = B {\bf R} + \tilde{B} \tilde{{\bf R}},
$$
where $B, \tilde{B} $ are real $\mathfrak{p} \times \mathfrak{r}$  matrices, with a condition $B A = \mathcal{I}$ which guarantees that the l.u. property is fulfilled.
As a result we obtain a l.u.  unbiased estimator whose mean square error effectively depends on $\tilde{B}$ and the choice of the ancillary Gaussian state $\tilde{\rho}$:
\begin{equation}\label{eq.risk.w}
\mathcal{C}(\tilde{B},\tilde{\rho})= {\rm Tr}(\G BVB^T) + {\rm Tr}(\G \tilde B\tilde V\tilde B^T).
\end{equation}
Since we require all the $Y_{i}$ to commute with each other, we have
\begin{equation}
\label{eq:commutegauss}
[ \tilde{\bf X},  \tilde{\bf X}^T ] = - [ {\bf X} ,  {\bf X}^T]  =
-   i BS B^T \mathbb{1}.
\end{equation}
Notice that we can trivially satisfy the above requirement if the symplectic matrix $\tilde{S} = -S$ and we take $\tilde{B}=B$. Physically, this corresponds to inverting the roles of position and momentum operators. Then total cost \eqref{eq.risk.w} simplifies to:
\begin{equation}
\mathcal{C}(\tilde{\rho})= {\rm Tr}(\G BVB^T) + {\rm Tr}(\G B\tilde V B^T)
\end{equation}
and what remains is to perform optimization over $\tilde\rho$ (or, effectively over $\tilde V$).
 The uncertainty principle \eqref{eq:heisenberggaussian} applied to the ancillary variables $\tilde{\bf X}$ gives the constraint
\begin{equation*}
\tilde V\succeq \frac{i}{2} S.
\end{equation*}
Using the same reasoning as the one leading to \eqref{step2} the above condition implies that:
\begin{equation*}
{\rm Tr}(\G B\tilde V B^T) \geq
\frac{1}{2}{\rm Tr}(|\sqrt{\G}B SB^T\sqrt{\G}|), 
\end{equation*}
with equality for $\tilde V=(\sqrt{\G}B)^{-1}|\sqrt{\G}B SB^T\sqrt{\G}|(B^T\sqrt{\G})^{-1}$, which satisfies the uncertainty condition \eqref{eq:heisenberggaussian}.
By choosing $\tilde{\rho}$ to be the corresponding Gaussian state,
we conclude that
\begin{equation*}
\CQ=  {\rm Tr }\left(\G BVB^T\right)+
\frac{1}{2}{\rm Tr}\left(|\sqrt{\G} B SB^T \sqrt{\G} |\right).
\end{equation*}
Since $B$ is the solution of \eqref{Bholevo}, we recover the HCR bound.

Note that the above construction of the optimal linear measurement is very similar in its spirit to the reasoning presented in Sec.~\ref{sec:hcrequivalent} leading to \eqref{matsu}. It utilizes an extended space in order to
make the measurement operators commuting on the extended space. However, unlike the reasoning presented here,
the derivation presented in Sec.~\ref{sec:hcrequivalent} does not necessarily
provide an explicit construction of a measurement that saturates the HCR bound, indeed it does so only in specific cases such as the pure states models discussed in Sec.~\ref{sec:saturability}.

Another special feature of the Gaussian shift model which stems from its covariance with respect to shifts, is the fact that the optimal measurement is independent of the actual value of $\bvar$.
To further emphasise the fundamental role of such models in quantum statistics, the QLAN theory described in Sec.~\ref{sec:qlan} shows that such models arise as asymptotic limits of quantum i.i.d. models where Gaussian shifts emerge from collective local unitary rotations in i.i.d. models.

Let us finish this section by considering three basic examples, which in the light of the QLAN discussed in Sec.~\ref{sec:qlan} will
be related with the three qubit model examples presented in Sec.~\ref{sec:examplequbit}.

\subsubsection{Two quantum variables.}
We first consider the standard joint position and momentum estimation problem on a single quantum mode with no classical variables, which corresponds to the following choice of Gaussian shift model parameters:
 $\mathfrak{p}=2$, $\mathfrak{r}=2$ ($\mathfrak{q}=1$, $\mathfrak{c}=0$), $\bvar = (q,p)$, $A = \t{diag}[1,1]$, $V= \t{diag}[\sigma^2_q, \sigma^2_p]$ (we assume no $q$, $p$ correlations for simplicity).
 Since in this case $\mathfrak{r}=\mathfrak{p}$ we can use Eq.~\eqref{eq:hcrgauss} and for the cost matrix $C=\t{diag}[1,1]$ we obtain the cost of the joint estimation of momentum and position exceeding the SLD CR bound by an amount equal to twice the vacuum fluctuation contribution:
 \begin{equation}
 \CH_{(q,p)} = \sigma_q^2 + \sigma_p^2 + 1,   \quad \mathcal{C}^{\t{SLD}}_{(q,p)}=\sigma_q^2 + \sigma_p^2.
 \end{equation}
In particular, when $\varrho$ is the minimum uncertainty state, $\sigma_p^2=\sigma_q^2=1/2$,
and we rescale the estimation parameters by choosing  $A = \frac{1}{\sqrt{2}}\t{diag}[1,1]$, the bounds take exactly the same values as for the pure qubit state estimation example, see Eqs.~(\ref{eq:hcrqubit1},\ref{eq:sldqubit1}).

\subsubsection{One quantum + one classical variable.}
Second, consider a situation when apart from a value of a single quantum canonical variable $Q$ the goal is to estimate an independent classical variable $Z$. Formally this correspond to the choice: $\mathfrak{p}=2$, $\mathfrak{r}=3$ ($\mathfrak{q}=1$, $\mathfrak{c}=1$),
$\bvar = (q,z)$,
\begin{equation}
A = \begin{bmatrix} 1 & 0 \\ 0 & 0 \\  0 & 1 \end{bmatrix}, \quad V = \t{diag}[\sigma_q^2,\sigma_p^2,\sigma_z^2].
\end{equation}
Even though we do not deal here with the $\mathfrak{p}=\mathfrak{r}$ case, if we choose the cost matrix to be diagonal
$C = \t{diag}[1,1]$, we can still use \eqref{eq:hcrgauss} since there are no correlations between $(Q,Z)$ and $P$  and hence we can simply ignore the latter---formally this corresponds to choosing $B$ equal to the pseudoinverse of $A$ which in this case corresponds to $B=A^T$.  As a result we get
\begin{equation}
 \CH_{(q,z)} = \sigma_q^2 + \sigma_z^2 = \mathcal{C}^{\t{SLD}}_{(q,z)}
 \end{equation}
and since HCR coincides with the SLD CR bound it implies that as expected there is no measurement incompatibility issue here.
Moreover, if we choose $\varrho$ to have the following $q,p$, variances , $\sigma_p^2=\sigma_q^2=\frac{1}{2 r}$,
rescale the estimation parameters $q,p$ by $\frac{1}{\sqrt{2 r}}$ and choose $\sigma_z^2 =1-r^2$ we
get the same bound for the cost as in the $(r,\theta)$ qubit estimation example, for the Euclidean cost choice $c(r)=1$, see Eq.~\eqref{eq:hcrqubit2}---we may also obtain the cost corresponding to arbitrary $c(r)$ function, by simply choosing the cost matrix in the Gaussian model by
$C = \t{diag}[1,c(r)]$.
\begin{center}
\begin{table}[t]
\begin{tabular}{|c|c|c|c|c|c|c|}
\hline
 &  \multicolumn{3}{c|}{qubit}& \multicolumn{3}{c|}{Gaussian}  \\
 &  \multicolumn{1}{c}{$(\theta,\varphi)$} &  \multicolumn{1}{c}{$(r, \theta)$ }&  $(r,\theta,\varphi)$ & \multicolumn{1}{c}{$(q,p)$} &  \multicolumn{1}{c}{$(q,z)$}&  $(q,p,z)$ \\
 \hline
 $\mathcal{C}^{\t{SLD}}$ & $2$  & $1 +c(r)(1-r^2)$ & $2 + c(r)(1-r^2)$  & $\sigma_q^2+\sigma_p^2$   & $\sigma_q^2+\sigma_z^2$ & $\sigma_q^2+\sigma_p^2 +\sigma_z^2$ \\
 \hline
 $\mathcal{C}^{\t{H}} - \mathcal{C}^{\t{SLD}} $ &  $2$ &  $ 0 $   & $2r$ & $1$ &$0$ & $1$   \\
 \hline
 {\tiny \begin{tabular}{c}
 asymptotic\\
measurement \\
incompatibility
 \end{tabular}}
  & $+$ & $-$ & $+$& $+$  &$-$ & $+$\\
 \hline
 {\tiny
\begin{tabular}{c}
 advantage of \\
 collective\\
 measurements
\end{tabular}}
 & $-$ &$+$  & $+$ &   $-$ & $-$& $-$\\
 \hline
\end{tabular}
\caption{Summary of the results for the SLD CR as well as HCR bounds obtained for the qubit as well as Gaussian shift model examples.
In the Gaussian models the cost matrix is chosen to be $\G = \t{diag}[1,1,1]$, for $(q,p,z)$ variables, while in the qubit models
 $\G = \t{diag}[c(r), r^2, r^2 \sin^2\theta]$, for $(r,\theta,\varphi)$ variables. If we choose the Gaussian state to have
the following variances of the canonical variables $\sigma_q^2= \sigma_p^2 = \frac{1}{2r}$, $\sigma_z^2 = 1-r^2$ and furthermore we rescale the  $q$ and $p$ parameters by a factor $\tfrac{1}{\sqrt{2r}}$ the bounds for the Gaussian models will coincide with the corresponding qubit models
for the Euclidean distance cost $c(r)=1$
---a manifestation of the general QLAN theorem discussed in Sec.~\ref{sec:qlan}
}
\label{tab:examples}
\end{table}
\end{center}

\subsubsection{Two quantum + one classical variable. }
Finally, consider the model which combines the two above cases and
corresponds to  $\mathfrak{p}=3$, $\mathfrak{r}=3$ ($\mathfrak{q}=1$, $\mathfrak{c}=1$), $\bvar = (q,p, z)$, $A = \t{diag}[1,1,1]$,
$V= \t{diag}[\sigma^2_q, \sigma^2_p, \sigma^2_z]$. Using \eqref{eq:hcrgauss} and \eqref{eq:sldgauss} we get:
\begin{equation}
\label{eq:hcrgauss3}
 \CH_{(q,p,z)} = \sigma_q^2 + \sigma_p^2 + \sigma_z^2 +1, \qquad \mathcal{C}^{\t{SLD}}_{(q,p,z)}=\sigma_p^2 + \sigma_q^2 + \sigma_z^2.
 \end{equation}
If we again choose $\varrho$ to have $\sigma_p^2=\sigma_q^2=\frac{1}{2r}$, rescale the estimation parameters $q,p$ by $\frac{1}{\sqrt{2 r}}$ and choose $\sigma_z^2 = 1-r^2$ we get the same bounds for the cost as in the $(r,\theta,\varphi)$ qubit estimation example for $c(r)=1$, see Eq.~\eqref{eq:hcrqubit3}.

Tab.~\ref{tab:examples} summarizes the results obtained in this subsection, and may be regarded as a take home message
that allows to understand the difference between various multi-parameter models in terms of how the achievable precision deviates from the one predicted by the SLD CR bound and the role of collective measurements in achieving the fundamental bound.
The similarity between the three qubit and three Gaussian examples is no coincidence and will become clear in the light if the QLAN
considerations presented in the next section.

\section{Quantum local asymptotic normality}
\label{sec:qlan}
We have ended the previous section with a list of examples of qubit and Gaussian shift models  that illustrated the essential features of multi-parameter quantum estimation. In this section we will see that the link between qubit and Gaussian estimation problems is stronger than one might expect at first sight, and the relation between these models is captured by the concept of quantum local asymptotic normality (QLAN)
\cite{GutaKahn, GutaJanssensKahn, KahnGuta, GutaJencova, KahnGuta2, Gill2011, yamagata2013quantum, Yang2019}.
Informally, QLAN states that in the limit of large $n$, the statistical model describing
independent ensembles of $n$ identically prepared finite dimensional systems can be approximated (locally in the parameter space) by a certain Gaussian shift model. This has three important consequences:
\begin{enumerate}
\item{It provides an asymptotically optimal estimation strategy for independent ensembles, which amounts to pulling back
the optimal Gaussian measurement to a collective measurement on the ensemble, by means of quantum channels.}
\item{When combined with the universal saturability of the HCR bound for Gaussian shift models, see Sec.~\ref{sec:examplegaussian}, QLAN implies that the HCR bound is asymptotically saturable on any multiple-copy models that satisfy certain regularity assumptions.}
\item{The optimal measurement of point (i) has asymptotically normal distribution, which provides asymptotic confidence regions for the estimator.}
\end{enumerate}

For a better understanding of QLAN, we first provide some intuition regarding the classical local asymptotic normality (LAN) concept. Classical LAN  \cite{vanderVaart} has very broad applicability including non-parametric estimation (estimation of infinite dimensional parameters, as in density estimation problems), and statistical problems involving non-i.i.d. data such as (hidden) Markov processes and time series. Here we will focus on parametric (finite dimensional) models with independent identically distributed (i.i.d.) samples, which serve us as a guide towards understanding the structure of quantum multi-copy models and the problem of optimal quantum state estimation.

\subsection{LAN in classical statistics}
Let us consider an i.i.d setting, where $n$ independent samples $m_1, \dots ,m_n$ are drawn from the probability distribution $p_\bvar(m)$ which depends smoothly on $\bvar\in \Theta\subset\mathbb{R}^{\mathfrak{p}}$. Since we expect the statistical uncertainty to scale as $n^{-1/2}$ with the increasing number of
samples, we will analyse this model at the local level and express parameters $\bvar$ in the neighbourhood of a fixed point $\bvar_0$ as
\begin{equation}
\bvar = \bvar_0 + \mathbf{u}/\sqrt{n}.
\end{equation}
Thanks to this reparametrization, we expect the asymptotic formulas
for the estimation precision of $\mathbf{u}$ to be independent of $n$.


Furthermore, let us denote by $\mathcal{N}(\mathbf{u}, V)$ a classical Gaussian shift model, which consists of drawing a single sample $\mathbf{x} \in \mathbb{R}^{\mathfrak{p}} $ from a normal distribution with mean $\mathbf{u}$ and the covariance matrix $V$.
Informally, LAN states that for large $n$, the i.i.d. model  $p^n_{\bf u}:= p^n_{\bvar_0 +\mathbf{u}/\sqrt{n}}$ is close to the Gaussian shift
model $\mathcal{N}({\bf u} , F^{-1}_{\bvar_0})$, where $F_{\bvar_0}$ is the FI matrix for the $p_\bvar$ distribution calculated at $\bvar=\bvar_0$.
Note that both models have the same Fisher information, and the CR inequality is attained in the Gaussian case by simply taking $\mathbf{x}$ as the estimator of the mean.

In order to understand in what sense the two models are \emph{close} to each other, consider the likelihood process
defined as the `random function' $\bvar\mapsto p_\bvar$ (a random variable with values $p_\bvar(m)$ for each $\bvar\in \Theta$).
For our purposes it is more interesting to look at the \emph{log-likelihood process}, which is defined with respect to a fixed reference point $\bvar_0$
\begin{equation}
\bvar\mapsto l_\bvar:= \log \frac {p_\bvar}{p_{\bvar_0}}.
\end{equation}
This is in fact a \emph{sufficient statistic}, which means that it captures the entire statistical information contained in the original samples.

%

In the specific case of the i.i.d sequence $p^n_{\bf u}$ with parameter
${\bf u}$, the log-likelihood ratio (with respect to ${\bf u}=0$) is
\begin{equation}
 l^n_{\bf u}
:= \sum_{i=1}^n l_{\bvar_0 + {\bf u}/\sqrt{n}}(m_i)=
\sum_{i=1}^n \log \frac{p_{\bvar_0 + {\bf u} /\sqrt{n}}}{p_{\bvar_0}} (m_i).
\end{equation}
By expanding $l_{\bvar_0 + {\bf u}/\sqrt{n}}$ to the second order with respect to $ {\bf u}/\sqrt{n}$  we obtain
\begin{equation}
l^n_{\bf u} = \frac{{\bf u}^T }{\sqrt{n}}\sum_{i=1}^n \boldsymbol{\nabla}{l}(m_i) +
\frac{1}{2n} \sum_{i=1}^n {\bf u}^T   \left[\boldsymbol{\nabla}\boldsymbol{\nabla}^T{l}(m_i)\right] {\bf u} + o(n^{-1}),
\end{equation}
where $\boldsymbol{\nabla}$ is the gradient operator with respect to $\bvar$ taken at $\bvar=\bvar_0$, while
$\boldsymbol{\nabla}\boldsymbol{\nabla}^T$ represent the  matrix of second derivatives (Hessian) at $\bvar=\bvar_0$.
By applying the central limit theorem (CLT) to the first sum and the law of large numbers to the second sum we obtain the (joint) convergence in distribution
\begin{equation}\label{eq.weak.lan}
l^n_{\bf u} \xrightarrow[p^n_{\mathbf{u}}]{~~n\to\infty~~}
{\bf u}^T F_{\bvar_0} \mathbf{x} - \frac{1}{2} {\bf u}^T F_{\bvar_0}{\bf u},
\end{equation}
where $\mathbf{x}$ is a real random variable with distribution $\mathcal{N}(0, F^{-1}_{\bvar_0})$. Note that the right hand side is the log-likelihood ratio of the Gaussian shift model $\mathcal{N}({\bf u}, F^{-1}_{\bvar_0})$ with respect to the reference point ${\bf u}=0$.
A similar result can be shown for an arbitrary local parameter as reference. This amounts to what is called \emph{weak convergence} of the i.i.d. model $p^n_{\bf u}$  to the Gaussian limit model
$\mathcal{N}({\bf u}, F^{-1}_{\bvar_0})$.
In the next subsection we will describe a quantum version of weak LAN; we will then introduce the notion of \emph{strong} LAN which allows for a more complete understanding of the Gaussian approximation, and the solution of the optimal estimation problem in the asymptotic regime. 

\subsection{Weak convergence approach to QLAN}
\label{sec:weakLAN}
A quantum i.i.d. version of the weak LAN convergence has been established in \cite{GutaJencova} and a different approach was taken in \cite{yamagata2013quantum,FujiwaraYamagata}. In the specific setup of pure state models weak convergence corresponds roughly to the geometric idea of convergence of state overlaps and can be used to derive LAN for correlated states such as outputs of quantum Markov processes (or stationary, purely-generated finitely correlated states) \cite{GutaKiukas1,GutaKiukas2}. However, for mixed states models, the theory of weak convergence is currently still in its infancy and the notion of strong convergence, discussed in Sec.~\ref{sec:QLANstrong}, appears to be a
more versatile tool which yields operationally meaningful statements.

\subsubsection{Single parameter pure state model.}

For an intuitive illustration we will start by considering the special case of a single parameter pure state model consisting of a unitary rotation family $| \psi_\theta\rangle$ with $\theta\in \mathbb{R}$ and a selfadjoint generator $G$
\begin{equation}
| \psi_\theta\rangle := e^{i \var G} | \psi_0 \rangle,\qquad
\langle \psi_0 |G| \psi_0\rangle =0.
\end{equation}
The corresponding QFI is
$F_Q= 4 {\rm Var} (G) = 4\langle \psi_0 |G^2 |\psi_0\rangle$ and does not depend on $\theta$.
We consider an ensemble of $n$ independent systems, and assume that the parameter is of the order of the statistical uncertainty, so that $\theta =\theta_0+ u/\sqrt{n}$ with $\theta_0$ fixed and known and $u$ an unknown `local parameter'.
The joint state of the ensemble is
\begin{equation}
|\psi^n_{u}\rangle := |\psi^{\otimes n}_{\theta_0 + u/\sqrt{n}}\rangle.
\end{equation}
Since the QFI is additive and the parameter has been rescaled accordingly, the model $|\psi^n_{u}\rangle$ has Fisher information $F_Q$.

In addition to the i.i.d. model, we consider the \emph{quantum Gaussian shift model} consisting of coherent states of a one-mode continuous variables system with canonical coordinates $(Q,P)$
\begin{equation}
\left|\sqrt{\tfrac{F_Q}{2}}u\right\rangle := e^{-i u \sqrt{\tfrac{F_Q}{2}} P} |{\bf 0}\rangle,
\end{equation}
where $ |{\bf 0}\rangle\in\mathcal{F}$ denotes the vacuum state. The model is parametrised by $u$, such that the expectations of $(Q,P)$
are $(\sqrt{\tfrac{F_Q}{2}}u, 0)$, and it has quantum Fisher information $F_Q$.

Since a pure state model is a family of Hilbert space vectors, its structure is uniquely determined by the inner products of pairs of vectors with different parameters. Therefore it is natural to say that a sequence of models converges to a limit model if such overlaps converge pointwise
(see \cite{GutaKiukas1} for a more general discussion taking into account the phase ambiguity).
We will call this notion the \emph{weak convergence} of quantum statistical models.
\begin{figure}
\begin{center}
\includegraphics[width=10cm]{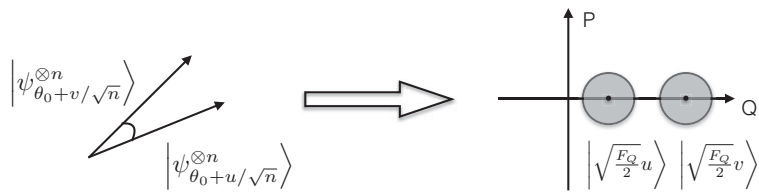}
\end{center}
\caption{In the case of one-parameter pure state models, local asymptotic normality can be understood as convergence of inner products of states with local parameters $u$ and $v$ to the inner product of the corresponding coherent states in a one-mode Gaussian shift model, whose mean encodes the unknown local parameter. A similar result holds for multidimensional models.}
\label{fig.weak.convergence}
\end{figure}

The following calculation shows that  the sequence of models
$|\psi_u^n\rangle$  converges weakly to the limit model
$ | \sqrt{F_Q/2} u\rangle\ $, as illustrated in Fig.~\ref{fig.weak.convergence}
\begin{multline}
\langle\psi^n_{u} | \psi^n_{v} \rangle =
\left\langle \psi_0 \left| e^{i \frac{(u-v)}{\sqrt{n}}G} \right|\psi_0\right\rangle^n =
\left(1 -\tfrac{(u-v)^2 F_Q }{8n} + O(n^{-3/2})\right)^n \\
\xrightarrow{~~n\to\infty~~}
\exp\left(- \tfrac{(u-v)^2 F_Q }{8}\right) =
\left\langle \sqrt{\tfrac{F_Q}{2}} u \right|\left. \sqrt{\tfrac{F_Q}{2}} v\right\rangle.
\label{eq.weak.qlan}
\end{multline}
Note that even though we deal with a one-dimensional pure states model, the limit model is not classical as one might expect but another pure state quantum model. This reflects the fact that the limit model may be used for different statistical problems (e.g. parameter estimation, testing) whose optimal measurements are incompatible, and is related to the fact that the SLD is not $\mathcal{D}$ invariant (see Sec.~\ref{sec:Dinvariance}).



\subsubsection{Two-parameter pure qubit model.}

In order to understand the measurement incompatibility from the QLAN perspective, we will now consider a two-dimensional qubit model obtained by applying a small rotation to one of the basis vectors
\begin{equation}
\ket{\psi_{\bvar}} = e^{\frac{i}{2}( \var_1 \sigma_x -   \var_2 \sigma_y)} |0\rangle, \qquad \bvar= (\var_1, \var_2).
\end{equation}
Note that up to a unitary rotation this model is locally equivalent to the pure qubit state
estimation model $(\theta,\varphi)$ discussed in Sec.~\ref{sec:examplequbit1}.

The joint state of an i.i.d. ensemble of $n$ qubits is expressed in terms of the local parameter $ {\bf u} = (u_1, u_2)^T\in \mathbb{R}^2$ around $\bvar_0 =0$ as
\begin{equation}
|\psi^n_{{\bf u}} \rangle = |\psi^{\otimes n}_{{\bf u}/\sqrt{n}}\rangle:= \left( e^{\frac{i}{2}( u_2\sigma_x -  u_1\sigma_y)/\sqrt{n} } | 0 \rangle\right)^{\otimes n}.
\end{equation}
The corresponding SLDs $(L_1^n, L_2^n)$ at ${\bf u}=0$ and the generators $(G_1,G_2)$ are given by the collective spin observables
\begin{equation}
L^n_1 = 2 G_2 = \frac{1}{\sqrt{n}}\sum_{i=1}^n  \sigma^{(i)}_x = \frac{2}{\sqrt{n}} J_x , \qquad
L^n_2 = - 2 G_1 = \frac{1}{\sqrt{n}}\sum_{i=1}^n  \sigma^{(i)}_y = \frac{2}{\sqrt{n}} J_y.
\end{equation}
Since $\langle \psi_{\bf 0} | [L_1^1, L_2^1] |  \psi_{\bf 0} \rangle \neq 0$, the SLD CR bound is not achievable even in the asymptotic sense. This was reflected in the discussion in Sec.~\ref{sec:examplequbit} where we found that the HCR bound was strictly larger than the SLD CR bound.

In the same vein as the calculation \eqref{eq.weak.qlan}, it can be shown that the following `weak convergence' holds
$$
\left\langle \psi^n_{{\bf u}} | \psi^n_{{\bf v}}\right\rangle \xrightarrow{~~n\to\infty~~}
 \left\langle \tfrac{1}{\sqrt{2}}{\bf u} \right.  \left| \tfrac{1}{\sqrt{2}}{\bf v} \right\rangle,
$$
where $ | \tfrac{1}{\sqrt{2}}{\bf u} \rangle$ are coherent states forming a quantum Gaussian shift model
\begin{equation}
\varrho_{\mathbf{u}} =
\left| \tfrac{1}{\sqrt{2}} {\bf u} \right\rangle  \left\langle \tfrac{1}{\sqrt{2}} {\bf u}\right|,
\qquad
\ket{\psi_\mathbf{u}}
=
\left| \tfrac{1}{\sqrt{2}}{\bf u} \right\rangle= e^{\frac{i}{\sqrt{2}}( u_2 Q -u_1 P)} \ket{{\bf 0}}
\end{equation}
with $\ket{{\bf 0}}\in \mathcal{F}$ again denoting the vacuum state.
The same conclusion can be reached by using the quantum central limit theorem \cite{Petz} to show that the `joint distribution' (or more precisely the joint moments) of the SLDs of $|\psi^n_{\mathbf{u}} \rangle$ converges to that of the SLDs of the Gaussian model, which are equal to $(\sqrt{2}Q, \sqrt{2}P )$ at ${\bf u}=0$. Similarly, the generators of collective spin rotations converge to those of phase translations for the Gaussian model. This is in fact a statistical take on the well known Holstein-Primakov theory of coherent spin states \cite{HolsteinPrimakov,Radcliffe}, but holds generally for any i.i.d. pure state model.

Using the above relation we may now use
the known results for the minimal cost of estimation in the Gaussian shift model, as discussed in Sec.~\ref{sec:examplegaussian},
to infer the analogous results for the asymptotic qubit model.
For simplicity, as well as in order to stay in accordance with the corresponding  qubit example from Sec.~\ref{sec:examplequbit1},
we choose the cost matrix $C=\mathcal{I}$ so the corresponding cost
function  is  $\mathcal{C}(\mathbf{u},\mathbf{\tilde{u}}) =  \| \tilde{\bf u}-{\bf u} \|^2$. Using the notations of Sec.~\ref{sec:examplegaussian}, the Gaussian shift model considered here is a single mode ($\mathfrak{q}=1$),
two parameter ($\mathfrak{p}=2$) estimation model with matrices $A=\frac{1}{\sqrt{2}} \mathcal{I}$, $V=\frac{1}{2}\mathcal{I}$, and  according to \eqref{eq:hcrgauss} the resulting cost reads:
\begin{equation}
\CQ^{\varrho} =4.
\end{equation}
Note that half of the contribution is from the inherent incompatibilty of simultaneous measurement of $Q$ and $P$, as the SLD CR bound would yield $\mathcal{C} \geq \mathcal{C}^{\t{SLD}} =  2$. As expected, this result coincides with the HCR bound formula for the corresponding qubit model, see Eq.~\eqref{eq:hcrqubit1}.

The general optimal linear measurement construction discussed in Sec.~\ref{sec:examplegaussian} corresponds here to the standard heterodyne measurement \cite{Leonhardt} which can be implemented as follows. Consider an ancillary cv system $(\tilde{Q}, \tilde{P})$ prepared in the vacuum state $|{\bf 0}\rangle$, and measure the commuting pair of linear combinations $(Q+ \tilde{Q})/\sqrt{2}$ and $(P-\tilde{P})/\sqrt{2}$ for the joint state $|\frac{1}{\sqrt{2}}{\bf u}\rangle \otimes |{\bf 0}\rangle$. Physically, this can be interpreted as splitting the coherent state with a balanced beamsplitter and measuring different coordinates of the outgoing beams. The outcomes have  normal distribution
with covariance matrix $\mathcal{I}/2$ and mean $\mathbf{u}/2$, and when multiplied by $2$ yield the optimal unbiased estimator
$\tilde{\mathbf{u}}$ with the corresponding cost $\mathcal{C} = \mathbb{E} \|\tilde{\bf u}-{\bf u}\|^2=4$.

Based on the central limit argument and the optimality of the heterodyne measurement, we can now devise an asymptotically optimal measurement scheme for the original qubit estimation problem. 
In a first step (corresponding to the beamsplitter action)  the qubits ensemble is separated into two equal parts; then the two collective spins are measured on each sub-ensemble. The (appropriately rescaled) outcomes
$\tilde{\mathbf{u}}^n$ have asymptotically normal distribution $\mathcal{N}({\bf u}, 2 \mathcal{I})$ and
\begin{equation}
\lim_{n\to\infty} n \mathcal{C}^n =  \lim_{n\to\infty} n \mathbb{E} \|  \tilde{\bf u}^n - {\bf u}\|^2 = 4 = \mathcal{C}^{\varrho}.
\end{equation}
Note that even though formally the collective spins are measured, this measurement is equivalent to the local measurement strategy as discussed in Sec.~\ref{sec:examplequbit1} that saturated the HCR bound for the corresponding qubit model.

However,  our heuristic arguments rely on the vague assumption that we deal with
small rotations around a given state, rather than a completely unknown pure state. We will continue to ignore this for the moment but will revisit the issue in the context of strong QLAN where we detail a rigorous two step adaptive procedure for the optimal state estimation.


\subsection{Central limit argument for mixed qubit states} \label{sec.clt.mixed}

Let us now consider the extension of the previous qubit model to mixed states.
We are interested in the structure of the quantum model i.i.d. in a neighbourhood of size $1/\sqrt{n}$ of a given mixed state.  Without loss of generality, the latter can be chosen to be the state $\rho_{{\bf r}_0}$ with Bloch vector
${\bf r}_0 = (0,0, r_0)$ with $0 < r <1$.


Adopting the notation of Sec.~\ref{sec:examplequbit} we parameterise the neighbourhood of $\rho_{{\bf r}_ 0}$ using the `local parameter'
${\bf u}$  as follows
\begin{equation}
\rho_{{\bf r}_0 + {\bf u}/\sqrt{n}} =
\rho_{{\bf r}_0} +
\tfrac{1}{2\sqrt{n}}{\bf u} \boldsymbol{\sigma}=
\frac{1}{2}\left(
\begin{array}{cc}
1 + r_0 + \frac{u_3}{\sqrt{n}}  & \frac{u_1 -i u_2}{\sqrt{n}}\\
\frac{u_1+ iu_2}{\sqrt{n}} &1-r_0 - \frac{u_3}{\sqrt{n}}
\end{array}
\right), \qquad {\bf u} = (u_1, u_2, u_3)^T\in \mathbb{R}^3.
\end{equation}
The off-diagonal parameters $(u_1,u_2)$ describe a unitary rotation of $\rho_{{\bf r}_0}$, while
the diagonal parameter $u_3$ describes the change in eigenvalues.
The local i.i.d. model is given by $n$ independent qubits with the joint state
\begin{equation}\label{eq.qubits.mixed.ensemble}
\rho_{\bf u}^n = \rho_{{\bf r}_0 + {\bf u}/\sqrt{n}}^{\otimes n}
\end{equation}

As above, we use the quantum central limit theorem to uncover the structure of the limit Gaussian model. The three SLDs at
${\bf u}=0$ are again given by collective observables 
$$
L_1^n = \frac{1}{\sqrt{n}} \sum_{i=1}^n \sigma_x^{(i)}, \quad
L_2^n = \frac{1}{\sqrt{n}} \sum_{i=1}^n \sigma_y^{(i)}, \quad
L_3^n = \frac{1}{\sqrt{n}(1-r_0^2)}
\sum_{i=1}^n ( \sigma_z^{(i)} - r_0 \openone).
$$
By applying the quantum CLT with respect to the state $\rho_{\bf u}^n$ we obtain the joint convergence (in moments or characteristic function)
$$
(L_1^n, L_2^n, L_3^n)  \xrightarrow{~~n\to\infty~~} (L_1, L_2, L_3),
$$
where the limit observables $(L_1, L_2, L_3)$ are canonical variables of a Gaussian shift model whose state is denoted $\varrho_{\bf u}$.
To completely identify the Gaussian model we first compute the commutations relations of the SLDs at $\mathbf{u}=0$. By the CLT, the commutators of the limit SLDs are proportional to the identity, with coefficients given by the expectations of the commutators of one qubit SLDs
\begin{equation}\label{eq.commutations.sld}
[L_1, L_2] = {\rm Tr} (\rho_{{\bf r}_0} [L_1^1, L_2^1]) \openone= 2 i r_0 \openone,  \qquad
[L_1, L_3] = [L_2, L_3] = 0.
\end{equation}

This means that the first two coordinates can be identified (up to a constant) with those of a one-mode cv system
$(L_1, L_2)= (\sqrt{2r_0}Q, \sqrt{2r_0} P )$ while $L_3$ is a classical real valued random variable, as it commutes with all the others. The
corresponding FI matrix at ${\bf u}=0$ equals
\begin{equation}
\label{eq:QFIqubitmixed}
  (F_Q)_{ij} =\trace (\rho_{{\bf r}_0} L_i^1 \circ L_j^1 ) =
{\rm diag} \left[1,1, \frac{1}{1-r_0^2}\right].
\end{equation}
Finally, since
\begin{equation}
\trace \left( \left.\frac{\partial \rho_{{\bf r}_0 +{\bf u}}}{\partial u_i} \right|_{{\bf u} =0}  L^1_j \right) = (F_Q)_{ij},
\end{equation}
then in order for the Gaussian model to properly account for the shifts in the $\mathbf{u}$ parameter we need to have:
\begin{equation}\label{eq.means.sld}
\trace(\varrho_{\mathbf{u}} L_1 ) = u_1,\quad
\trace(\varrho_{\mathbf{u}} L_2 ) = u_2,\quad
\trace(\varrho_{\mathbf{u}} L_3 ) =  \frac{u_3}{1-r_0^2}.
\end{equation}
Based on equations \eqref{eq.commutations.sld}, \eqref{eq:QFIqubitmixed} and \eqref{eq.means.sld}, we conclude that the  Gaussian model is given by the quantum-classical state
\begin{equation}
\label{eq.classical.quantum.gaussian}
\varrho_{\bf u}= q_{\bf u} \otimes p_{\bf u},
\end{equation}
where $p_{\bf u}$ is the probability density of the normal random variable
$Z:=(1-r_0^2) L_3$  and $q_{\bf u}$
 is a single mode `displaced thermal state'. The mean and the covariance matrix of the canonical variables
 $(Q, P, Z)$ with respect to $\varrho_{\bf u}$  are given by
\begin{equation}
A \mathbf{u} = [\tfrac{1}{\sqrt{2r_0}}u_1,  \tfrac{1}{\sqrt{2r_0}}u_2, u_3]^T,
\qquad
V =\t{diag}[\tfrac{1}{2r_0}, \tfrac{1}{2r_0}, 1-r_0^2].
\end{equation}
Note that using the formula \eqref{eq:gaussfisher} for the QFI matrix in the Gaussian shift model we indeed recover
\eqref{eq:QFIqubitmixed}.

Consider now the estimation problem, and for simplicity let the cost function be the square Euclidean distance $\|\tilde{\bf u} - {\bf u}\|^2$ with the cost matrix $C= \mathcal{I}$. The optimal measurement for the quantum component is the heterodyne which provides independent unbiased estimators $(\tilde{u}_1, \tilde{u}_2)$ with distribution $\mathcal{N}((u_1, u_2), (1+r_0)\mathcal{I})$, while the estimator for $u_3$ is given by the classical component $\tilde{u}_3 := Z$.
The resulting cost will be
\begin{equation}
\label{eq:gauss3param}
\mathcal{C}^{\varrho} = 2(1+r_0) + (1-r_0^2) = 3 + 2r_0 -r_0^2,
\end{equation}
which according to Eqs.~(\ref{eq:hcrqubit3},\ref{eq:hcrgauss3}) coincides with the HCR bound for the qubit model as
well as for the corresponding Gaussian model.

However, when we now look back at the multi-copy qubit model it is not clear how to construct the qubits measurement corresponding to the limiting heterodyne, and how the classical variable emerges in the limit.
In the next section we introduce an alternative approach to QLAN which can answer these questions.

\subsection{Strong convergence approach to QLAN for qubits}\label{sec:QLANstrong}

Let us return to the i.i.d. qubit model $\rho^n_{\bf u}$ introduced in equation \eqref{eq.qubits.mixed.ensemble}.
We will analyse its structure using group representation theory, and proceed to define an operational notion of
strong convergence to the limit Gaussian model.

The qubits space
$(\mathbb{C}^2)^{\otimes n}$ carries commuting unitary representations of the symmetric group
$S_n$ and the unitary group SU(2)
\begin{eqnarray*}
~\pi(\sigma) : |\psi_1\rangle \otimes \dots \otimes |\psi_n\rangle &\mapsto &
|\psi_{\sigma^{-1}(1)}\rangle \otimes \dots \otimes |\psi_{\sigma^{-1}(n)} \rangle,\qquad \sigma\in S_n,\\
\pi^\prime(u) : |\psi_1\rangle \otimes \dots \otimes |\psi_n\rangle &\mapsto &
u|\psi_1\rangle \otimes \dots \otimes u|\psi_n\rangle  \qquad ~~~~~~~~u\in \t{SU(2)}.
\end{eqnarray*}
According to Weyl's Theorem \cite{Fulton1991} the space decomposes as a direct sum of tensor products of irreducible representations indexed by the total spin $j\in \mathcal{J}_n:= \{0 (\tfrac{1}{2}), \dots , \tfrac{n}{2}\}$
\begin{equation}
(\mathbb{C}^2)^{\otimes n} =
\bigoplus_{j\in \mathcal{J}_n} \mathbb{C}^{2j+1} \otimes \mathbb{C}^{m_j}, \quad
\pi(\sigma) = \bigoplus_{j\in \mathcal{J}_n} \openone_{2j+1} \otimes \pi_j(\sigma), \quad
\pi^\prime(u) = \bigoplus_{j\in \mathcal{J}_n} \pi^\prime_j(u)  \otimes\openone_{m_j},
\end{equation}
where $m_j$ is the dimension of the corresponding irreducible representation of $S_n$. By permutation symmetry, the joint state has a block-diagonal decomposition
\begin{equation}\label{eq.irrep.decomp}
\rho_{\bf u}^n = \bigoplus_{j\in \mathcal{J}_n} p_{\bf u}^{n,j} \rho_{\bf u}^{n,j} \otimes \frac{\openone_{m_j}}{m_j},
\end{equation}
where $p_{\bf u}^{n,j}$ is a probability distribution over $j\in \mathcal{J}_n$ and $\rho_{\bf u}^{n,j}$ are density matrices, both depending on ${\bf u}$. Let $P_j$ denote the projection onto the subspace $\mathbb{C}^{j+1} \otimes \mathbb{C}^{m_j}$. The decomposition \eqref{eq.irrep.decomp}
 implies that  the projective measurement $\{P_j\}_{j\in\mathcal{J}_n }$  is non-demolition for the family $\{\rho_{\bf u}^n\}$. The classical outcome $j$ and the corresponding quantum conditional state are the two components of a classical-quantum model which in the limit of large $n$ converges to the Gaussian model in equation \eqref{eq.classical.quantum.gaussian}, as shown below. Let us describe these components in more detail.


\subsubsection{Classical part.}
Let us denote by $J^{(n)}$ the classical random variable with probability distribution  $p_{\bf u}^{n,j}$ over $j \in \mathcal{J}_n$. This provides information about the eigenvalue parameter $u_3$ and (in the first order of approximation) does not depend on the `rotation parameters' $u_1,u_2$. Note that the measurement $\{P_j\}_{j\in\mathcal{J}_n }$  does not amount to measuring the $z$ component of the collective spin, but is rather to measuring the total spin
$$
J= \sqrt{J_x^2 + J_y^2 +J_z^2}, \qquad {\rm where} \qquad
J_j = \frac{1}{2}\sum_{i=1}^n\sigma_j^{(i)}.
$$
Intuitively however, we expect that for large $n$ the distributions of the two observables will be similar on the basis of the fact that $J_x$ and $J_y$ are centred and have standard deviations of the order $n^{1/2}$ while $J_z$ has mean of order $n$. Indeed, one can show \cite{GutaJanssensKahn} that $J$ satisfies the same central limit as $J_z$:
\begin{equation}\label{eq.zn}
Z_n:= \frac{1}{\sqrt{n}} (2J^{(n)} - r n) \xrightarrow{n\to\infty} \mathcal{N}(u_3, 1-r_0^2).
\end{equation}
Therefore, by defining the estimator  of $u_3$ as $\tilde{u}^n_3= Z_n $ we obtain the asymptotic MSE
$$
\lim_{n\to\infty}\mathbb{E} (\tilde{u}^n_3 - u_3)^2 = (1-r_0^2).
$$
Additionally,  $J^{(n)}$ satisfies a concentration property, namely it belongs to the interval
$\mathcal{J}_n^\delta :=[\frac{r n}{2} - n^{1/2 +\delta} ,\frac{r n}{2} + n^{1/2 +\delta}] $ with probability converging exponentially fast to one \cite{GutaJanssensKahn}.
Since the asymptotic estimation cost typically scales as $n^{-1}$, we can safely ignore events of
exponentially small probability and assume that $J^{(n)}\in \mathcal{J}_n^\delta$. We now proceed to
analysis of the quantum component of the statistical model under this assumption.

\subsubsection{Quantum part.}
Conditional on the measurement outcome $J^{(n)}$ taking the value $j\in \mathcal{J}_n^\delta$, the state of the
qubits ensemble is $\rho_{\bf u}^{n,j}\otimes \mathbb{1}/m_j$; since the right-side of the tensor product is trivial, it can be traced over and we remain with the  state $\rho_{\bf u}^{n,j}$ on
$\mathbb{C}^{2j+1}$. We now construct an explicit isometric embedding of $\mathbb{C}^{2j+1}$ into the Fock space $\mathcal{F}$ of a one-mode cv system, and use this to map the qubit states into a cv state.
Let $\{|j, m \rangle : m =-j,\dots ,j\}$ be the orthonormal basis of $\mathbb{C}^{2j+1}$ consisting of eigenvectors of $J_z$ (i.e.
$J_z |j, m \rangle = m |j, m \rangle $), and let $\{|k\rangle : k=0, 1,\dots\}$ be the Fock basis in
$\mathcal{F}$. Then the map
\begin{equation}
W^{j} : \mathbb{C}^{2j+1} \to \mathcal{F}, \qquad \qquad
 | j, m \rangle \mapsto  | j-m \rangle
\end{equation}
extends to an isometry, and defines a quantum channel
\begin{equation}\label{eq.Tj}
T^j: M(\mathbb{C}^{2j+1}) \to \mathcal{T}^1(\mathcal{F}) , \qquad
T^j (\rho) = W^j \rho W^{j^\dagger},
\end{equation}
where $\mathcal{T}^1(\mathcal{F})$ denotes the trace-class operators on $\mathcal{F}$. On the other hand, a reverse channel can be defined as
\begin{equation}\label{eq.Sj}
S^j :\mathcal{T}^1(\mathcal{F})\to M(\mathbb{C}^{j+1}) , \qquad S^j(\varrho) = W^{j\dagger} \varrho W^j +
{\rm Tr}( {Q}^j_\perp \varrho) \frac{\mathbb{1}_j}{j+1},
\end{equation}
whereby the state is first measured with projections $({Q^j := W^jW^{j\dagger}}, Q^j_\perp)$ and conditionally on the outcome, the isometry is reversed or the trivial state $\mathbb{1}_j/(2j+1)$ is prepared.
Now, the essence of the strong formulation of QLAN is that for \emph{all} typical values of $j$, the embedded state is well approximated by a Gaussian state, \emph{uniformly} over the local parameter $\bf{u}$. Formally this can be expressed as the following convergence statement that holds for $\epsilon$ small enough  \cite{GutaJanssensKahn}
\begin{align}
\label{theorem.lan.1}
&\lim_{n\to\infty} \, \max_{j\in \mathcal{J}_n^\delta} \, \sup_{\|{\bf u} \| \leq n^{\epsilon}}\,
\left\| T^j (\rho_{\bf u}^{n,j}) - q_{\bf u} \right\|_1 = 0, \\ \nonumber
&\lim_{n\to\infty} \, \max_{j\in \mathcal{J}_n^\delta} \, \sup_{\|{\bf u} \| \leq n^{\epsilon}}\,
\left\| S^j (q_{\bf u}) - \rho_{\bf u}^{n,j} \right\|_1 = 0,
\end{align}
where  $\rho_{\bf u}^{n,j}$ is the conditional state in the decomposition \eqref{eq.irrep.decomp}, $T^j$ and $S^j$ are the channels defined in \eqref{eq.Tj} and respectively \eqref{eq.Sj} and $q_{\bf u}$ is the density matrix of the quantum part of the Gaussian state
as defined in \eqref{eq.classical.quantum.gaussian}.

What is the difference between this statement and the central limit approach to LAN of the previous section? The latter is a statement about pointwise convergence (for fixed ${\bf u}$)  of the `joint distribution' of SLDs from the multi-copy model to those of the Gaussian model. The former is an \emph{operational procedure} for mapping the first model into the second one (and backwards) while controlling the trace-norm approximation errors uniformly with respect to the parameter ${\bf u}$. As we will see in the next section, this is crucial in defining the measurement strategy and proving its optimality.

\subsubsection{Strong QLAN for classical and quantum components}
The convergence in \eqref{theorem.lan.1} concerns only the `quantum part' of the i.i.d. model, but a similar statement can be made for the full classical-quantum model as follows. For each $n$ we can define a `quantum to quantum-classical' channel
$$
T_n: M((\mathbb{C}^{2})^{\otimes n} ) \to \mathcal{T}^1(\mathcal{F}) \otimes L^1(\mathbb{R})
$$
whose action is described by the following sequence of operations \cite{GutaJanssensKahn}:


\begin{enumerate}
\item{
Measure $\{P_j\}_{j\in\mathcal{J}_n }$ to obtain outcome $J^{(n)}=j$ and conditional state $\rho^{n,j}_{\bf u}$.
}
\item{
Rescale $J^{(n)}$ to obtain $Z_n$ defined in equation \eqref{eq.zn}}.
\item{
Randomise $Z_n$ by adding an independent sample from a centred normal distribution with variance $1/(2{\sqrt{n}})$.
to obtain the classical output of the channel $T_n$
}
\item{
Map $\rho^{n,j}_{\bf u}$ through the channel $T^j$ to obtain the quantum output of the channel $T_n$}.
\end{enumerate}
Step (3) requires some justification. While $Z_n$ converges \emph{in distribution} to the desired Gaussian
$\mathcal{N}(u_3, 1-r_0^2)$ (cf. equation \eqref{eq.zn}), the fact that it takes discrete values prevents it from converging also in the norm-one sense used in \eqref{theorem.lan.1}. This can be remedied by adding a `small' continuous noise without spoiling the statistical information (see \cite{Kahnthesis} for details). A similar procedure can be used to define the reverse channel $S_n$ by discretising the Gaussian variable to obtain a sample $\tilde{j} \in \mathcal{J}_n$ and mapping the quantum Gaussian through the corresponding channel $S_{\tilde{j}}$. The following result is similar to Eq.~\eqref{theorem.lan.1} but captures  the convergence to classical-quantum Gaussian model more clearly:
\begin{align}
\label{th.lan.2}
&\lim_{n\to\infty}  \, \sup_{\|{\bf u} \| \leq n^{\epsilon}}\,
\left\| T_n (\rho_{\bf u}^{n}) - \varrho_{\bf u} \right\|_1 = 0 \\ \nonumber
&\lim_{n\to\infty}  \, \sup_{\|{\bf u} \| \leq n^{\epsilon}}\,
\left\| S_n (\varrho_{\bf u}) - \rho_{\bf u}^{n} \right\|_1 = 0.
\end{align}
where $\rho_{\bf u}^{n}$ be the i.i.d. state \eqref{eq.qubits.mixed.ensemble}, $T_n$ and $S_n$ are the channels defined above
and $\varrho_{\bf u}  = q_\mathbf{u} \otimes p_{\bf u}$ is the quantum-classical Gaussian state as defined in \eqref{eq.classical.quantum.gaussian}.

\subsection{Asymptotically optimal estimation strategy and the region of applicability}
\label{sec:estimation.strategyLAN}

We now come to the key issue demonstrating the power of the strong QLAN formulation. In  Sec.~\ref{sec:hcr},
while discussing the HCR and SLD CR bounds as well as the optimal estimation strategies that saturated the bounds, we have stayed within the l.u. estimation paradigm.
This approach has a significant deficiency in that the derived optimal estimation strategies are guaranteed to
perform as expected only in the direct vicinity of a fixed parameter, and is a priori unclear what parameter region may be covered by an estimation strategy derived within this paradigm. This in some extreme cases may lead
 to controversies, such as e.g. that surrounding the  correct scaling constant in the actually achievable Heisenberg limit \cite{Gorecki2019a}.

Fortunately, this problem is absent for the Gaussian shift models, see Sec.~\ref{sec:examplegaussian}, where the optimal linear measurement strategy does not depend on the unknown parameter.
Since the essential message of  QLAN is the equivalence of asymptotic multi-copy models with the Gaussian shift models
it is highly relevant to understand to what extent the uniform achievability for Gaussian models translates to multi-copy models. We will show that this is indeed the case in the asymptotic setting, as is already hinted at by the convergence results \eqref{theorem.lan.1} and \eqref{th.lan.2}.

Let us recall that the original parameter ${\bf r}$ (the Bloch vector)
is related to the local one as ${\bf r} = {\bf r}_0 + {\bf u}/\sqrt{n}$ where ${\bf r}_0$ is a fixed and known. Therefore, Eqs.~\eqref{theorem.lan.1} and \eqref{th.lan.2} say that if we restrict ourselves to regions in the parameter space of size $n^{-1/2+\epsilon}$, then the i.i.d. model $\rho_{\bf r}^{\otimes n}$ can be approximated by a simple Gaussian shift model. Since the approximation has an operational meaning in terms of quantum channels, this promises to simplify the estimation task, at least in the limit of large $n$. But can we claim that this is consistent with the setting where ${\bf r}$ is assumed to be completely unknown? We will show that this can be done by using the following two step adaptive procedure.

\begin{enumerate}
\item{ \emph{Localise parameter:} use  $\tilde{n}= n^{1-\epsilon}$ samples to produce a rough estimator $\tilde{\rho}$ such that
the probability that $\|\tilde{\rho} - \rho\| > n^{-1/2 +\epsilon}$ is exponentially small;
for instance, the combination of Pauli measurements and maximum likelihood has this property \cite{GutaKahn}. Since the `large deviation' event has small probability, it does not contribute to the asymptotic cost (which scales as $1/n$) and can be ignored.
}
\item{\emph{rotate to diagonal state:}
the remaining $n^\prime = n-\tilde{n}$ samples are  rotated by means of a unitary $U$ which diagonalises $\tilde{\rho}$, i.e.  $U\tilde{\rho} U^\dagger = \rho_{{\bf r}_0}= \frac{1}{2}(\openone + r_0 \sigma_z)$. Here $r_0$ is not known in advance but can be considered fixed and known from this point on. The rotated qubits can be parametrised as $\rho_{\bf u}$ with local parameter $\|{\bf u}\|\leq n^\epsilon$.
}
\end{enumerate}

After this localisation procedure we can assume that the unknown parameter is in the local region in which the QLAN approximation is valid. In the next steps we describe an asymptotically optimal measurement strategy. Since $n^\prime/n \to 1$, replacing $n^\prime$ by $n$ will
not change the asymptotic analysis and so for simplicity we will continue to use $n$.

\begin{enumerate}
\addtocounter{enumi}{2}
\item{\emph{Estimate eigenvalue parameter:} perform the measurement $\{P_j\}_{j\in\mathcal{J}_n }$ which projects onto a tensor products of irreducible representations
$\mathbb{C}^{2j+1}\otimes \mathbb{C}^{m_j}$. Rescale outcome $J^{(n)}$ to obtain estimator $\tilde{u}^n_3 = Z_n$ as defined in equation \eqref{eq.zn}.
}
\item{\emph{Embed into Fock space:} map the conditional state $\rho^{n,j}_{\bf u}$ through the channel $T^j$, which
 according to \eqref{theorem.lan.1} is close to the Gaussian state $q_\mathbf{u}$}
\item{ \emph{Estimate the rotation parameters:} apply the heterodyne measurement.
The rescaled outcome is the estimator
$(\tilde{u}^n_1, \tilde{u}^n_2)$ which according to the convergence result \eqref{theorem.lan.1}, has asymptotic distribution $\mathcal{N}((u_1, u_2), (1+r_0)\mathcal{I})$
}
\item{\emph{Compute the final estimator:} using $\tilde{\bf u}^n := (\tilde{u}^n_1, \tilde{u}^n_2, \tilde{u}^n_3)^T$, we define the final estimator of the qubit state
\begin{equation}
\tilde{\rho}_n : =U^\dagger \rho_{ {\bf r}_0 + \tilde{\bf u}^n/\sqrt{n}} U.
\end{equation}}
\end{enumerate}
Note that steps (iii) and (iv) amount to mapping the i.i.d. state through the channel $T_n$ defined earlier, with the exception of the randomisation step which is only needed for `technical'
reasons in \eqref{th.lan.2}.

For the rest of this section we discuss in what sense this procedure is asymptotically optimal. We define our figure of merit in terms of the maximum cost (risk) and minimax estimators, as is customary in mathematical statistics. As before, we consider the standard  cost function
given by the Euclidean distance  $ C(\tilde{\bf r}, {\bf r} ) = \| \tilde{\bf r} -{\bf r} \|^2$.
Let us fix the Bloch vector ${\bf r}_0$ and consider the estimation error for states in the neighbourhood of $\rho_{{\bf r}_0}$.
We will use the \emph{local maximum} cost of
$\tilde{\bf r}_n$ (or $\tilde{\rho}_n = \rho_{\tilde{\bf r}_n}$) which is defined as
$$
\mathcal{C}^n_{\textrm{max},c}(\tilde{\bf r}_n) = \sup_{ \|{\bf r}-{\bf r}_0\|^2  \leq c/n }
\mathbb{E} \| \tilde{\bf r}_n - {\bf r} \|^2,
$$
where $c$
is a positive constant, and the expectation is computed with respect to the state
$\rho^{\otimes n}_{\bf r}$. This reflects the hardness of the estimation problem around
${\bf r}_0$ and is both more operationally meaningful than a single point cost for an l.u. strategy as well as
more informative than the maximum cost over all parameters, or the Bayes cost for a particular prior distribution.
Since $\mathcal{C}^n_{\textrm{max},c}$ scales as $n^{-1}$ for reasonable estimators, the asymptotic behaviour of the best estimator is determined by the constant called the \emph{local asymptotic minimax} (LAM) cost at ${\bf r}_0$
\begin{equation}
\mathcal{C}_{\rm minmax} :=
\sup_{c>0}
\underset{n\to\infty}{\lim\sup}  \,\inf_{\tilde{\bf r}_n} \,
n \mathcal{C}^n_{\textrm{max},c} (\tilde{\bf r}_n),
\end{equation}
where the dependence on the constant $c$ is lifted in the last step. A sequence of estimators $\tilde{\rho}_n= \rho_{\tilde{\bf r}_n}$ is LAM if it achieves the LAM cost in the limit of large $n$
\begin{equation}
\sup_{c>0}\, \underset{n\to\infty}{\lim\sup} \, n \mathcal{C}^n_{\textrm{max},c}(\tilde{\bf r}_n) =
 \mathcal{C}_{\rm minmax}.
\end{equation}

The strong QLAN convergence \eqref{th.lan.2} implies that the asymptotic minimax cost $ \mathcal{C}_{\rm minmax}$ of the multi-copy model is equal to the minimax cost
$\mathcal{C}^\varrho_{\rm minmax}$ of the limit Gaussian model, and the sequence of estimators $\tilde{\rho}_n$ defined in steps (i) to (vi) above is LAM. The proof of the first statement follows the same lines as that of its classical counterpart and we refer to \cite{GutaJanssensKahn} for the details. The key idea is that strong LAN controls the norm-one distance between the i.i.d. models and the Gaussian one, so that the optimal Gaussian measurement can be pulled back into a LAM measurement for the i.i.d. model. The second statement follows from the fact that the measurement performed in step (v) is minimax for the quantum part of the Gaussian model. Indeed, it can be shown more generally that thanks to the covariance properties of the model, the optimal measurement for l.u. estimators discussed in Sec.~\ref{sec:examplegaussian} is also a minimax estimator. Therefore, $\mathcal{C}^\varrho_{\rm minmax} =
 \mathcal{C}^\varrho$, where the latter is the HCR bound computed in \eqref{eq:gauss3param}, and by combining with strong LAN we get
\begin{equation}
 \mathcal{C}_{\rm minmax}  =
 \mathcal{C}^\varrho_{\rm minmax} =
 \mathcal{C}^\varrho   = 3 + 2 r_0 - r_0^2.
\end{equation}

Moreover, since the estimator for the Gaussian model has normal distribution, this will also be true asymptotically, for the i.i.d. model
$$
\sqrt{n} \left(\tilde{\bf r}_n - {\bf r}\right) \xrightarrow{n\to\infty}
\mathcal{N}(0,\Sigma), \qquad \Sigma= {\rm diag}[1+r_0, 1+r_0,1-r_0^2].
$$
This means that the estimator $\tilde{\bf r}_n$ is not only LAM optimal, but is normally distributed around the true parameter, which equips us with asymptoticcally exact confidence regions (error bars).

\subsection{Optimal estimation for i.i.d. ensembles via QLAN}


In this section we go beyond qubit models and discuss a general state estimation problem for multi-copy models with finite dimensional systems.
The reasoning follows the same line as in the qubits case and therefore we will not repeat all technical considerations and proofs but rather focus on the key steps and results.


\subsubsection{The qudit model.}
Let $\rho_0 = {\rm diag}[\mu_1, \dots,  \mu_d]$ be a $d$-dimensional mixed state
with $\mu_1>\dots >\mu_d>0$, and let us parametrise the states around $\rho_0$ as
\begin{equation}\label{rho.theta.tilde}
\rho_{\bf u}
:=
\begin{bmatrix}
\mu_1 + h_1 & \zeta_{1,2}^* & \dots & \zeta_{1,d}^*
\\
\zeta_{1,2} & \mu_2  + h_2 & \ddots& \vdots
\\
\vdots & \ddots & \ddots & \zeta_{d-1,d}^* \\
\zeta_{1,d} & \dots & \zeta_{d-1,d} &\mu_d - \sum_{i=1}^{d-1} h_i
\end{bmatrix},
\end{equation}
where ${\bf u} = ({\bf h}, \boldsymbol{\zeta})\in \mathbb{R}^{d-1}\times \mathbb{C}^{d(d-1)/2}$
 represent eigenvalue changes and rotations respectively.
 The multi-copy local model for an ensemble of $n$ qudits has a joint state
$$
\rho_{\bf u}^n = \rho_{{\bf u} /\sqrt{n}}^{\otimes n}.
$$

\subsubsection{Gaussian shift model.}
The limiting Gaussian model can be identified by applying the same CLT arguments of the qubits case. This shows that
it is a product  of
independent classical and quantum Gaussian shifts
\begin{equation}
\varrho_{\bf u}:= q_{\bf u} \otimes p_{\bf u},
\end{equation}
where the  classical component is the $(d-1)$-dimensional Gaussian
\begin{equation}
\label{eq:QLANgaussian}
p_{\bf u} = \mathcal{N}({\bf h} , V_c(\boldsymbol{\mu})), \quad V_c(\boldsymbol{\mu})_{ij} :=  \delta_{ij} \mu_{i}  -\mu_{i}\mu_{j}.
\end{equation}
and is the limit of the classical multinomial model obtained by restricting the attention to diagonal states.
The quantum component is a tensor product
\begin{equation}
\label{eq:quantumQLAN}
q_{\bf u} = \bigotimes_{i<j} \varrho \left(\sqrt{\frac{2}{\mu_i-\mu_j}} \zeta_{ij}, \, \frac{\mu_i+\mu_j}{2(\mu_i-\mu_j)}\right)
\end{equation}
of `displaced thermal states', each carrying information about one of the off-diagonal elements
$\zeta_{ij}$ and implicitly about rotations with respect to the standard basis---here we used a short-hand notation
where $\varrho(\zeta, v)$ denotes a single mode Gaussian state with mean $\zeta$ and covariance matrix $v \mathcal{I}$.

\subsubsection{QLAN for finite dimensional states.}
We can now state the general QLAN result for finite dimensional i.i.d. quantum models. It is important to note that the convergence to the Gaussian model holds under the assumption that $\rho_0$ is \emph{fully mixed}, and is generally not valid for rank-deficient states which lie on the boundary of the parameter space. Although QLAN results can be proved for restricted models around such states (e.g. pure state models), the general asymptotic analysis for boundary states needs to be dealt with separately (see \cite{AcharyaGuta} for the qubits minimax theory) and will not be discussed here.

The construction is similar in spirit to that of \eqref{th.lan.2}, but since we deal here with $d$-dimensional systems we use Weyl's Theorem \cite{Fulton1991} to write the i.i.d. state as a block diagonal matrix with blocks indexed by Young diagrams $\boldsymbol{\lambda}$ with $d$ rows and $n$ boxes
\begin{align*}
\left(\mathbb{C}^d \right)^{\otimes n} &= \bigoplus_{\boldsymbol{\lambda}}
\mathbb{C}^{d_{\boldsymbol{\lambda}} } \otimes
\mathbb{C}^{m_{\boldsymbol{\lambda},n}},\\
\rho^{n}_{\bf u} &= \bigoplus_{\boldsymbol{\lambda}}
p^{n, \boldsymbol{\lambda}}_{\bf u} \rho^{n, \boldsymbol{\lambda}}_{\bf u}  \otimes
\frac{\mathbb{1} }{m_{\boldsymbol{\lambda},n}} ,
\end{align*}
where $d_{\boldsymbol{\lambda}}$ and $m_{\boldsymbol{\lambda},n}$ are the dimensions of the irreducible representations of $\t{SU}(d)$ and $S(n)$. The classical part is a probability distribution $p^{n, \boldsymbol{\lambda}}_{\bf u} $ over Young diagrams which concentrates on `typical' diagrams with rows lengths $\lambda_i \approx n \mu_i$, and it converges to the Gaussian distribution $p_{\bf u}$  \eqref{eq:QLANgaussian} after rescaling, similarly to \eqref{eq.zn}.
The conditional quantum state $\rho^{n, \boldsymbol{\lambda}}_{\bf u}$ can be mapped isometrically into the multimode Fock space
$\mathcal{F}^{\otimes d(d-1)/2}$ so that it approximates the Gaussian state $q_{\bf u}$ defined in \eqref{eq:quantumQLAN}. Unlike the qubit case, there isn't a natural orthonormal basis that can be used to define the isometry, but rather an `approximate' one. Indeed, the irreducible representation
$\mathbb{C}^{d_{\boldsymbol{\lambda}}}$ carries a natural basis
$| \boldsymbol{\lambda}, {\bf m} \rangle$ indexed by (semistandard) Young tableaux $t_{\bf m}$. These are Young diagrams whose boxes have labels in $\{1,\dots, d\}$ such that each row is increasing from left to right and each column is strictly increasing from top to bottom. A tableau is completely determined by the multiplicities ${\bf m}= \{m_{i,j} : 1\leq i<j\leq d \}$ where $m_{i,j}$ is the number of $j$s in the $i$-th row, for instance
$$
t_{\bf m}= \young(1111111122333,222223,333) ~, \qquad {\rm with}~ m_{1,2}=2, m_{1,3}=3,m_{2,3}=1.
$$
The basis $| \boldsymbol{\lambda}, {\bf m} \rangle$ is not orthogonal (except when $d=2$), but for large $n$ the states $\rho^{n, \boldsymbol{\lambda}}_{\bf u}$ concentrate on `low excitation' basis vectors ($|{\bf m}|\ll n$), which are approximately orthogonal and can be mapped approximately into  Fock basis vectors
$|{\bf m}\rangle := \otimes_{i<j} |m_{ij}\rangle$.

This leads to the general form of QLAN  \cite{KahnGuta}
which states that there exist quantum channels
\begin{eqnarray}
T_n &:&
M(\mathbb{C} ^{d})^{\otimes n} \to L^{1}(\mathbb{R}^{d-1})
\otimes \mathcal{T}_{1}(\mathcal{F}^{\otimes d(d-1)/2}),\\
S_n &:& L^{1}(\mathbb{R}^{d-1}) \otimes \mathcal{T}_{1}(\mathcal{F}^{\otimes d(d-1)/2})
\to M(\mathbb{C} ^{d})^{\otimes n}
\end{eqnarray}
such that
\begin{align}
\label{th.main.lan}
\sup_{ \| {\bf u}\|\in   \Theta^{(\beta, \gamma)}_n }  \left\lVert \varrho_{\bf u} - T_n(\rho^{n}_{\bf u}) \right\rVert_1  =
O (n^{- \epsilon } ) , \\ \nonumber
\label{Sn}
\sup_{ \| {\bf u}\|\in  \Theta^{(\beta, \gamma)}_n }  \left\lVert S_n(\varrho_{\bf u}) - \rho^{n}_{\bf u} \right\rVert_1  =
O (n^{- \epsilon } ).
\end{align}
for some constant $\epsilon = \epsilon(\beta, \gamma, \boldsymbol{\mu}) > 0$, and where
local parameters are restricted to the `slowly growing' balls $\Theta^{(\beta, \gamma)}_n :=\left\{ {\bf u} = ({\bf h}, \boldsymbol{\zeta})\,: \,
\| {\bf h}\| \leq n^\gamma, \| \boldsymbol{\zeta} \| \leq n^\beta  \right\}$ with technical restrictions $\beta < 1 / 9$ and $\gamma < 1/4$.

\subsubsection{Local asymptotic minimax estimation.}
The QLAN theorem \eqref{th.main.lan} can be used to construct LAM estimators in a two step procedure, as detailed in the qubit case. A subsample $\tilde{n}= n^{1-\epsilon}$ is used to localise the state within a local neighbourhood of the type $\Theta_n^{(\beta, \gamma)}$. After an appropriate unitary rotations, the remaining qudits are mapped via the channel $T_n$ into a classical-quantum state which is close to $\varrho_{\bf u}$. In general, the optimal measurement depends on the chosen cost function.
 Let us consider the following cost function
\begin{equation}
\mathcal{C}(\mathbf{u}, \tilde{\mathbf{u}}) = (\mathbf{h} - \tilde{\mathbf{h}})^T C_c (\mathbf{h} - \tilde{\mathbf{h}}) + \sum_{i<j} C_q^{ij} |{\zeta}_{ij} - \tilde{\zeta}_{ij}|^2,
\end{equation}
where $C_c \geq 0$ and $C_q^{ij}\geq 0$ for all $i<j$.
 This is a general quadratic cost function, where we have separated the contribution to the cost coming  from the eigenvalue changes (classical) and rotations (quantum) and for simplicity we have assumed all potential cross-terms are zero.
In this case, the optimal measurement is the heterodyne for each Gaussian mode, which provides (asymptotically) independent unbiased estimator of the off-diagonal parameters
$\tilde{\zeta}^n_{ij}\sim \mathcal{N}(\zeta_{ij}, \mu_i/2 )$. On the other hand, the classical component is an (asymptotically) unbiased estimator of the diagonal parameters $\tilde{\bf h}^n\sim \mathcal{N}({\bf h}, V_c(\boldsymbol{\mu}))$. Therefore, the LAM risk
reads explicitly
\begin{equation}
\mathcal{C}_{\rm minmax} =
\sup_{c>0}\, \underset{n\to\infty}{\lim\sup} \inf_{\tilde{\mathbf{u}}^n} \, n \sup_{\|{\bf u} \|\leq c}
\mathcal{C}(\tilde{\mathbf{u}}^n , \mathbf{u}) = {\rm Tr} (V_c C_c) +  \sum_{i<j} C_q^{ij} \mu_i.
\end{equation}
Let us look at two often encountered examples of cost function.

If we consider the Frobenius (norm-two square) distance in the space of density matrices $\rho_{\mathbf{u}}$ this results in the
corresponding cost in the parameter space:
\begin{equation}
\mathcal{C}^F(\mathbf{u}, \tilde{\mathbf{u}})  = d_F(\rho_{\mathbf{u}},\rho_{\tilde{\mathbf{u}}}) = \|\rho_{\bf u}- \rho_{\tilde{\bf u}}\|_2^2 =
(\mathbf{h}-\tilde{\mathbf{h}})^T(\mathbf{h}-\tilde{\mathbf{h}}) +2  \sum_{i<j} |\zeta_{ij}- \tilde{\zeta}_{ij}|^2,
\end{equation}
which gives
\begin{equation}
\mathcal{C}^{F}_{\rm minmax} =\sum_{i=1}^d
\mu_i (1-\mu_i) + 2 \sum_{i=1}^d (d-i) \mu_i \leq 2d-1.
\end{equation}
In contrast to this, the optimal Frobenius cost for \emph{separate} measurements scales as
$d^2$ \cite{HaahHarrow} which shows that collective measurements outperform separate
measurements by a factor $d$.

 If instead of the Frobenius distance we take the  Bures distance \cite{Bures, Bengtsson2006} the corresponding cost function is
\begin{multline}
\mathcal{C}^B(\mathbf{u}, \tilde{\mathbf{u}})  = d_B (\rho_{\bf u}, \rho_{\tilde{{\bf u}}}) =
2 -2 {\rm Tr} \left( \sqrt{\rho_{\bf u} \sqrt{\rho_{\tilde{{\bf u}}}} \rho_{\bf u}} \right)
\overset{ o(\|{\bf u} - \tilde{{\bf u}}\|^2) }{=}  \\
({\bf u} - \tilde{{\bf u}})^T F(\boldsymbol{\mu})  ({\bf u} - \tilde{{\bf u}}) =
({\bf h} - \tilde{{\bf h}})^T F_{c} (\boldsymbol{\mu}) ({\bf h} - \tilde{{\bf h}}) +
\sum_{i<j}
(\boldsymbol{\zeta} - \tilde{\boldsymbol{\zeta}})^T F_q (\boldsymbol{\mu})(\boldsymbol{\zeta} - \tilde{\boldsymbol{\zeta}}),
\end{multline}
which corresponds to the QFI and has a block diagonal form with respect to the classical set of parameters  and each pair
$({\rm Re} \zeta_{ij}, {\rm Im} \zeta_{ij})$ of quantum  parameters
\begin{equation}
F_c = V_c(\boldsymbol{\mu}), \qquad
F^{ij}_q(\boldsymbol{\mu}) = \frac{4}{\mu_i+\mu_j} \mathcal{I}.
\end{equation}
The LAM Bures cost is then
\begin{equation}
\mathcal{C}^{B}_{\rm minmax} =(d-1) + 4 \sum_{i<j} \frac{\mu_i}{\mu_i+\mu_j}.
\end{equation}
It follows immediately that $ d^2-1\leq \mathcal{C}^{B}_{\rm minmax} \leq (d-1)(2d+1)$.
The inequality confirms the expectation that the cost is larger than the
quantum SLD CR lower bound, which is achieved only for the completely incoherent state
$\rho_0 = \openone/d$. Strictly speaking however, the QLAN results as stated is not valid at this state since all eigenvalues are equal. One can check that at this point the limit model is \emph{completely classical} which explains why the Cram\'{e}r-Rao bound is
achievable for this state.
On the other hand, the upper bound is approached in the limit of almost pure states with $1\approx \mu_1\gg \mu_2\gg \dots \gg \mu_d$.

\section{Bayesian approach}
\label{sec:bayes}
In this section we follow the Bayesian paradigm and describe methods that allow to find the optimal measurement and estimation strategies
in multi-parameter metrological problems.  Similarly as in the frequentist approach, a Bayesian quantum estimation model consists of a family of states $\rho_\bvar$, but it is additionally supplemented by the prior distribution $p(\bvar)$ representing prior knowledge on the set of parameters to be estimated. Given a cost function $\mathcal{C}(\bvar, \tilde{\bvar})$, the goal is to minimize the average Bayesian cost $\overline{\mathcal{C}}$
as defined in  Eq.~\eqref{eq:bayescost} over measurement $\{\M_m\}$ and estimators $\tilde{\bvar}(m)$.

Note that we can formally coarse grain the measurement operators $\M_m$ and relabel operators by the estimated value of parameter $\M_{\tilde{\bvar}(m)}$  with $\M_{\tilde{\bvar}} = \int \t{d}m\, \M_m \delta(\tilde{\bvar} - \tilde{\bvar}(m))$. Thanks to this we can combine the double minimization over the measurement and the estimator to a single optimization over the measurements only:
\begin{align}
\min_{\{\M_{\tilde{\bvar}}\}} \overline{\mathcal{\G}}, \quad \M_{\tilde{\bvar}} \geq 0, \quad  \int\t{d}\tilde{\bvar}\,
\M_{\tilde{\bvar}} = \openone, \quad
\overline{\mathcal{\G}} = \int \t{d} \bvar \t{d}\tilde{\bvar} p(\bvar) \trace(\rho_\bvar \M_{\tilde{\bvar}}) \mathcal{C}(\bvar,\tilde{\bvar}).
\end{align}
Of course this in general is an untractable problem, as the space of all allowed generalized measurements is enormous.
Still, as demonstrated below with some additional assumptions on the cost function or the set of states, the problem may be solved.

\subsection{Single parameter case}
Let us start with the simplest exactly solvable case, namely \emph{single} parameter Bayesian estimation with a quadratic cost function
$\mathcal{C}(\var, \tilde{\var}) = (\var-\tilde{\var})^2$.
For simplicity of the formulas that follow we redefine the parameter $\var$ so that the expectation value of the prior
distribution is zero, $\int \t{d} \theta\, p(\var) \var = 0$.
The Bayesian variance to be minimized takes the form:
\begin{multline}
\label{eq:costquadratic}
\overline{\costs} = \int \t{d} \var\, \t{d}\tilde{\var}\, p(\var)\trace[\rho_\var \M_{\tilde{\var}}(\var - \tilde{\var})^2] =
\int \t{d} \var\, p(\var) \var^2 + \trace \left[\int \t{d}\var\, p(\var) \rho_\var \int \t{d}\tilde{\var} \M_{\tilde{\var}} \tilde{\var}^2   \right] + \\ - 2 \trace\left[\int \t{d}\var\, p(\var)  \var \rho_\var  \int \t{d}\tilde{\var}\M_{\tilde{\var}} \tilde{\var} \right]
= \priorvar + \trace(\bar{\rho} \Lambda_2) - 2 \trace(\bar{\rho}^\prime \Lambda_1),
\end{multline}
where $\priorvar = \int \t{d} \var\, p(\var) \var^2$ represents the variance of the prior distribution,
$\bar{\rho}= \int\t{d}\var \, p(\var)\rho_\var$ is  the average state, $\bar{\rho}^\prime = \int \t{d}\var\, p(\var)  \var \rho_\var$
and $\Lambda_k = \int \t{d}\tilde{\var}\, \M_{\tilde{\var}} \tilde{\var}^k$.

Let us first prove that if a given POVM measurement $\{\M_{\tilde{\var}}\}$ is optimal, then
we may find a projective measurement yielding the same cost.
Let us perform eigen-decomposition of $\Lambda_1$ operator:
\begin{equation}
\Lambda_1 = \int \t{d}\tilde{\var}\, \M_{\tilde{\var}} \tilde{\var} = \sum_i \tilde{\var}_i \ket{\tilde{\var}_i}\bra{\tilde{\var}_i}.
\end{equation}
Consider now the following inequality:
\begin{equation}
\label{eq:ineqBayesbound}
\int\t{d}\tilde{\var}\,(\tilde{\var} - \Lambda_1) \M_{\tilde{\var}} (\tilde{\var} - \Lambda_1) \geq 0,
\end{equation}
which is true since $\M_{\tilde{\var}} \geq 0$ while $\Lambda_1$ is hermitian.
This implies:
\begin{equation}
\int\t{d}\tilde{\var}\, \M_{\tilde{\var}} \tilde{\var^2}  + \Lambda_1^2 - 2
\Lambda_1^2 \geq 0
\end{equation}
and hence
\begin{equation}
\label{eq:l1l2}
\Lambda_2 \geq \Lambda_1^2.
\end{equation}

Let us now replace the measurement $\{\M_{\tilde{\var}}\}$ with the projective measurement, corresponding to the
projection on the eigenbasis $\ket{\tilde{\theta}_i}$ of $\Lambda_1$.
For this choice $\Lambda_2 = \Lambda_1^2$, which according to \eqref{eq:l1l2} is the smallest operator possible.
Inspecting \eqref{eq:costquadratic} we see that we want the term $\trace (\bar{\rho}\Lambda_2)$ to be as small as possible, and hence
it is always optimal to choose the projective measurement in the eigenbasis of $\Lambda_1$.

Assuming the measurement is projective, we may now introduce a single operator variable write $\bar{\Lambda} =\Lambda_1$, $
\Lambda_2 = \bar{\Lambda}^2$ and the optimization problem amounts to minimization of the following cost function over a single hermitian operator $\bar{\Lambda}$:
\begin{equation}
\overline{\costs} = \priorvar + \trace(\bar{\rho} \bar{\Lambda}^2) - 2 \trace(\bar{\rho}^\prime \bar{\Lambda}).
\end{equation}
Since the above  formula is quadratic in matrix $\bar{\Lambda}$, the minimization can be performed explicitly and
the condition for vanishing first derivative amounts to the following linear equation:
\begin{equation}
\label{eq:lambdabayes}
\bar{\Lambda} \bar{\rho} + \bar{\rho}\bar{\Lambda} -2 \bar{\rho}^\prime = 0.
\end{equation}
Multiplying the above equality by $\bar{\Lambda}$ and taking the trace of both sides we get
that $\trace(\bar{\rho}^\prime \bar{\Lambda}) = \trace(\bar{\rho}\bar{\Lambda}^2)$ and therefore we find that the minimal
Bayesian cost reads \cite{Helstrom1976, Macieszczak2014}:
\begin{equation}
\label{eq:bayescostopt}
\overline{\costs} = \priorvar - \trace\left(\bar{\rho} \bar{\Lambda}^2\right),
\end{equation}
where $\bar{\Lambda}$ is defined by \eqref{eq:lambdabayes}.

Let us note here an interesting observation that the above formula can be related with the formula for the QFI in case of Gaussian priors,
where the $\bar{\rho}^\prime$ operator is related with the derivative of the $\bar{\rho}$ operator with respect to the shift of the center of the prior  \cite{Macieszczak2014}. \
Indeed, consider the prior $p_{\var_0}(\var)=\frac{1}{\sqrt{2\pi\priorvar}}e^{-(\var-\var_0)^2/(2 \priorvar)}$ which center $\theta_0$
we treat as a parameter to estimate by performing a measurement on an effectively averaged state
$\bar{\rho}_{\theta_0}= \int\t{d}\var \, p_{\var_0}(\var)\rho_\var$. We can now regard $\bar{\rho}_{\theta_0}$ as a family of states
as considered in the frequentist estimation approach. Notice the following mathematical identity:
\begin{equation}
\label{deriv}
\bar{\rho}^\prime=\int \t{d}\var\, p_{\var_0=0}(\var)\var\rho_\var=\int \t{d}\var\, \priorvar\frac{\t{d}p_{\var_0}(\var)}{\t{d}\var_0}\Big|_{\var_0=0}\rho_\var=\priorvar \frac{\t{d}}{\t{d}\var_0}\int \t{d}\var\, p_{\var_0}(\var)\rho_\var\Big|_{\var_0=0}=\priorvar\frac{\t{d}\bar{\rho}_{\theta_0}}{\t{d}\var_0}\Big|_{\var_0=0}
\end{equation}
and therefore from \eqref{eq:lambdabayes}
\begin{equation}
\bar{\rho}^\prime = \frac{1}{2}(\bar\rho \Lambda+\bar\Lambda\bar\rho)=\priorvar\left.\frac{\t{d}\bar\rho_{\var_0}}{\t{d}\var_0}\right|_{\var_0=0}.
\end{equation}
It means that $\bar\Lambda$ is proportional to SLD in the freuquentist estimation problem of estimating $\var_0$ on states $\bar{\rho}_{\theta_0}$: $\bar\Lambda=\priorvar\cdot L$. Therefore, the mean Bayesian cost may be written as:
\begin{equation}
\label{eq:bayescostopt}
\overline{\costs} = \priorvar\left[ 1 -\priorvar \left(F_Q(\bar\rho_{\theta_0})\right)\right],
\end{equation}
 where $F_Q(\bar\rho_{\var_0})$ is the QFI for the $\var_0$ estimation problem for the family of states $\bar{\rho}_{\var_0}$.

\subsection{Multi-parameter case}
\label{sec:bayesquadratic}
Let us now turn to a  multi-parameter scenario  and see whether the reasoning from the previous subsection can be
generalized to work in this case.
For a given choice of the cost matrix $C$ the average Bayesian cost reads:
\begin{equation}
\label{eq:bayescostmulti}
\overline{\costs} = \int \t{d} \bvar\, \t{d} \tilde{\bvar}\, p(\bvar) \trace\left[\rho_{\bvar} \M_{\tilde{\bvar}} (\bvar - \tilde{\bvar})^T C (\bvar - \tilde{\bvar}) \right] = \Delta^2_{\G} \bvar  + \trace\left(\bar{\rho} \Lambda_2 \right)
- 2 \trace \left({\boldsymbol{\bar{\rho}}}^{\prime T} C \boldsymbol{\Lambda_{1}}  \right),
\end{equation}
where analogously as in the single parameter case,
$\Delta^2_{\G} \bvar = \int \t{d}\bvar \, p(\bvar) \bvar^T  C \bvar$ represents the cost corresponding to the  prior distribution,
$\bar{\rho} = \int \t{d}\bvar \, p(\bvar) \rho_{\bvar}$ is the average state,
$\Lambda_2 = \int \t{d} \tilde{\bvar}\, \M_{\tilde{\bvar}} \tilde{\bvar}^T C \tilde{\bvar}$,
${\boldsymbol{\bar{\rho}}^\prime} =\int \t{d}\bvar \, p(\bvar) \bvar\rho_{\bvar} $ while
$\boldsymbol{\Lambda_1} =\int \t{d}\tilde{\bvar} \, \M_{\tilde{\bvar}} \tilde{\bvar} $---note that
$\boldsymbol{\bar{\rho}^\prime}$, $\boldsymbol{\Lambda_1}$ are now operator vectors.

Consider now an inequality which is a multi-parameter generalization of \eqref{eq:ineqBayesbound}:
\begin{equation}
\int\t{d}\tilde{\var}\,  \M_{\tilde{\var}} (\tilde{\bvar} - \boldsymbol{\Lambda_1})^T C (\tilde{\bvar} - \boldsymbol{\Lambda_1}) \geq 0.
\end{equation}
From this it follows that:
\begin{equation}
\label{eq:ineql2l1}
\Lambda_2 \geq  \boldsymbol{\Lambda_1}^T C \boldsymbol{\Lambda_1}.
\end{equation}
Replacing $\Lambda_2$ in Eq.~\eqref{eq:bayescostmulti} with the r.h.s. of the above inequality we get
\begin{equation}
\overline{\costs} = \Delta^2_{\G} \bvar  + \tracep\left(\G\left(\trace(\bar{\rho} \boldsymbol{\Lambda_1}\boldsymbol{\Lambda_1}^T)-2\trace( \boldsymbol{\Lambda_1}\boldsymbol{\bar{\rho}_\bvar^\prime}^T) \right)\right)
\end{equation}
We can now perform minimization over
$\boldsymbol{\Lambda_1}$ (treating formally different vector components of $\boldsymbol{\Lambda_1}$ as independent operators) and obtain effectively a lower bound on the cost in the form:
\begin{equation}
  \overline{\costs} \geq \Delta^2_{\G} \bvar - \tracep\left(\G\trace\left(\bar{\rho} \boldsymbol{\bar{\Lambda}}\boldsymbol{\bar{\Lambda}^T}\right)\right),
\end{equation}
where $\bar{\boldsymbol{\Lambda}}$ is the solution of the following equation:
\begin{equation}
  \boldsymbol{\bar{\Lambda}} \bar{\rho} +  \bar{\rho}\boldsymbol{\bar{\Lambda}} = 2 \boldsymbol{\bar{\rho}^\prime}.
\end{equation}
Unfortunately, unlike in the single parameter case this bound is not always saturable. This is due to the fact that there is in general
no single eigenbasis that would diagonalize all operators forming the optimal vector $\boldsymbol{\bar{\Lambda}}$.
Had there been such an eigenbasis, we could write $\boldsymbol{\bar{\Lambda}} = \sum_i \tilde{\bvar}_{i} \ket{\tilde{\bvar}_i}\bra{\tilde{\bvar}_i}$, where $\ket{\tilde{\bvar}_i}$ form an orthonormal eigenbasis, and $\tilde{\bvar}_i$ vectors of eigenvalues. In this case inequality \eqref{eq:ineql2l1} would be saturated and hence the bound would be saturated
provided the measurement is performed in this eigebasis and the estimated values correspond to $\tilde{\bvar}_i$.

Analogously to the one-parameter case, in case of multiparmeter Gaussian prior the above formula may be
related with the QFI matrix of the corresponding frequentist estimation problem of estimating the mean $\bvar_0$ of the prior $p_{\bvar_0}(\bvar)$.
 Consider
 \begin{equation}
 p_{\bvar_0}(\bvar)=\frac{1}{\sqrt{2\pi}^{\mathfrak{p}}\sqrt{\det V}}e^{-\tfrac{1}{2}(\bvar-\bvar_0)^T V^{-1} (\bvar-\bvar_0)},
 \end{equation}
 where $V$ is the positive symmetric covariance matrix of the prior---in this case the prior cost reads simply: $\Delta^2_{\G} \bvar = \tracep(C V)$. Repeating the reasoning from \eqref{deriv} we have
$\boldsymbol{\bar\rho^{\prime}}\big|_{\bvar_0=0}=V\boldsymbol\nabla \bar\rho_{\bvar_0}|_{\bvar_0=0}$ (where $\boldsymbol{\nabla}$ here denotes gradient operator with respect to $\bvar_0$), and from that $\boldsymbol{\bar{\Lambda}}=V\cdot\bold{L}$. Keeping in mind that $F_Q(\bar\rho_{\bvar_0})=\trace(\bar\rho\bold{L}\bold{L}^T)$ we finally get:
\begin{equation}
  \overline{\costs} \geq \tracep(C V)  - \tracep\left(\G V F_Q(\bar\rho_{\bvar_0}) V \right).
\end{equation}
An analogous result was derived in a slightly different way in \cite{Sidhu2019, Rubio2019}.  It is appealing as it yields a simple bound on multi-parameter Bayesian estimation cost utilizing the QFI matrix. One needs to keep in mind, however, that due to the use of the QFI, this bound will not properly
address the potential optimal measurement incompatibility issue which may affect its tightness.

\subsection{Covariant estimation}
\label{sec:bayescovariant}
The simplicity of the quadratic cost function made it possible to solve a general single-parameter Bayesian estimation problem, while in the multi-parameter case allowed to derive non-trivial bounds. There is no universal way, however, to find an exact solution in case of a general multi-parameter Bayesian estimation problem in this way. Still, provided the problem enjoys certain symmetry one may utilize
the powerful method of covariant measurements in order to find the solution. We will start by defining the class of covariant problems.


A covariant estimation problem involves two group actions which act in a covariant fashion. On the hand, the parameter space $\Theta$ carries the action of a Lie group $G$: for each $\bvar\in \Theta$, the action of $g\in G$ is denoted $\bvar\mapsto g\bvar$. On the other hand, the Hilbert space of the quantum statistical model $\{\rho_\bvar\}_{\bvar\in \Theta}$ carries a unitary representation $U_g$ of $G$. We say that the estimation problem is covariant with respect to the two actions if and only if the following conditions are satisfied:
\begin{enumerate}
\item{The parameter to be estimated is an element of the group $g \in G$.}
\item{The family of states is covariant with respect to the group representation
$$
\rho_{g\bvar} = U_g \rho_\bvar U_g^\dagger
$$}
\item{The cost function is left invariant with respect to the action of the group:
$ \mathcal{C}(g \bvar_1, g \bvar_2)= \mathcal{C}(\bvar_1,\bvar_2)$ for all $g\in G$. }
\item{The prior distribution is invariant with respect to the  group action:
$ p(gd\bvar) =  p(d\bvar)$--- This represent in a formal sense the maximal prior ignorance about the parameter.}
\end{enumerate}

For a covariant estimation problems the Bayesian cost is given by:
\begin{equation}
\overline{\mathcal{C}} = \int \t{d}g\,  \t{d}\tilde{g} \, \trace(U_g \rho_e U_g^\dagger \, \M_{\tilde{g}}) \mathcal{C}(g,\tilde{g}),
\end{equation}
where we assume that $\t{d}g$ is the normalized Haar measure on the group, $\int \t{d} g =1$, with respect to which the prior is trivial
$p(g)=1$.

Thanks to the covariance property, it can be proven that the one can restrict the search for optimal measurements to the
class of covariant measurements \cite{Holevo1982}. A measurement $\M_{\tilde{g}}$ is said to be covariant with respect to the action of the group representation if and only if
\begin{equation}
 U_h \M_{\tilde{g}}U_h^\dagger = \M_{h \tilde{g}}, \qquad {\rm for~all~}\qquad \tilde{g}, h.
\end{equation}
In particular for a covariant measurement
\begin{equation}
\M_{\tilde{g}} = U_{\tilde{g}} \M_e U_{\tilde{g}}^\dagger,
\end{equation}
so that all measurement operators are determined by a single \emph{seed} operator $\M_e$.
Note that thanks to the covariance property of the measurement and the group invariance of the Haar measure we have:
\begin{multline}
\label{eq:covproblem}
\overline{\mathcal{C}} = \int \t{d}g\, \t{d}\tilde{g}\, \trace\left(\M_{\tilde{g}} \rho_g \right) \mathcal{C}(g,\tilde{g})
=  \int \t{d}g\, \t{d}\tilde{g}\, \trace\left(U^\dagger_{\tilde{g}^{-1} g} \M_e U_{\tilde{g}^{-1}g} \rho_e \right) \mathcal{C}(g,\tilde{g}) = \\
\overset{g\rightarrow \tilde{g} g}{=} \int \t{d}g\, \t{d}\tilde{g}\, \trace\left(U^\dagger_g \M_e U_g \rho_e  \right) \mathcal{C}(g,e)
= \int \t{d}g\, \trace\left(\M_e \rho_g  \right) \mathcal{C}(g,e).
\end{multline}
As such, the whole problem now amounts to minimization of the above quantity over a \emph{single} operator $\M_e$ with constraints
$\M_e\geq 0 $, $\int \t{d}g\, U_g \M_e U_g^\dagger = \openone$.
This is a huge simplification of the original problem and often the optimal operator $\M_e$ may be found analytically, as is demonstrated below.

\subsection{Qubit models}
\label{sec:bayesianqubit}

\subsubsection{Pure qubit case.}
First, we consider an estimation model in which we are given $n$ copies of a completely unknowns qubit state.
Using the standard Bloch sphere parametrization, we write the state as
\begin{equation}
\rho_{\Omega}^{n} = \ket{\psi_\Omega}\bra{\psi_\Omega}^{\otimes n}, \ \ket{\psi_\Omega} = \cos(\theta/2)\ket{0} +
e^{i\varphi }\sin(\theta/2) \ket{1},
\end{equation}
where for compactness of notation we have introduced  $\Omega = (\theta,\varphi)$. As a cost function we choose:
\begin{equation}
\mathcal{C}(\psi,\tilde{\psi}) = 4(1-|\braket{\psi}{\tilde{\psi}}|^2) \overset{\tilde{\psi} \approx \psi+ \t{d}\psi}{=}
\t{d}\theta^2 + \sin^2\theta \t{d}\varphi^2,
\end{equation}
which in the first order approximation reduces to the standard metric on the sphere---a useful property that will let us relate the asymptotic Bayesian cost with the costs obtained within the frequentist approach.

In order to think of this problem as a covariant estimation problem we may view $\ket{\psi_\Omega}$ as
obtained by a rotation a fixed state $\ket{0}$ using the defining representation of the SU(2) group.
 More precisely, since the initial state $\ket{0}$ will not change under rotations around the $z$ axis, the parameter set corresponds to the $\t{SU}(2)/\t{U}(1)$ (i.e. the set of left cosets of $\t{U}(1)$ in $\t{SU}(2)$). However, $\t{SU}(2)/\t{U}(1)$ is not a group itself. Therefore, to use \eqref{eq:covproblem} directly, we still need to refer to the full $\t{SU}(2)$ group. 
 Consider SU(2) parameterizations using Euler's angles $(\uppsi,\theta,\varphi)$:
\begin{equation}
\ket{\psi_{(\theta,\varphi)}}=U_{(\uppsi,\theta,\varphi)}\ket{0},\quad U_{(\uppsi,\theta,\varphi)} = e^{i \varphi \sigma_z/2}e^{i \theta \sigma_y/2}  e^{i \uppsi \sigma_z/2},\quad \varphi\in [0,2\pi[,\theta\in]0,\pi[,
 \uppsi\in[0,4\pi[, 
\end{equation}
with the corresponding Haar measure:
\begin{equation}
\t{d}g=\frac{\t{d}\uppsi}{4\pi} \frac{\sin(\theta)\t{d}\theta}{2}\frac{\t{d}\varphi}{2\pi}.
\end{equation}
Using \eqref{eq:covproblem} (and after performing a trivial integration over $\uppsi$) we have
\begin{equation}
\overline{\mathcal{C}}^{n} = \int \t{d}\Omega \, \trace\left(\M_e \ket{\psi_\Omega}\bra{\psi_\Omega}^{\otimes n} \right)4(1-|\braket{\psi_\Omega}{0}|^2),
\end{equation}
where
\begin{equation}
\t{d}\Omega = \frac{1}{4\pi}\t{d} \theta \t{d}\varphi \sin\theta
\end{equation}
is the measure on $\t{SU}(2)/\t{U}(1)$, induced by the Haar measure on $\t{SU}(2)$. 

In this way we have now formulated the problem as a covariant estimation problem, which can be solved explicitly \cite{Massar1995}.
We need to minimize $\overline{\mathcal{C}}^{n}$ over $\M_e$, keeping in mind $\M_e \geq 0$, $\int\t{d}\Omega \, U_\Omega^{\otimes n} \M_e U_{\Omega}^{\dagger \otimes n } = \openone$ (we assume without loss that $M_e$ is invariant for rotation around the $z$ axis). We can rewrite the expression for the cost as
\begin{equation}
\overline{\mathcal{C}}^{n} =  4\left[ 1 - \int\t{d}\Omega \,\trace\left(\M_e \otimes \ket{0}\bra{0} \cdot \ket{\psi_\Omega}\bra{\psi_\Omega}^{\otimes n+1}  \right) \right].
\end{equation}
We may then take the integration over $\t{d} \Omega$ under the trace and make use of the following property
\begin{equation}
\int \t{d} \Omega \ket{\psi_\Omega}\bra{\psi_\Omega}^{\otimes n+1} = \frac{1}{n+2} \openone_{\mathcal{H}_S^{\otimes n+1}},
 \end{equation}
 where $\mathcal{H}_S^{\otimes n+1}$ is the fully symmetric subspace of $n+1$ qubits---this fact follows from the Schur Lemma and the irreduciblity of the SU(2) representation that acts on the fully symmetric space of qubits.

Therefore the optimal $\M_e$ is the one that maximizes $\trace\left(\M_e \otimes \ket{0}\bra{0} \, \openone_{\mathcal{H}_S^{\otimes n+1}}   \right)$. We may restrict the $\M_e$ operator to act solely on the symmetric subspace $\mathcal{H}_S^{\otimes n}$ as
this is the subspace where states $\ket{\psi_\Omega}^{\otimes n}$ live. Let us denote $U_g^{j=n/2}$ to be the irreducible representation of SU(2) acting on this subspace. Taking into account the completeness condition for $\M_e$:
\begin{equation}
\int \t{d} \Omega U_\Omega^{j=n/2} \M_e U_\Omega^{j=n/2 \dagger} = \openone_{\mathcal{H}_S^{\otimes n}}
\end{equation}
we see that $\trace\M_e = n+1$. It is clear that in order to have the largest overlap between $\M_e \otimes \ket{0}\bra{0}$
and  $\openone_{\mathcal{H}_S^{\otimes n+1}}$, we would like to have  $\M_e \otimes \ket{0}\bra{0}$ operator fully supported on $\mathcal{H}_S^{\otimes n+1}$. This will be so provided we choose
\begin{equation}
\M_e = \ket{0}\bra{0}^{\otimes n} (n+1).
\end{equation}
As a result we get
\begin{equation}
\bar{\mathcal{C}}^{n} = 4\left(1-\frac{n+1}{n+2} \right) \overset{n \rightarrow \infty}{\approx} \frac{4}{n}
\end{equation}
for which the asymptotic form indeed agrees with the saturable HCR bound derived in Sec.~\ref{sec:examplequbit1}. This proves asymptotic consistency between the Bayesian and the frequentist approaches.

\subsubsection{Mixed qubit case.}
Consider now a mixed qubit state estimation problem, as in Sec.~\ref{sec:examplequbit3}, with the $n$-copy state
\begin{equation}
\rho_{\mathbf{r}}^n = \rho_{\mathbf{r}}^{\otimes n}, \quad  \rho_\mathbf{r} = \frac{1}{2}\left(\openone + \boldsymbol{\sigma} \cdot \mathbf{r} \right).
\end{equation}
Here, we assume that the  prior distribution of the $\mathbf{r}$ parameter when written using spherical coordinates takes the form:
\begin{equation}
p(\mathbf{r}) d\mathbf{r} = w(r)\t{d}r \t{d} \Omega,
\end{equation}
where $w(r)$ is an arbitrary nonnegative function of the Bloch vector length for $0 \leq r \leq 1$.
This prior is invariant with respect to Bloch ball rotations, and hence the problem will be covariant with respect to $\theta,\varphi$ estimation,  but not with respect to the $r$ parameter for which there is no corresponding group structure. Still, this partial group covariance
leads to a significant simplification in obtaining the final solution \cite{Vidall1999, Bagan2006a}.
Let us choose the cost function to be:
\begin{equation}
\mathcal{C}(\mathbf{r},\tilde{\mathbf{r}}) = 4\left(1- \trace(\sqrt{\sqrt{\rho_\mathbf{r}}\rho_{\tilde{\mathbf{r}}} \sqrt{\rho_\mathbf{r}}})^2  \right) \overset{\tilde{\mathbf{r}} = \mathbf{r} + \t{d} \tilde{\mathbf{r}}}{=}  \frac{\t{d}r^2}{1-r^2} + r^2(\t{d} \theta^2 + \sin^2\theta \t{d}\varphi^2),
\end{equation}
which is the most natural choice, as it is directly related with quantum state fidelity \cite{Uhlmann1976, Jozsa1994, Bengtsson2006}. Furthermore, for neighbouring states it reduces to the quadratic cost function equivalent to the Bures metric---this implies that in order to make a meaningful asymptotic
comparison with the results obtained within the  frequentist approach, we should set
the cost matrix to be $C_{\t{Bures}} = \t{diag}[\tfrac{1}{1-r^2}, r^2, r^2 \sin^2\theta]$ in the frequentist formulas.

As discussed in detail in Sec.~\ref{sec:QLANstrong} the $\rho_\mathbf{r}^n$ state may be decomposed in terms of irreducible SU(2) and $S_n$ subspaces
as
\begin{equation}
\rho_{\bf r}^n = \bigoplus_{j\in \mathcal{J}_n} p_{\bf r}^{n,j} \rho_{\bf r}^{n,j} \otimes \frac{\openone_{m_j}}{m_j}.
\end{equation}
It can be shown \cite{Bagan2006a} that the above structure of the state together with the partial covariant nature of the problem implies that the optimal
measurement is  a measurement covariant with respect to the action of the SU(2):
\begin{equation}
M_{j,\tilde{\Omega}} = (2j+1) U_{\tilde{\Omega}}^j\ket{j,j}\bra{j,j}U_{\tilde{\Omega}}^{j\dagger} \otimes \openone_{m_j},  \quad\bigoplus_{j \in \mathcal{J}_n} \int\t{d} \tilde{\Omega} M_{j,\tilde{\Omega}} = \openone.
\end{equation}
This measurement may be understood as acting trivially on multiplicity space, as it carries no information on the state, yields information on the total angular momentum $j$ and, moreover, within each irreducible subspace corresponding to a given $j$ performs a covariant measurement
obtained by rotating a state with the maximum angular momentum projection on the $z$ axis $\ket{j,j}$. The Bloch vector direction estimate is $\tilde{\Omega}$, while all the information
on the Bloch vector length comes from $j$ and the explicit optimal estimator of $\tilde{r}$
reads \cite{Bagan2006a}
\begin{equation}
\tilde{r}(j) = \frac{|v_j^z|}{\sqrt{{v_j^{0}}^2 + v_j^{z2}}}  \overset{n \rightarrow \infty}{\approx} \frac{j}{n/2},
\end{equation}
where
\begin{align}
v_j^0&= \int_0^1 \t{d}r\, w(r) \sqrt{1-r^2} \left(\frac{1-r^2}{4}\right)^{\frac{n}{2}}\sum_{m=-j}^j \left(\frac{1+r}{1-r}\right)^m, \\ \nonumber
v_j^z&= \int_0^1 \t{d}r\, \frac{w(r) r}{j+1}  \left(\frac{1-r^2}{4}\right)^{\frac{n}{2}} \sum_{m=-j}^j m  \left(\frac{1+r}{1-r}\right)^m.
\end{align}
The resulting optimal cost reads
\begin{equation}
\label{eq:examplequbit3bayes}
\bar{\mathcal{C}}^n = 2\left(1 - \sum_{j=0(\frac{1}{2})}^{\frac{n}{2}} m_j \sqrt{{v_j^{0}}^2 + v_j^{z2}}\right)
\overset{n \rightarrow \infty}{\approx} \int_0^1\t{d}r\, w(r)\frac{3 + 2r}{n}.
\end{equation}
Note that the asymptotic formula is in agreement with the HCR bound
for the corresponding frequentist model, see Sec.~\ref{sec:examplequbit3}---if in Eq.~\eqref{eq:hcrqubit3} we choose $c(r)=1/(1-r^2)$ then the cost matrix becomes the Bures cost matrix $C=C_{\t{Bures}}$ and we get $\mathcal{C}^n = (3 + 2r)/n$.  If we now perform the averaging of the HCR bound over $w(r)$ we indeed obtain \eqref{eq:examplequbit3bayes} (up to the $1/n$ rescaling due to the number of copies).

Similar analysis can been performed for the case of $(r,\varphi)$ estimation \cite{Bagan2006a}, with the natural prior
$w(r) \frac{\t{d}\varphi}{2\pi}$ guaranteeing $U(1)$ covariance of the problem.
Without going into details, we just mention that in this case the asymptotic Bayesian cost reads:
\begin{equation}
\bar{\mathcal{C}}^n   \overset{n \rightarrow \infty}{\approx} \frac{2}{n},
\end{equation}
which is independent of the prior distribution $w(r)$ and is in agreement with the frequentist bounds for the corresponding cost matrix $C=\t{diag}[\tfrac{1}{1-r^2}, 1]$, see Eq.~\eqref{eq:hcrqubit2}.

\subsection{Bayesian Cram{\'e}r-Rao like bounds}
\label{sec:bayescr}
The examples of the previous subsection indicate that
while the Bayesian approach is typically technically more demanding than the frequentist approach,
it offers a deeper understanding of the actually achievable cost using finite resources. Moreover, the obtained results
coincide with the latter in the limit of multiple copy estimation. Unfortunately,
Bayesian models can only be solved exactly in special cases and therefore
one may wonder whether an efficiently calculable Bayesian bounds exist which would
combine the simplicity of the computation characteristic for the frequentist bounds while taking into account the impact of prior information as well as the finite resources.

In the single parameter case the quantum Bayesian CR bound can be derived using the classical Van Trees inequality \cite{Trees1968}
by simply replacing the classical FI with the QFI in the formulas.
The direct multi-parameter generalization (see the proof below) leads to  \cite{Gill1995, Tsang2011}
\begin{equation}
\label{eq:bayesianCRbound}
\overline{\costs} \geq \tracep[ C (\overline{F_Q} + I)^{-1}],
\end{equation}
where $\overline{F_Q} = \int \t{d}\bvar \, p(\bvar)F_Q(\rho_\bvar)$ is the QFI matrix averaged over the prior $p(\bvar)$ while
\begin{equation}
I = \int \frac{1}{p(\bvar)}\boldsymbol{\nabla}p(\bvar)\boldsymbol{\nabla}^Tp(\bvar)\t{d}\bvar
\end{equation}
is a matrix representing  the information contribution coming from the prior.

This bound, while providing some insight into the impact of the prior knowledge and finite data, is insensitive to the measurement incompatibility issue as it is based on the QFI matrix. One might try to derive an analogous bound using the HCR bound
rather than SLD CR bound, while retaining the Bayesian framework and the prior information contribution in the bound.
 This task is somehow challenging, as unlike the SLD CR bound the HCR bound can not be reduced to a single matrix inequality
 and therefore we do not have a matrix with which we can replace the QFI matrix in \eqref{eq:bayesianCRbound} in order to strengthen the bound.

Below we provide a derivation of the Bayesian bound that involves the HCR bound following along the lines presented
in \cite{Gill1995, Gill2005}. At the same time this will also lead us to a proof  of \eqref{eq:bayesianCRbound} as a corrolary.

Let us first focus on a purely classical Bayesian model with prior $p(\bvar)$ and conditional probability
 $p_{\bvar}(\tilde{\bvar})$ of measuring outcome $\tilde{\bvar}$ given the true value is $\bvar$.
 We will write $p(\tilde{\bvar},\bvar) = p_{\bvar}(\tilde{\bvar}) p(\bvar)$ to denote the joint probability distribution---note that we avoid using $p(\tilde{\bvar}|\bvar))$ for conditional probability in order not to have different notation in Bayesian and frequentist approaches for the same quantity.

In what follows  $\mathbb{E}$ denotes expectation value with respect to $p(\tilde{\bvar},\bvar)$  so
\begin{equation}
\mathbb{E}[f]=\iint  f(\tilde\bvar,\bvar)  p(\tilde{\bvar},\bvar) d\tilde\bvar d\bvar.
\end{equation}
Let us introduce a $\mathfrak{p}\times\mathfrak{p}$ matrix $Q(\bvar)$ and define vectors:
\begin{equation}
\bA=\tilde\bvar-\bvar, \quad \bB=\frac{1}{p(\tilde{\bvar},\bvar)}\left[\boldsymbol{\nabla}^T\left(Q(\bvar) p(\tilde{\bvar},\bvar)\right)\right]^T.
\end{equation}
Using the Cauchy-Schwarz inequality for vector functions $\sqrt{\G}\bold{A},\sqrt{\G^{-1}}\bold{B}$ we get:
\begin{equation}
\begin{split}
\label{CSb}
\mathbb{E}[\bA^T\G \bA]\mathbb{E}[\bB^T\G^{-1} \bB]&\geq\mathbb{E}[\bB^T\bA]^2.
\end{split}
\end{equation}
The first term is simply the Bayesian cost:
\begin{equation}
\mathbb{E}[\bA^T\G\bold{A}]=\overline\CQ.
\end{equation}
The second one:
\begin{multline}
\mathbb{E}[\bold{B}^T\G^{-1}\bold{B}]=
\iint d\bvar d\tilde\bvar \frac{1}{p(\tilde{\bvar},\bvar)}\boldsymbol{\nabla}^T\left(Q(\bvar)
p(\tilde{\bvar},\bvar)\right)\G^{-1}\left[\boldsymbol{\nabla}^T\left(Q(\bvar)p(\tilde{\bvar},\bvar)\right)\right]^T=\\
\int \tracep\left(Q(\bvar)\G^{-1} Q(\bvar)^T
\underbrace{\int\frac{1}{p_{\bvar}(\tilde{\bvar})}\boldsymbol{\nabla}p_{\bvar}(\tilde{\bvar})\boldsymbol{\nabla}^Tp_{\bvar}(\tilde{\bvar}) \t{d}\tilde\bvar}_{F(\bvar)}
\right) p(\bvar)\t{d}\bvar+\\
\tracep\left(\G^{-1}\underbrace{\int \frac{1}{p(\bvar)}\left[\boldsymbol{\nabla}^T(Q(\bvar)p(\bvar))\right]^T[\boldsymbol{\nabla}^T (Q(\bvar)p(\bvar))]d\bvar}_{I_Q}\right),
\end{multline}
where we have used the Leibniz rule and the fact that thanks to the normalization of conditional probability
$\int \t{d} \tilde{\bvar} \nabla p_{\bvar}(\tilde{\bvar}) = 0$  (which makes cross-terms vanish). Here $F(\bvar)$ is the classical FI calculated for a given value of $\bvar$. Finally,
\begin{equation}
\mathbb{E}[\bold{B}^T\bold{A}]=\iint d\bvar d\tilde\bvar\boldsymbol{\nabla}^T\left(Q(\bvar)p(\tilde{\bvar},\bvar)\right)(\tilde\bvar-\bvar)
=\int d\bvar
\tracep(Q(\bvar))p(\bvar)d \bvar = \overline{\tracep(Q)},
\end{equation}
where we have applied integration by parts assuming the prior distribution vanishes at the boundaries.
Therefore, from \eqref{CSb} we have:
\begin{equation}
\label{baygen}
\overline{\CQ} \geq \frac{\left(\overline{\tracep(Q)}\right)^2}{\overline{\tracep(\G^{-1} Q^T F Q )}+\tracep(\G^{-1}
 I_Q)}.
\end{equation}
For $Q(\bvar)=\sqrt{\G}(\overline{F} + I)^{-1}\sqrt{\G}$
we recover the standard Bayesian CR inequality \cite{Gill1995}
\begin{equation}
\label{vantree}
\overline{\CQ} = \tracep(\overline{\Sigma} C)\geq\tracep [\G\left(\overline{F}+I\right)^{-1}],\quad I=\int \frac{1}{p(\bvar)}\boldsymbol{\nabla}p(\bvar)\boldsymbol{\nabla}^Tp(\bvar)d\bvar.
\end{equation}
As the inequality works for any cost matrix $\G$, it is in fact equivalent to a matrix inequality:
\begin{equation}
\overline{\Sigma} \geq\left(\overline{F}+I\right)^{-1}.
\end{equation}

In  a quantum model we have $p_{\bvar}(\tilde{\bvar}) = \trace(\rho_{\bvar} M_{\tilde{\bvar}})$ and hence
 $F(\bvar)$ depends on the measurement $\{M_{\tilde\bvar}\}$ chosen.
 In order to arrive at an universally valid  bound independent on choice of measurement, one may use
 matrix inequality involving the QFI matrix $F_Q(\bvar)\succeq F(\bvar)$ and arrive at \eqref{eq:bayesianCRbound}.

If instead we take $Q(\bvar)=F(\bvar)^{-1}\G$ then from  \eqref{baygen} we have
\begin{equation}
\label{eq:bayesianhcr}
\overline{\CQ} \geq  \frac{\left(\overline{\tracep(\G F^{-1})}\right)^2}{\overline{\tracep(\G F^{-1})}+R} \geq
\overline{\tracep(\G F^{-1})}-R,
\end{equation}
where
\begin{equation}
R=
\int \frac{1}{p(\bvar)}\tracep\left(\G \left[\boldsymbol{\nabla}^T(F(\bvar)^{-1}p(\bvar))\right]^T[\boldsymbol{\nabla}^T (F(\bvar)^{-1}p(\bvar))]\right)\t{d}\bvar
\end{equation}
and where in the last inequality in \eqref{eq:bayesianhcr} we have used $\frac{a^2}{a+b}\geq a-b$.
Finally, we may use the HCR bound $\tracep(\G F(\bvar)^{-1}) \geq  \mathcal{C}^{\t{H}}(\bvar)$ to obtain a `Bayesian Holevo CR bound':
\begin{equation}
\label{eq:bayesianhcr2}
\overline{\CQ} \geq
\overline{C^{\t{H}}(\bvar)}-R.
\end{equation}
Note that this bound is still measurement dependent since $R$ depends on the FI matrix $F(\bvar)$ corresponding to some measurement. The
bound will be valid whatever the measurement chosen yet the choice may affect the tightness of the bound---see \cite{Gill2005} where some further improvement of the bound was proposed.

Now consider many copies of a system. Then we have $\C^H(\bvar)^n =\frac{1}{n}\C^H(\bvar)^n$,  $F^n(\bvar)=\frac{1}{n}F(\bvar)$ and from that $R^n=\frac{1}{n^2}R$. Therefore
\begin{equation}
\CQ^n \geq  \frac{1}{n}\overline{\C^H(\bvar)}-\frac{1}{n^2}R
\end{equation}
and in the limit of large $n$ the second term may be neglected resulting in the asymptotic bound \cite{Gill2005}
\begin{equation}\label{eq.BayesH}
\lim_{n\to \infty} n\CQ^n  \geq \overline{\C^H(\bvar)}.
\end{equation}
One can see that for large sample size the impact of prior knowledge becomes negligible, and the asymptotic Bayesian cost is bounded by the mean value of the HCR, which is in agreement with our observation from Sec.~\ref{sec:bayesianqubit}  that
 the asymptotic form of the Bayesian cost was actually \emph{equal} to the average HCR bound in the qubit models we have studied.
We conjecture that under appropriate model and prior regularity conditions, the asymptotic equivalence of Bayes and frequentist costs holds generally for finite dimensional systems with fully mixed states. The QLAN theory described in Sec. \ref{sec:qlan} has already been used to establish the achievability of the bound \eqref{eq.BayesH} for the qubit case \cite{GutaKahn, GutaJanssensKahn} and should be the appropriate tool for obtaining similar results for more general finite dimensional models.


\section{Summary and outlook}
\label{sec:summary}
In this review we discussed various theoretical methods for quantum multi-parameter estimation going beyond the mere computation of the QFI matrix and the related SLD CR bound. We have seen than that the HCR bound, the QLAN theory as well as Bayesian methods offer an advantage over the QFI by being able to address the issue of potential incompatiblity of measurements that are optimal from the point of view of extracting information on different parameters. We have also pointed out the key contribution of the QLAN approach which led to realization that the HCR bound (which predates QLAN by many years), is the actual asymptotically saturable bound in a quantum estimation problem involving many independent copies. Moreover, when discussing the Bayesian approach we stressed that it offers more insight into the finite number of samples regime and is capable of incorporating the prior knowledge into the estimation process. Importantly, both Bayesian and frequentist approaches agree in the asymptotic limit of multiple copy estimation under appropriate regularity conditions.

An often overlooked assumption which is crucial for both the CR theory and LAN is
that the unknown parameter is an interior point of the parameter space, and does not lie on its boundary. However,  interesting states such as low rank states often do lie on the boundary, in which case asymptotic normality generally fails \cite{BlumeKohout}. Although QLAN may still be useful, the corresponding asymptotic theory theory is less well understood \cite{GutaAcharya}. A related problem is that certain natural distances such as the Bures distance do not have a quadratic approximation, which leads to estimation cost scaling with the `anomalous' rate $1/\sqrt{n}$ for
\emph{fixed, separate} measurements and $1/n$ for \emph{adaptive} measurements \cite{Mahler,Ferrie,GranadeFerrie}. In parallel to the developments in `optimality' theory, there has been significant interest in practical estimation methods
for estimation of large dimensional systems under sparsity assumptions such as \emph{low rank} \cite{gross_2010, flammia_2012,kueng_2017,guta_2018}, finite correlations \cite{cramer_2010}, or permutational symmetry \cite{toth_2010}. Another important developing direction with relevance for quantum metrology concerns the finite sample (non-asymptotic) theory, e.g. estimation bounds \cite{Haah2017,odonnel2016} and confidence regions \cite{christandl_reliable_2012, blume-kohout2012, faist_practical_2016}.


We have focused on methods of multi-parameter estimation and have not entered into the more
physical and practical aspects characteristic to the field of quantum metrology \cite{Giovaennetti2006, Paris2009, Toth2014, Demkowicz2015, Dowling2015, Pezze2018, Pirandola2018, Braun2018}. In particular, the optimal $n$-probe states that
 appear in quantum metrological considerations are often entangled and therefore go beyond the i.i.d. setting on which we have focused our attention in this review.
 In general the QLAN methods cannot be directly applied in these cases, the asymptotic limit does not necessary imply saturability of the
  HCR bound, whereas the Bayesian methods, while in principle applicable, are typically too challenging to yield a closed rigorous solution.

 In the single parameter case, most of these issues have been resolved in recent years. Efficient methods to compute asymptotically saturable bounds in quantum metrologial scenarios in presence of uncorrelated decoherence models have been developed \cite{Fujiwara2008, Escher2011, Demkowicz2012, Kolodynski2013, Demkowicz2014, Knysh2014, Demkowicz2017, zhou2018achieving}.
 Interestingly, in presence of typical decoherence models the asymptotic optimal cost
 in quantum metrological protocols will scale as $1/n$ in a similarly manner as in the multi-copy (independent) setups and hence some of the claims proven in the latter case generalize to theses models as well. This includes asymptotic saturability as well as asymptotic equivalence between Bayesian and frequentist approaches \cite{Jarzyna2015, Yan2018}.
 The optimal states require only short-range entanglement structures and can be to some extent regarded as close to i.i.d. models. In particular they may be effectively described using the matrix product states formalism \cite{Jarzyna2013, Chabuda2020}.
 Interestingly, matrix product states, are also closely related with the input-output formalism describing the Markov evolution of an open system interacting with a bath modelled as quantum Bosonic noise.
 The  quantum Fisher information of the output process has been studied in \cite{Molmer2014,GutaGarrahan} and the QLAN and information geometry theory were established in \cite{GutaKiukas1,GutaKiukas2,GutaBoutenCatana}.

 On the other hand, for a special class of quantum metrological models involving unitary parameter estimation (e.g. phase), where the effect of noise  can be either neglected or  effectively mitigated via application of e.g. quantum error correction techniques
 \cite{Kessler2014a, Arrad2014, Dur2014, Demkowicz2017, sekatski2017quantum, zhou2018achieving, Layden2019, Gorecki2019},
  one may in principle reach the Heisenberg scaling where the effective quadratic cost scales as $1/n^2$.
  In this case, the relevant states and results are far from the i.i.d. setup and there is even no guarantee of  asymptotic saturability of the SLD CR bound in the single parameter case. In fact, in an effectively noiseless unitary single-parameter estimation model there is a $\pi$ constant factor discrepancy between asymptotically achievable Bayesian cost and the frequentist bound \cite{Gorecki2019a}.

 Very few of the above mentioned methods and results in quantum metrology have been satisfactorily generalized to multi-parameter scenarios.
There are cases, where the character of the problem allows for a direct generalization of single-parameter metrological bounds \cite{Tsang2011, Ragy2016, Baumgratz2016, Vidrighin2014, Pezze2017, Gessner2018}  but in general such a procedure will typically lead to loose bounds that do not account for measurement incompatibility as well as trade-offs between the probes states optimally sensing various parameters.
An example when such a single-to-multiple parameter generalization has been succesfully realized is the generaliztion of quantum error-correction schemes that yield optimal quantum metrological protocols provided the character of decoherence  allows for the Heisenberg scaling to be
preserved \cite{Gorecki2019}.
The open question is whether the same can be done for  generic multi-parameter metrological models where the character of decoherence allows only for a constant factor improvement over the i.i.d. scenarios. We speculate that that  matrix product states and QLAN may offer some deeper insight into the structure of optimal probe states and the resulting achievable precision. The other open avenue, which is challenging already on the single parameter level, is to develop efficient methods to deal with quantum metrological models involving spatially or temporally correlated noise---some single parameter models of this kind have been analyzed in the literature \cite{Jeske2014, Frowis2014, Layden2019, Chabuda2020}, but this research is far from complete, not to mention its generalization to multi-parameter case.

Let us finish this review with a more lightweight and a slightly philosophical remark. There is an anecdote related with the Bohr's obsessive use of the notion of complementarity in quantum mechanics. At some point, von Neumann made a remark
wondering why Bohr keeps on talking about \emph{two} non-commuting variables, saying:
 `Well, there are many things which do not commute and you can  easily find three operators which do not commute' \cite{Jammer1966}.
 In light of asymptotic normality, we see that maybe Bohr was right after all! (at least asymptotically).
Indeed in the limit of large ensembles, local transformations of quantum states may be equivalently viewed as either being classical (change in eigenvalues) or generated by complementary observable equivalent to position and momentum operators of quantum harmonic oscillators.
We may just wonder, whether this insight has some deeper implications for our understanding of quantum mechanics as a whole as well as the problem of quantum-to-classical transition.

\subsection*{Acknowledgements}
We thank Francesco Albarelli and Richard D. Gill for fruitful discussions. WG and RDD acknowledge support from the National Science Center (Poland) grant No.\ 2016/22/E/ST2/00559.

\bibliographystyle{iopart-num}
\providecommand{\newblock}{}

\end{document}